\documentclass[aps,prd,twocolumn,showkeys,amsmath,amssymb]{revtex4}
\usepackage{graphicx}
\usepackage{epstopdf}   
\usepackage{multirow}
\usepackage{subfigure}
\usepackage{extarrows}
\usepackage{feynmf}
\usepackage[colorlinks,citecolor=blue,anchorcolor=red,menucolor=red,linkcolor=red,filecolor=red,runcolor=red,urlcolor=blue,frenchlinks=red]{hyperref}

\newcommand{\feynp}[1]{#1\kern-0.45em/}
\def\dab{\int^{\alpha_{max}}_{\alpha_{min}}d\alpha\int^{\beta_{max}}_{\beta_{min}}d\beta}
\def\qq{\langle\bar qq\rangle}
\def\GGa{\langle GG\rangle}
\def\GGb{\langle g_s^2GG\rangle}

\def\qGqa{\langle\bar qGq\rangle}
\def\qGqb{\langle g_s\bar q\sigma Gq\rangle}
\def\FF(s){\left[(\alpha+\beta)m_c^2-\alpha\beta s\right]}
\def\HH(s){\left[m_c^2-\alpha(1-\alpha) s\right]}
\def\non{\\ \nonumber}

\allowdisplaybreaks[3]

\begin{document}

\title{Hidden-charm pentaquark states through the current algebra:\\ From their productions to decays}

\author{Hua-Xing Chen}
\email{hxchen@seu.edu.cn}
\affiliation{
School of Physics, Southeast University, Nanjing 210094, China
}

\begin{abstract}
There may exist seven $\bar D^{(*)} \Sigma_c^{(*)}$ hadronic molecular states. We construct their corresponding interpolating currents, and calculate their masses and decay constants using QCD sum rules. Based on these results, we calculate their relative production rates in $\Lambda_b^0$ decays through the current algebra, {\it i.e.}, $\mathcal{B}(\Lambda_b^0 \to P_c K^-):\mathcal{B}(\Lambda_b^0 \to P_c^\prime K^-)$ with $P_c$ and $P_c^\prime$ two different states. We also study their decay properties through the Fierz rearrangement, and further calculate these ratios in the $J/\psi p$ mass spectrum, {\it i.e.}, $\mathcal{B}(\Lambda_b^0 \to P_c K^- \to J/\psi p K^-):\mathcal{B}(\Lambda_b^0 \to P_c^\prime K^- \to J/\psi p K^-)$. Our results suggest that the $\bar D^{*} \Sigma_c^{*}$ molecular states of $J^P = 1/2^-$ and $3/2^-$ are possible to be observed in future experiments.
\end{abstract}

\maketitle

\section{Introduction}
\label{sec:intro}

Since the discovery of the $X(3872)$ by Belle in 2003~\cite{Choi:2003ue}, many charmonium-like $XYZ$ states were discovered in the past two decades~\cite{pdg}. Some of these structures may contain four quarks, $\bar c c \bar q q$ ($q=u/d$), so they are good candidates of hidden-charm tetraquark states.

In recent years the LHCb Collaboration continuously observed as many as five interesting exotic structures:
\begin{itemize}

\item In 2015 LHCb observed two structures $P_c(4380)^+$ and $P_c(4450)^+$ in the $J/\psi p$ invariant mass spectrum of the $\Lambda_b^0\to J/\psi p K^-$ decays~\cite{Aaij:2015tga}:
\begin{eqnarray}
P_c(4380)^+:    &M=& 4380 \pm 8 \pm 29 \mbox{ MeV} \, ,
\\ \nonumber                  &\Gamma=& 205 \pm 18 \pm 86  \mbox{ MeV} \, ,
\\[1mm]            P_c(4450)^+:    &M=& 4449.8 \pm 1.7 \pm 2.5  \mbox{ MeV} \, ,
\\ \nonumber                  &\Gamma=&   39 \pm 5 \pm 19 \mbox{ MeV} \, .
\end{eqnarray}
This observation is supported by the later LHCb experiment investigating the $J/\psi p$ invariant mass spectrum of the $\Lambda_b^0\to J/\psi p \pi^-$ decays~\cite{Aaij:2016ymb}.

\item In 2019 LHCb observed a new structure $P_c(4312)^+$ and further separated the $P_c(4450)^+$ into two substructures $P_c(4440)^+$ and $P_c(4457)^+$, still in the $J/\psi p$ invariant mass spectrum of the $\Lambda_b^0\to J/\psi p K^-$ decays~\cite{Aaij:2019vzc}:
\begin{eqnarray}
     P_c(4312)^+:    &M=& 4311.9 \pm  0.7 ^{+6.8}_{-0.6}  \mbox{ MeV} \, ,
\\ \nonumber                  &\Gamma=&    9.8 \pm  2.7 ^{+3.7}_{-4.5}  \mbox{ MeV} \, ,
\\[1mm]            P_c(4440)^+:    &M=& 4440.3 \pm  1.3 ^{+4.1}_{-4.7}  \mbox{ MeV} \, ,
\\ \nonumber                  &\Gamma=&   20.6 \pm  4.9 ^{+8.7}_{-10.1} \mbox{ MeV} \, ,
\\[1mm]  P_c(4457)^+:    &M=& 4457.3 \pm  0.6 ^{+4.1}_{-1.7}  \mbox{ MeV} \, ,
\\ \nonumber                  &\Gamma=&    6.4 \pm  2.0 ^{+5.7}_{-1.9}  \mbox{ MeV} \, .
\end{eqnarray}

\item In 2020 LHCb reported the evidence of a hidden-charm pentaquark state with strangeness, $P_{cs}(4459)^0$, in the $J/\psi \Lambda$ invariant mass spectrum of the $\Xi_b^- \to J/\psi \Lambda K^-$ decays~\cite{lhcb}:
\begin{eqnarray}
P_{cs}(4459)^0:    &M=& 4458.8 \pm  2.9^{+4.7}_{-1.1}  \mbox{ MeV} \, ,
\\ \nonumber                &\Gamma=& 17.3 \pm  6.5^{+8.0}_{-5.7}    \mbox{ MeV} \, .
\end{eqnarray}

\end{itemize}
These structures contain at least five quarks, $\bar c c u u d$ or $\bar c c u d s$, so they are perfect candidates of hidden-charm pentaquark states. The charmonium-like $XYZ$ and hidden-charm pentaquark states have attracted lots of attentions, and their studies have significantly improved our understanding of the non-perturbative behaviors of the strong interaction at the low energy region~\cite{Chen:2016qju,Liu:2019zoy,Chen:2022asf,Lebed:2016hpi,Esposito:2016noz,Guo:2017jvc,Ali:2017jda,Olsen:2017bmm,Karliner:2017qhf,Brambilla:2019esw,Guo:2019twa,Meng:2022ozq}.

To understand the above $P_c$ and $P_{cs}$ states, various theoretical interpretations were proposed, such as loosely-bound hadronic molecular states~\cite{Chen:2019bip,Chen:2019asm,Liu:2019tjn,He:2019ify,Huang:2019jlf,Guo:2019kdc,Fernandez-Ramirez:2019koa,Xiao:2019aya,Meng:2019ilv,Wu:2019adv,Yamaguchi:2019seo,Valderrama:2019chc,Liu:2019zvb,Burns:2019iih,Wang:2019ato,Gutsche:2019mkg,Du:2019pij,Xiao:2020frg,Ozdem:2018qeh,Wang:2019hyc,Zhang:2019xtu,Azizi:2016dhy}, tightly-bound compact pentaquark states~\cite{Maiani:2015vwa,Lebed:2015tna,Stancu:2019qga,Giron:2019bcs,Ali:2019npk,Weng:2019ynv,Eides:2019tgv,Cheng:2019obk,Ali:2019clg,Wang:2019got,Wang:2020eep}, and kinematical effects~\cite{Guo:2015umn,Liu:2015fea,Bayar:2016ftu,Kuang:2020bnk}, etc. Especially, the three narrow states $P_c(4312)^+$, $P_c(4440)^+$, and $P_c(4457)^+$ are just below the $\bar D \Sigma_c$ and $\bar D^{*} \Sigma_c$ thresholds, so it is natural to explain them as the $\bar D^{(*)} \Sigma_c$ hadronic molecular states, whose existence had been predicted in Refs.~\cite{Wu:2010jy,Wang:2011rga,Yang:2011wz,Karliner:2015ina,Wu:2012md} before the LHCb experiment performed in 2015~\cite{Aaij:2015tga}; the other narrow state $P_{cs}(4459)^0$ is just below the $\bar D^{*} \Xi_c$ threshold, so it is natural to explain it as the $\bar D^{*} \Xi_c$ molecular state~\cite{Chen:2020uif,Peng:2020hql}.

However, these exotic structures were only observed by LHCb~\cite{Aaij:2015tga,Aaij:2016ymb,Aaij:2019vzc,lhcb}. It is crucial to search for their partner states as well as some other potential decay channels, in order to further understand their nature. There have been some theoretical studies on this subject, using effective approaches~\cite{Guo:2019fdo,Xiao:2019mst,Cao:2019kst,Lin:2019qiv}, the quark interchange model~\cite{Wang:2019spc,Dong:2020nwk}, the heavy quark symmetry~\cite{Voloshin:2019aut,Sakai:2019qph}, and QCD sum rules~\cite{Xu:2019zme}, etc. We refer to reviews~\cite{Chen:2016qju,Liu:2019zoy,Chen:2022asf,Lebed:2016hpi,Esposito:2016noz,Guo:2017jvc,Ali:2017jda,Olsen:2017bmm,Karliner:2017qhf,Brambilla:2019esw,Guo:2019twa,Meng:2022ozq} and references therein for detailed discussions.

In this paper we shall systematically investigate hidden-charm pentaquark states as $\bar D^{(*)} \Sigma_c^{(*)}$ hadronic molecular states through their corresponding hidden-charm pentaquark interpolating currents. We shall systemically construct all the relevant currents, and apply the method of QCD sum rules to calculate their masses and decay constants. The obtained results will be used to further study their production and decay properties.

Our strategy is quite straightforward. Firstly, we construct a hidden-charm pentaquark current, such as
\begin{eqnarray}
\sqrt2 \xi_1(x) &=& [\delta^{ab} \bar c_a(x) \gamma_5 d_b(x)]
\\ \nonumber && ~~~~~ \times [\epsilon^{cde} u_c^T(x) \mathbb{C} \gamma_\mu u_d(x) \gamma^\mu \gamma_5 c_e(x)] \, ,
\end{eqnarray}
where $a \cdots e$ are color indices. It is the current best coupling to the $D^- \Sigma_c^{++}$ molecular state of $J^P = 1/2^-$, through
\begin{equation}
\langle 0 | \xi_1 | D^- \Sigma_c^{++}; 1/2^-(q) \rangle = f_1 u(q) \, ,
\end{equation}
where $u(q)$ is the Dirac spinor of $| D^- \Sigma_c^{++}; 1/2^- \rangle$. Its decay constant $f_1$ can be calculated using QCD sum rules.

Secondly, we investigate the three-body $\Lambda_b^0 \to J/\psi p K^-$ decays. The total quark content of the final states is $udc \bar c s \bar u u$, where the intermediate states $D^{(*)-} \Sigma_c^{(*)++} K^-$ can be produced. We apply the Fierz rearrangement to carefully examine the combination of these seven quarks, from which we select the current $\xi_1$ and evaluate the relative production rate of $| D^- \Sigma_c^{++}; 1/2^- \rangle$.

Thirdly, we apply the Fierz rearrangement of the Dirac and color indices to transform the current $\xi_1$ into
\begin{equation}
\sqrt2 \xi_1 \rightarrow {1\over6} ~ [\bar c_a \gamma_5 c_a]~N
- {1\over12} ~ [\bar c_a \gamma_\mu c_a]~\gamma^\mu \gamma_5 N + \cdots \, ,
\end{equation}
where $N = \epsilon^{abc} (u_a^T \mathbb{C} d_b) \gamma_5 u_c - \epsilon^{abc} (u_a^T \mathbb{C} \gamma_5 d_b) u_c$ is the Ioffe's light baryon field well coupling to the proton~\cite{Ioffe:1981kw,Ioffe:1982ce,Espriu:1983hu}. Accordingly, $\xi_1$ couples to the $\eta_c p$ and $J/\psi p$ channels simultaneously:
\begin{eqnarray}
\langle 0 | \xi_1 | \eta_c p \rangle &\approx& {\sqrt2\over12} \langle 0 | \bar c_{a} \gamma_5 c_a | \eta_c \rangle~\langle 0 | N | p \rangle + \cdots \, ,
\\ \nonumber \langle 0 | \xi_1 | \psi p \rangle &\approx& -{\sqrt2\over24} \langle 0 | \bar c_{a} \gamma_\mu c_a | \psi \rangle~\gamma^\mu \gamma_5 \langle 0 | N | p \rangle + \cdots \, .
\end{eqnarray}
We can use these two equations to straightforwardly calculate the relative branching ratio of the $| D^- \Sigma_c^{++}; 1/2^- \rangle$ decay into $\eta_c p$ to its decay into $J/\psi p$~\cite{Yu:2017zst}. We refer to Ref.~\cite{Chen:2020pac} for detailed discussions. There we have applied the same method to study decay properties of $P_c(4312)^+$, $P_c(4440)^+$, and $P_c(4457)^+$ as $\bar D^{(*)} \Sigma_c$ molecular states, and in this paper we shall apply it to study decay properties of $\bar D^{(*)} \Sigma_c^*$ molecular states.

This paper is organized as follows. In Sec.~\ref{sec:current} we systematically construct the hidden-charm pentaquark currents corresponding to $\bar D^{(*)} \Sigma_c^{(*)}$ hadronic molecular states. We use them to perform QCD sum rule analyses in Sec.~\ref{sec:sumrule}, and calculate their masses and decay constants. The obtained results are used in Sec.~\ref{sec:production} to study their productions in $\Lambda_b^0$ decays through the current algebra. In Sec.~\ref{sec:decay} we use the Fierz rearrangement of the Dirac and color indices to study decay properties of $\bar D^{(*)} \Sigma_c^{*}$ molecular states, and calculate some of their relative branching ratios. The obtained results are summarized and discussed in Sec.~\ref{sec:summary}.

\section{Hidden-charm pentaquark interpolating currents}
\label{sec:current}

%
\begin{figure*}[hbt]
\begin{center}
\subfigure[~\mbox{$\theta(x)$ currents: $[\bar c c][u u d]$}]{\includegraphics[width=0.3\textwidth]{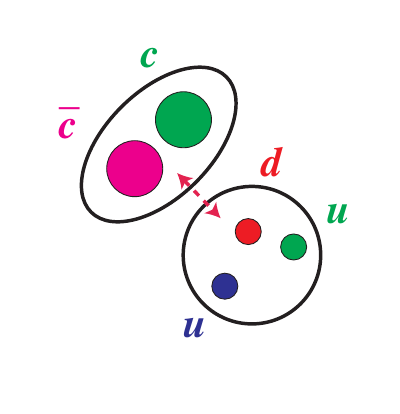}}
~~~~~~
\subfigure[~\mbox{$\eta(x)$ currents: $[\bar c u][u d c]$}]{\includegraphics[width=0.3\textwidth]{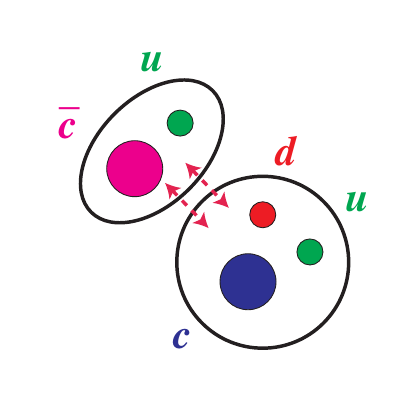}}
~~~~~~
\subfigure[~\mbox{$\xi(x)$ currents: $[\bar c d][u u c]$}]{\includegraphics[width=0.3\textwidth]{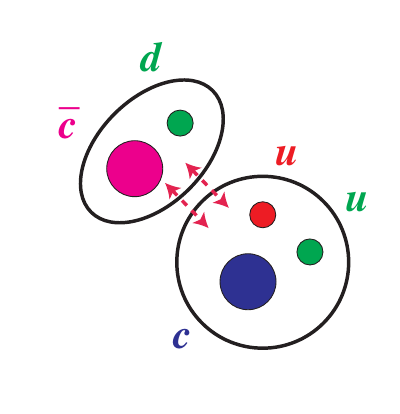}}
\caption{Three types of hidden-charm pentaquark interpolating currents, that are the $\theta(x)$, $\eta(x)$, and $\xi(x)$ currents. Quarks are shown in red/green/blue color, and antiquarks are shown in cyan/magenta/yellow color. Taken from Ref.~\cite{Chen:2020pac}.}
\label{fig:current}
\end{center}
\end{figure*}
%

In this section we use the $\bar c$, $c$, $u$, $u$, and $d$ ($q=u/d$) quarks to construct hidden-charm pentaquark interpolating currents. We consider the following three types of currents:
\begin{eqnarray}
\nonumber    \theta(x) &=& [\bar c_a(x) \Gamma^\theta_1 c_b(x)] ~ \Big[[q^T_c(x) \mathbb{C} \Gamma^\theta_2 q_d(x)] ~ \Gamma^\theta_3 q_e(x)\Big] \, ,
\\ \nonumber \eta(x)   &=& [\bar c_a(x) \Gamma^\eta_1 u_b(x)]    ~ \Big[[u^T_c(x) \mathbb{C} \Gamma^\eta_2 d_d(x)]    ~ \Gamma^\eta_3 c_e(x)   \Big]   \, ,
\\ \nonumber \xi(x)    &=& [\bar c_a(x) \Gamma^\xi_1 d_b(x)]   ~ \Big[[u^T_c(x) \mathbb{C} \Gamma^\xi_2 u_d(x)]   ~ \Gamma^\xi_3 c_e(x)  \Big]    \, ,
\\
\end{eqnarray}
where $a \cdots e$ are color indices, $\Gamma_{1/2/3}^{\theta/\eta/\xi}$ are Dirac matrices, and $\mathbb{C} = i\gamma_2 \gamma_0$ is the charge-conjugation operator. We illustrate them in Fig.~\ref{fig:current}. These three configurations can be related together by the Fierz rearrangement in the Lorentz space as well as the color rearrangement:
\begin{equation}
\delta^{ab} \epsilon^{cde} = \delta^{ac} \epsilon^{bde} + \delta^{ad} \epsilon^{cbe} + \delta^{ae} \epsilon^{cdb} \, .
\label{eq:colorsinglet}
\end{equation}
This will be detailedly discussed in Sec.~\ref{sec:decay}. There we shall construct the $\theta(x)$ currents by combining charmonium operators and light baryon fields.

In this section we construct the $\eta(x)$ and $\xi(x)$ currents, and further use them to construct currents corresponding to $\bar D^{(*)} \Sigma_c^{(*)}$ hadronic molecular states. To do this we combine charmed meson operators and charmed baryon fields. There are five independent charmed meson operators:
\begin{eqnarray}
&\bar c_a q_a \, [0^+] \, , \, \bar c_a \gamma_5 q_a \, [0^-] \, ,&
\\ \nonumber &\bar c_a \gamma_\mu q_a \, [1^-] \, , \, \bar c_a \gamma_\mu \gamma_5 q_a \, [1^+] \, , \, \bar c_a \sigma_{\mu\nu} q_a \, [1^\pm] \, .&
\end{eqnarray}
Besides, there exists another one, $\bar c_d \sigma_{\mu\nu} \gamma_5 q_d$, but it is related to $\bar c_d \sigma_{\mu\nu} q_d$ through
\begin{equation}
\sigma_{\mu\nu} \gamma_5 = {i\over2} \epsilon_{\mu\nu\rho\sigma} \sigma^{\rho\sigma} \, .
\end{equation}
Especially, we need the $J^P = 0^-$ and $1^-$ ones to construct the $\eta(x)$ and $\xi(x)$ currents, which couple to the ground-state charmed mesons $\mathcal{D} = D/D^{*}$:
\begin{eqnarray}
J_{D} &=& \bar c_a \gamma_5 q_a \, ,
\\ \nonumber J_{D^{*}} &=& \bar c_a \gamma_\mu q_a \, .
\end{eqnarray}

Charmed baryon fields have been systematically constructed and studied in Refs.~\cite{Liu:2007fg,Chen:2017sci,Cui:2019dzj,Dmitrasinovic:2020wye} using the method of QCD sum rules~\cite{Shifman:1978bx,Reinders:1984sr} within the heavy quark effective theory~\cite{Grinstein:1990mj,Eichten:1989zv,Falk:1990yz}. In this paper we need the following charmed baryon fields $J_{\mathcal{B}}$, which couple to the ground-state charmed baryons $\mathcal{B} = \Lambda_c/\Sigma_c/\Sigma_c^{*}$:
\begin{eqnarray}
\nonumber J_{\Lambda_c^+} &=& \epsilon^{abc} [u_a^T \mathbb{C} \gamma_{5} d_b] c_c \, ,
\\ \nonumber
\sqrt2 J_{\Sigma_c^{++}} &=& \epsilon^{abc} [u_a^T \mathbb{C} \gamma_{\mu} u_b] \gamma^{\mu}\gamma_{5} c_c \, ,
\\ \nonumber
J_{\Sigma_c^{+}}  &=& \epsilon^{abc} [u_a^T \mathbb{C} \gamma_{\mu} d_b] \gamma^{\mu}\gamma_{5} c_c \, ,
\\
\sqrt2 J_{\Sigma_c^{0}}  &=& \epsilon^{abc} [d_a^T \mathbb{C} \gamma_{\mu} d_b] \gamma^{\mu}\gamma_{5} c_c \, ,
\label{eq:heavybaryon}
\\ \nonumber
\sqrt2 J^\alpha_{\Sigma_c^{*++}} &=& \epsilon^{abc} P_{3/2}^{\alpha\mu} [u_a^T \mathbb{C} \gamma_{\mu} u_b] c_c \, ,
\\ \nonumber
J^\alpha_{\Sigma_c^{*+}}  &=& \epsilon^{abc} P_{3/2}^{\alpha\mu} [u_a^T \mathbb{C} \gamma_{\mu} d_b] c_c \, ,
\\ \nonumber
\sqrt2 J^\alpha_{\Sigma_c^{*0}}  &=& \epsilon^{abc} P_{3/2}^{\alpha\mu} [d_a^T \mathbb{C} \gamma_{\mu} d_b] c_c \, .
\end{eqnarray}
Here $P_{3/2}^{\mu\nu}$ is the spin-3/2 projection operator
\begin{equation}
P_{3/2}^{\mu\nu} = g^{\mu\nu} - {1 \over 4} \gamma^\mu\gamma^\nu \, .
\end{equation}

In the molecular picture the $P_c(4312)^+$, $P_c(4440)^+$, and $P_c(4457)^+$ are usually interpreted as the $\bar D \Sigma_c$ and  $\bar D^* \Sigma_c$ hadronic molecular states~\cite{Wu:2012md,Chen:2019asm,Liu:2019tjn}. Their relevant currents have been constructed in Ref.~\cite{Chen:2020pac}. In this paper we further construct the $\bar D \Sigma_c^{*}$ and $\bar D^* \Sigma_c^{*}$ currents. We summarize all of them here, for completeness.

There can exist altogether seven $\bar D^{(*)} \Sigma_c^{(*)}$ hadronic molecular states, that are $\bar D \Sigma_c$ of $J^P = {1\over2}^-$, $\bar D^* \Sigma_c$ of $J^P = {1\over2}^-/{3\over2}^-$, $\bar D \Sigma_c^*$ of $J^P = {3\over2}^-$, and $\bar D^* \Sigma_c^*$ of $J^P = {1\over2}^-/{3\over2}^-/{5\over2}^-$:
\begin{eqnarray}
&& | \bar D \Sigma_c; {1/2}^- ; \theta \rangle
\label{def:molecule1}
\\ \nonumber && ~~~~ = \cos\theta~| \bar D^0 \Sigma_c^+;1/2^- \rangle + \sin\theta~| D^- \Sigma_c^{++};1/2^- \rangle \, ,
\\[1mm] && | \bar D^* \Sigma_c; {1/2}^- ; \theta \rangle
\label{def:molecule2}
\\ \nonumber && ~~~~ = \cos\theta~| \bar D^{*0} \Sigma_c^+;1/2^- \rangle + \sin\theta~| D^{*-} \Sigma_c^{++};1/2^- \rangle \, ,
\\[1mm] && | \bar D^* \Sigma_c; {3/2}^- ; \theta \rangle
\label{def:molecule3}
\\ \nonumber && ~~~~ = \cos\theta~| \bar D^{*0} \Sigma_c^+;3/2^- \rangle + \sin\theta~| D^{*-} \Sigma_c^{++};3/2^- \rangle \, ,
\\[1mm] && | \bar D \Sigma_c^*; {3/2}^- ; \theta \rangle
\label{def:molecule4}
\\ \nonumber && ~~~~ = \cos\theta~| \bar D^{0} \Sigma_c^{*+};3/2^- \rangle + \sin\theta~| D^{-} \Sigma_c^{*++};3/2^- \rangle \, ,
\\[1mm] && | \bar D^* \Sigma_c^*; {1/2}^- ; \theta \rangle
\label{def:molecule5}
\\ \nonumber && ~~~~ = \cos\theta~| \bar D^{*0} \Sigma_c^{*+};1/2^- \rangle + \sin\theta~| D^{*-} \Sigma_c^{*++};1/2^- \rangle \, ,
\\[1mm] && | \bar D^* \Sigma_c^*; {3/2}^- ; \theta \rangle
\label{def:molecule6}
\\ \nonumber && ~~~~ = \cos\theta~| \bar D^{*0} \Sigma_c^{*+};3/2^- \rangle + \sin\theta~| D^{*-} \Sigma_c^{*++};3/2^- \rangle \, ,
\\[1mm] && | \bar D^* \Sigma_c^*; {5/2}^- ; \theta \rangle
\label{def:molecule7}
\\ \nonumber && ~~~~ = \cos\theta~| \bar D^{*0} \Sigma_c^{*+};5/2^- \rangle + \sin\theta~| D^{*-} \Sigma_c^{*++};5/2^- \rangle \, ,
\end{eqnarray}
where $\theta$ is an isospin parameter, satisfying $\theta = -55^{\rm o}$ for $I=1/2$ and $\theta = 35^{\rm o}$ for $I=3/2$. In the present study we shall concentrate on the former $I=1/2$ states, so that we can simplify the notations to be
\begin{eqnarray}
&& | \bar D^{(*)} \Sigma_c^{(*)}; J^P \rangle
\label{def:molecule}
\\ \nonumber &=& {\sqrt{1/3}}~| \bar D^{(*)0} \Sigma_c^{(*)+}; J^P \rangle - {\sqrt{2/3}}~| D^{(*)-} \Sigma_c^{(*)++}; J^P \rangle .
\end{eqnarray}
Their relevant interpolating currents are:
\begin{equation}
J_i = \cos\theta~\eta_i + \sin\theta~\xi_i \, ,
\label{def:current}
\end{equation}
where
\begin{eqnarray}
\nonumber \eta_1 &=& [\delta^{ab} \bar c_a \gamma_5 u_b] ~ [\epsilon^{cde} u_c^T \mathbb{C} \gamma_\mu d_d \gamma^\mu \gamma_5 c_e]
\\ &=& \bar D^0 ~ \Sigma_c^+ \, ,
\label{def:eta1}
\\[1mm] \nonumber \eta_2 &=& [\delta^{ab} \bar c_a \gamma_\nu u_b] ~ \gamma^\nu \gamma_5 ~ [\epsilon^{cde} u_c^T \mathbb{C} \gamma_\mu d_d \gamma^\mu \gamma_5 c_e]
\\ &=& \bar D^{*0}_\nu ~ \gamma^\nu \gamma_5 ~ \Sigma_c^+ \, ,
\label{def:eta2}
\\[1mm] \nonumber \eta_3^\alpha &=& P_{3/2}^{\alpha\nu} ~ [\delta^{ab} \bar c_a \gamma_\nu u_b] ~ [\epsilon^{cde} u_c^T \mathbb{C} \gamma_\mu d_d \gamma^\mu \gamma_5 c_e]
\\ &=& P_{3/2}^{\alpha\nu} ~ \bar D^{*0}_\nu ~ \Sigma_c^+ \, ,
\label{def:eta3}
\\[1mm] \nonumber \eta_4^\alpha &=& [\delta^{ab} \bar c_a \gamma_5 u_b] ~ P_{3/2}^{\alpha\mu} [\epsilon^{cde} u_c^T \mathbb{C} \gamma_\mu d_d c_e]
\\ &=& \bar D^0 ~ \Sigma_c^{*+;\alpha} \, ,
\label{def:eta4}
\\[1mm] \nonumber \eta_5 &=& [\delta^{ab} \bar c_a \gamma_\nu u_b] ~ P_{3/2}^{\nu\mu} [\epsilon^{cde} u_c^T \mathbb{C} \gamma_\mu d_d c_e]
\\ &=& \bar D^{*0}_\nu ~ \Sigma_c^{*+;\nu} \, ,
\label{def:eta5}
\\[1mm] \nonumber \eta_6^{\alpha} &=& [\delta^{ab} \bar c_a \gamma_\nu u_b] ~ P_{3/2}^{\alpha\rho} ~ \gamma^\nu \gamma_5 ~ P^{3/2}_{\rho\mu} [\epsilon^{cde} u_c^T \mathbb{C} \gamma^\mu d_d c_e]
\\ &=& \bar D^{*0}_\nu ~ P_{3/2}^{\alpha\rho} ~  \gamma^\nu \gamma_5 ~ \Sigma_{c;\rho}^{*+} \, ,
\label{def:eta6}
\\[1mm] \nonumber \eta_7^{\alpha\beta} &=& P^{\alpha\beta,\nu\rho}_{5/2} ~ [\delta^{ab} \bar c_a \gamma_\nu u_b] ~ P^{3/2}_{\rho\mu} [\epsilon^{cde} u_c^T \mathbb{C} \gamma^\mu d_d c_e]
\\ &=& P^{\alpha\beta,\nu\rho}_{5/2} ~ \bar D^{*0}_\nu ~ \Sigma_{c;\rho}^{*+} \, ,
\label{def:eta7}
\end{eqnarray}
and
\begin{eqnarray}
\xi_1 \nonumber &=& {1\over\sqrt2} ~ [\delta^{ab} \bar c_a \gamma_5 d_b] ~ [\epsilon^{cde} u_c^T \mathbb{C} \gamma_\mu u_d \gamma^\mu \gamma_5 c_e]
\\ &=& D^- ~ \Sigma_c^{++} \, ,
\label{def:xi1}
\\[1mm] \xi_2 \nonumber &=& {1\over\sqrt2} ~ [\delta^{ab} \bar c_a \gamma_\nu d_b] ~ \gamma^\nu \gamma_5 ~ [\epsilon^{cde} u_c^T \mathbb{C} \gamma_\mu u_d \gamma^\mu \gamma_5 c_e]
\\ &=& D^{*-}_\nu ~ \gamma^\nu \gamma_5 ~ \Sigma_c^{++} \, ,
\label{def:xi2}
\\[1mm] \xi_3^\alpha \nonumber &=& {1\over\sqrt2} ~ P_{3/2}^{\alpha\nu} ~ [\delta^{ab} \bar c_a \gamma_\nu d_b] ~ [\epsilon^{cde} u_c^T \mathbb{C} \gamma_\mu u_d \gamma^\mu \gamma_5 c_e]
\\ &=& P_{3/2}^{\alpha\nu} ~ D^{*-}_\nu ~ \Sigma_c^{++} \, ,
\label{def:xi3}
\\[1mm] \nonumber \xi_4^\alpha &=& {1\over\sqrt2} ~ [\delta^{ab} \bar c_a \gamma_5 d_b] ~ P_{3/2}^{\alpha\mu} [\epsilon^{cde} u_c^T \mathbb{C} \gamma_\mu u_d c_e]
\\ &=& D^- ~ \Sigma_c^{*++;\alpha} \, ,
\label{def:xi4}
\\[1mm] \nonumber \xi_5 &=& {1\over\sqrt2} ~ [\delta^{ab} \bar c_a \gamma_\nu d_b] ~ P_{3/2}^{\nu\mu} [\epsilon^{cde} u_c^T \mathbb{C} \gamma_\mu u_d c_e]
\\ &=& D^{*-}_\nu ~ \Sigma_c^{*++;\nu} \, ,
\label{def:xi5}
\\[1mm] \nonumber \xi_6^{\alpha} &=& {1\over\sqrt2} ~ [\delta^{ab} \bar c_a \gamma_\nu d_b] ~ P_{3/2}^{\alpha\rho} \gamma^\nu \gamma_5 ~ P^{3/2}_{\rho\mu} [\epsilon^{cde} u_c^T \mathbb{C} \gamma^\mu u_d c_e]
\\ &=& D^{*-}_\nu ~ P_{3/2}^{\alpha\rho} ~  \gamma^\nu \gamma_5 ~ \Sigma_{c;\rho}^{*++} \, ,
\label{def:xi6}
\\[1mm] \nonumber \xi_7^{\alpha\beta} &=& {1\over\sqrt2} ~ P^{\alpha\beta,\nu\rho}_{5/2} ~ [\delta^{ab} \bar c_a \gamma_\nu d_b] ~ P^{3/2}_{\rho\mu} [\epsilon^{cde} u_c^T \mathbb{C} \gamma^\mu u_d c_e]
\\ &=& P^{\alpha\beta,\nu\rho}_{5/2} ~ D^{*-}_\nu ~ \Sigma_{c;\rho}^{*++} \, .
\label{def:xi7}
\end{eqnarray}
In the above expressions, we have used $\mathcal{D}$ and $\mathcal{B}$ to denote the charmed meson operators $J_{\mathcal{D}}$ and the charmed baryon fields $J_{\mathcal{B}}$ for simplicity; $P_{5/2}^{\mu\nu,\rho\sigma}$ is the spin-5/2 projection operator
\begin{eqnarray}
P_{5/2}^{\mu\nu,\rho\sigma} &=& {1\over2} g^{\mu\rho} g^{\nu\sigma} + {1\over2} g^{\mu\sigma} g^{\nu\rho} - {1 \over 6} g^{\mu\nu} g^{\rho\sigma}
\\ \nonumber && - {1 \over 12} g^{\mu\rho} \gamma^{\nu}\gamma^{\sigma} - {1 \over 12} g^{\mu\sigma} \gamma^{\nu}\gamma^{\rho}
\\ \nonumber && - {1 \over 12} g^{\nu\sigma} \gamma^{\mu}\gamma^{\rho} - {1 \over 12} g^{\nu\rho} \gamma^{\mu}\gamma^{\sigma} \, .
\end{eqnarray}

\section{Masses and decay constants through QCD sum rules}
\label{sec:sumrule}

In this section we use the method of QCD sum rules~\cite{Shifman:1978bx,Reinders:1984sr} to study $\bar D^{(*)} \Sigma_c^{(*)}$ molecular states through the currents $J_{1\cdots7}$, {\it i.e.}, $J_{1,2,5}$ of $J^P = 1/2^-$, $J^\alpha_{3,4,6}$ of $J^P = 3/2^-$, and $J_{7}^{\alpha\beta}$ of $J^P = 5/2^-$. We shall calculate their masses and decay constants, and the obtained results will be used in the next section to further calculate their relative production rates. Some of the calculations have been done in Refs.~\cite{Chen:2015moa,Chen:2016otp,Xiang:2017byz,Chen:2019bip}, and we refer to Refs.~\cite{Chen:2020uif,Wang:2019hyc,Zhang:2019xtu,Azizi:2016dhy,Wang:2019got} for more relevant QCD sum rule studies.

\subsection{Correlation functions}

We assume that the currents $J_{1\cdots7}$ couple to the $\bar D^{(*)} \Sigma_c^{(*)}$ molecular states $X_{1\cdots7}$ through
\begin{eqnarray}
\nonumber \langle 0 | J_{1,2,5} | X_{1,2,5}; 1/2^- \rangle &=& f_{X_{1,2,5}} u (p) \, ,
\\ \label{eq:gamma0} \langle 0 | J_{3,4,6}^\alpha | X_{3,4,6}; 3/2^- \rangle &=& f_{X_{3,4,6}} u^\alpha (p) \, ,
\\ \nonumber \langle 0 | J_{7}^{\alpha\beta} | X_{7}; 5/2^- \rangle &=& f_{X_7} u^{\alpha\beta} (p) \, ,
\end{eqnarray}
where $u(p)$, $u^\alpha(p)$, and $u^{\alpha\beta}(p)$ are spinors of the $X_{1\cdots7}$. The two-point correlation functions extracted from these currents can be written as:
\begin{eqnarray}
\nonumber \Pi_{1,2,5}\left(q^2\right) &=& i \int d^4x e^{iq\cdot x} \langle 0 | T\left[J_{1,2,5}(x) \bar J_{1,2,5}(0)\right] | 0 \rangle
\\ \label{pi:spin12} &=& (q\!\!\!\slash~ + M_{X_{1,2,5}}) ~ \Pi_{1,2,5}\left(q^2\right) \, ,
\\ \nonumber \Pi^{\alpha \alpha^\prime}_{3,4,6}\left(q^2\right) &=& i \int d^4x e^{iq\cdot x} \langle 0 | T\left[J^{\alpha}_{3,4,6}(x) \bar J^{\alpha^\prime}_{3,4,6}(0)\right] | 0 \rangle
\\ \label{pi:spin32} &=& \mathcal{G}_{3/2}^{\alpha \alpha^\prime} (q\!\!\!\slash~ + M_{X_{3,4,6}})~\Pi_{3,4,6}\left(q^2\right) \, ,
\\ \nonumber \Pi^{\alpha \beta,\alpha^\prime \beta^\prime}_7\left(q^2\right) &=& i \int d^4x e^{iq\cdot x} \langle 0 | T\left[J^{\alpha \beta}_7(x) \bar J^{\alpha^\prime \beta^\prime}_7(0)\right] | 0 \rangle
\\ \label{pi:spin52} &=& \mathcal{G}_{5/2}^{\alpha \beta,\alpha^\prime \beta^\prime} (q\!\!\!\slash~ + M_{X_7})~\Pi_7\left(q^2\right) \, ,
\end{eqnarray}
where $\mathcal{G}_{3/2}^{\mu\nu}$ and $\mathcal{G}_{5/2}^{\mu \nu,\rho \sigma}$ are coefficients of the spin-3/2 and spin-5/2 propagators, respectively:
\begin{eqnarray}
&& \mathcal{G}_{3/2}^{\mu\nu}(p)
\\ \nonumber &=& g^{\mu\nu} - {1\over3} \gamma^\mu \gamma^\nu - {p^\mu\gamma^\nu - p^\nu\gamma^{\mu} \over 3m} - {2p^{\mu}p^\nu \over 3m^2} \, ,
\\ && \mathcal{G}_{5/2}^{\mu \nu,\rho \sigma}(p)
\\ \nonumber &=& \frac{1}{2}(g^{\mu\rho}g^{\nu\sigma}+g^{\mu\sigma}g^{\nu\rho}) - \frac{1}{5}g^{\mu\nu}g^{\rho\sigma}
\\ \nonumber &-& \frac{1}{10}(g^{\mu\rho}\gamma^\nu\gamma^\sigma + g^{\mu\sigma}\gamma^\nu\gamma^\rho + g^{\nu\rho}\gamma^\mu\gamma^\sigma + g^{\nu\sigma}\gamma^\mu\gamma^\rho)
\label{propagator}
\\ \nonumber &+& \frac{1}{10m}\Bigl(g^{\mu\rho}(p^\nu\gamma^\sigma - p^\sigma\gamma^\nu) + g^{\mu\sigma}(p^\nu\gamma^\rho - p^\rho\gamma^\nu)
\\ \nonumber && ~~~~~~~ + g^{\nu\rho}(p^\mu\gamma^\sigma - p^\sigma\gamma^\mu) + g^{\nu\sigma}(p^\mu\gamma^\rho - p^\rho\gamma^\mu)\Bigr)
\\ \nonumber &+& \frac{1}{5m^2}(g^{\mu\nu}p^\rho p^\sigma + g^{\rho\sigma}p^\mu p^\nu)
\\ \nonumber &-& \frac{2}{5m^2}(g^{\mu\rho}p^\nu p^\sigma + g^{\mu\sigma}p^\nu p^\rho + g^{\nu\rho}p^\mu p^\sigma + g^{\nu\sigma}p^\mu p^\rho)
\\ \nonumber &+& \frac{1}{10m^2}\Bigl(\gamma^\mu p^\nu (\gamma^\rho p^\sigma + \gamma^\sigma p^\rho) + \gamma^\nu p^\mu (\gamma^\rho p^\sigma + \gamma^\sigma p^\rho)\Bigl)
\\ \nonumber &+& \frac{1}{5m^3}\Bigl(p^\rho p^\sigma (\gamma^\mu p^\nu  + \gamma^\nu p^\mu ) - p^\mu p^\nu (\gamma^\rho p^\sigma + \gamma^\sigma p^\rho ) \Bigl)
\\ \nonumber &+& \frac{2}{5m^4}p^\mu p^\nu p^\rho p^\sigma \, .
\end{eqnarray}
In the above expressions we have assumed that the states $X_{1\cdots7}$ have the same spin-parity quantum numbers as the currents $J_{1\cdots7}$, so that we can use the ``non-$\gamma_5$ coupling'' in Eqs.~(\ref{eq:gamma0}); while we need to use the ``$\gamma_5$ coupling'':
\begin{equation}
\langle 0 | J_{1\cdots7} | X_{1\cdots7}^\prime \rangle = f_{X^\prime_{1\cdots7}} \gamma_5 u (p) \, ,
\label{eq:gamma51}
\end{equation}
if the states $X^\prime_{1\cdots7}$ have the opposite parity from the currents $J_{1\cdots7}$; or we can use the partner currents $\gamma_5 J_{1\cdots7}$, which also have the opposite parity:
\begin{equation}
\langle 0 | \gamma_5 J_{1\cdots7} | X_{1\cdots7} \rangle = f_{X_{1\cdots7}} \gamma_5 u (p) \, .
\label{eq:gamma52}
\end{equation}
From Eqs.~(\ref{eq:gamma51}) and (\ref{eq:gamma52}) we can derive another ``non-$\gamma_5$ coupling'' between $\gamma_5 J_{1\cdots7}$ and $X^\prime_{1\cdots7}$:
\begin{equation}
\langle 0 | \gamma_5 J_{1\cdots7} | X_{1\cdots7}^\prime \rangle = f_{X^\prime_{1\cdots7}} u (p) \, .
\label{eq:gamma53}
\end{equation}
We refer to Refs.~\cite{Chung:1981cc,Jido:1996ia,Kondo:2005ur,Ohtani:2012ps} for detailed discussions.

The two-point correlation functions derived from Eqs.~(\ref{eq:gamma51}) and (\ref{eq:gamma52}) are similar to Eqs.~(\ref{pi:spin12})--(\ref{pi:spin52}), but just with $(q\!\!\!\slash~ + M_{X})$ replaced by $(- q\!\!\!\slash~ + M_{X})$. Based on this feature, we can extract parities of the $X_{1\cdots7}$: we use the terms proportional to $\bf 1$ to evaluate masses of the $X_{1\cdots7}$, which are then compared with those terms proportional to $q\!\!\!\slash~$ to extract their parities.

In QCD sum rule studies we need to calculate the two-point correlation function $\Pi\left(q^2\right)$ at both hadron and quark-gluon levels. At the hadron level, we use the dispersion relation to write it as
%
\begin{equation}
\Pi(q^2)={\frac{1}{\pi}}\int^\infty_{s_<}\frac{{\rm Im} \Pi(s)}{s-q^2-i\varepsilon}ds \, ,
\label{eq:disper}
\end{equation}
%
with $s_<$ the physical threshold. We define the imaginary part of the correlation function as the spectral density $\rho(s)$, which can be evaluated at the hadron level by inserting intermediate hadron states $\sum_n|n\rangle\langle n|$:
%
\begin{eqnarray}
\rho_{\rm phen}(s) &\equiv& {\rm Im}\Pi(s)/\pi
\label{eq:rho}
\\ \nonumber &=& \sum_n\delta(s-M^2_n)\langle 0|\eta|n\rangle\langle n|{\eta^\dagger}|0\rangle
\\ \nonumber &=& f_X^2\delta(s-m_X^2)+ \mbox{continuum}.
\end{eqnarray}
%
In the last step we adopt the usual parametrization of one-pole dominance for the ground state $X$ together with a continuum contribution.

At the quark-gluon level we calculate $\Pi\left(q^2\right)$ using the method of operator product expansion (OPE), and extract its corresponding spectral density $\rho_{\rm OPE}(s)$. After performing the Borel transformation at both hadron and quark-gluon levels, we approximate the continuum using the spectral density above a threshold value $s_0$ (quark-hadron duality), and arrive at the sum rule equation
%
\begin{equation}
\Pi(s_0, M_B^2) \equiv f^2_X e^{-M_X^2/M_B^2} = \int^{s_0}_{s_<} e^{-s/M_B^2}\rho_{\rm OPE}(s)ds \, .
\label{eq:borel}
\end{equation}
%
It can be used to further calculate $M_X$ and $f_X$ through
%
\begin{eqnarray}
M^2_X(s_0, M_B) &=& \frac{\int^{s_0}_{s_<} e^{-s/M_B^2}s\rho_{\rm OPE}(s)ds}{\int^{s_0}_{s_<} e^{-s/M_B^2}\rho_{\rm OPE}(s)ds} \, ,
\label{eq:mass}
\\ \nonumber
f_X^2(s_0, M_B) &=& e^{M_X^2(s_0, M_B) \over M_B^2} \int^{s_0}_{s_<} e^{-s/M_B^2}\rho_{\rm OPE}(s)ds \, .
\\
\label{eq:fx}
\end{eqnarray}
%

In the present study we have calculated OPEs at the leading order of $\alpha_s$ and up to the $D({\rm imension}) = 10$ terms, including the perturbative term, the charm quark mass, the quark condensate $\langle \bar q q \rangle$, the gluon condensate $\langle g_s^2 GG \rangle$, the quark-gluon mixed condensate $\langle g_s \bar q \sigma G q \rangle$, and their combinations $\langle \bar q q \rangle^2$, $\langle \bar q q \rangle\langle g_s \bar q \sigma G q \rangle$, $\langle \bar q q \rangle^3$, and $\langle g_s \bar q \sigma G q \rangle^2$. We summarized the obtained spectral densities $\rho_{1\cdots7}(s)$ in Appendix~\ref{app:ope}, which are extracted from the currents $J_{1\cdots7}$, respectively.

In the calculations we have ignored the chirally suppressed terms with the light quark mass, and adopt the factorization assumption of vacuum saturation for higher dimensional condensates, {\it i.e.}, $\langle (\bar q q)^2 \rangle = \langle \bar q q \rangle^2$, $\langle (\bar q q) (g_s \bar q \sigma G q) \rangle = \langle \bar q q \rangle\langle g_s \bar q \sigma G q \rangle$, $\langle (\bar q q)^3 \rangle = \langle \bar q q \rangle^3$, and $\langle (g_s \bar q \sigma G q)^2 \rangle = \langle g_s \bar q \sigma G q \rangle^2$. We find that the $D=3$ quark condensate $\qq$ and the $D=5$ mixed condensate $\langle g_s \bar q \sigma G q \rangle$ are both multiplied by the charm quark mass $m_c$, which are thus important power corrections.

In the next subsection we shall use the spectral densities $\rho_{1\cdots7}(s)$ to perform numerical analyses, and calculate masses and decay constants of the $X_{1\cdots7}$. Before doing this, let us investigate the current $J_1$ as an example. It has the quantum number $J^P = 1/2^-$ and couples to the $\bar D \Sigma_c$ molecular state $X_1$. Its spectral density $\rho_1(s)$ is given in Eqs.~(\ref{ope:J1}). We find that the terms timed by $m_c$ are almost positively proportional to those terms timed by $q\!\!\!\slash$. Hence, the parity of $X_1$ is extracted to be negative, that is the same as $J_1$; in other words, $J_1$ mainly couples to a negative-parity state. Similarly, all the $\bar D^{(*)} \Sigma_c^{(*)}$ molecular states defined in Eqs.~(\ref{def:molecule1}-\ref{def:molecule7}) are determined to have the negative
parity.

\subsection{Mass analyses}

In this subsection we use the spectral densities $\rho_{1\cdots7}(s)$ extracted from the currents $J_{1\cdots7}$ to perform numerical analyses, and calculate masses and decay constants of the $X_{1\cdots7}$. As discussed in the previous subsection, we only use the terms proportional to $m_c$ to do this.

We use the current $J_1$ as an example, whose spectral density $\rho_1(s)$ can be found in Eq.~(\ref{ope:J1}). We use the following values for various QCD sum rule parameters~\cite{Yang:1993bp,Eidemuller:2000rc,Narison:2002pw,Gimenez:2005nt,Jamin:2002ev,Ioffe:2002be,Ovchinnikov:1988gk,Ellis:1996xc,colangelo}:
\begin{eqnarray}
\nonumber m_c &=& 1.275 ^{+0.025}_{-0.035} \mbox{ GeV} \, ,
\\ \nonumber \langle \bar qq \rangle &=& - (0.24 \pm 0.01)^3 \mbox{ GeV}^3 \, ,
\\ \langle g_s^2GG\rangle &=& (0.48 \pm 0.14) \mbox{ GeV}^4\, ,
\label{paramaters}
\\ \nonumber \langle g_s \bar q \sigma G q \rangle &=& M_0^2 \times \langle \bar qq \rangle\, ,
\\ \nonumber M_0^2 &=& (0.8 \pm 0.2) \mbox{ GeV}^2 \, ,
\end{eqnarray}
where the running mass in the $\overline{MS}$ scheme is used for the charm quark.

\begin{figure}[hbt]
\begin{center}
\includegraphics[width=0.47\textwidth]{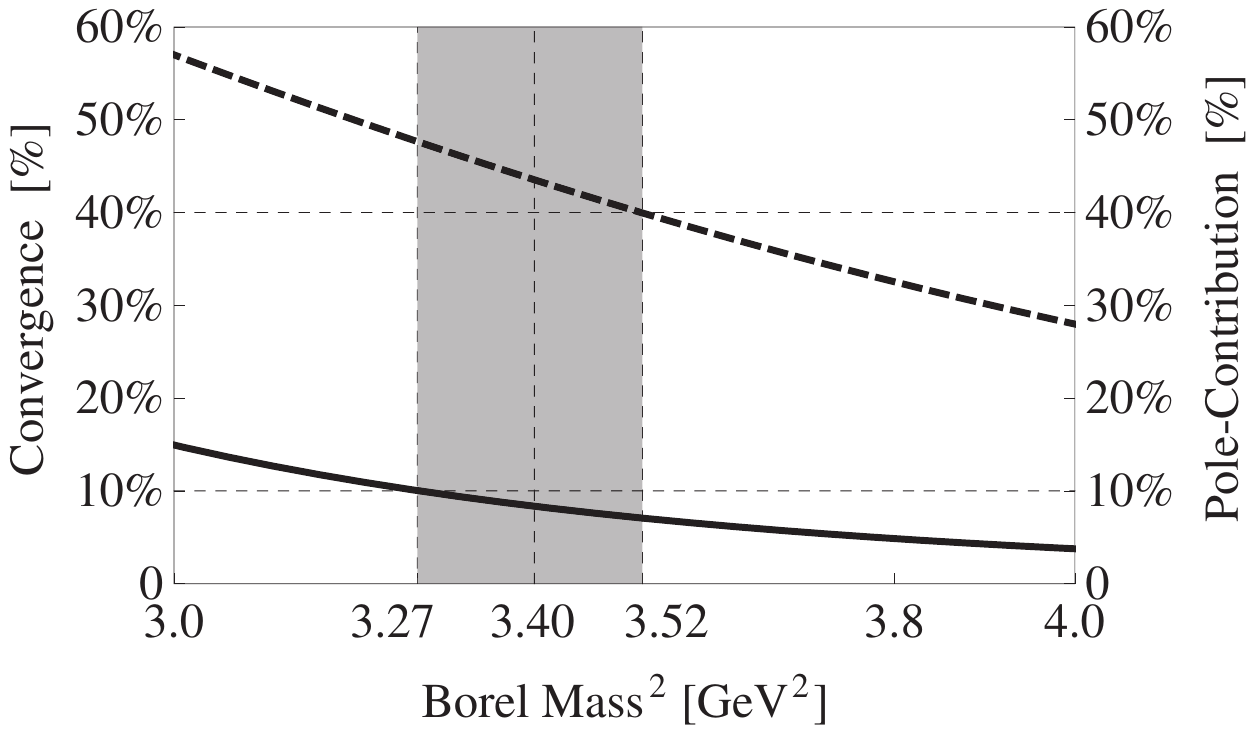}
\caption{Convergence (solid curve, defined in Eq.~(\ref{eq:cvg})) and Pole-Contribution (dashed curve, defined in Eq.~(\ref{eq:pole})) as functions of the Borel mass $M_B$. These curves are obtained using the current $J_1$ when setting $s_0 = 24$~GeV$^2$.}
\label{fig:cvgpole}
\end{center}
\end{figure}

There are two free parameters in Eqs.~(\ref{eq:mass}) and (\ref{eq:fx}): the Borel mass $M_B$ and the threshold value $s_0$. We use two criteria to constrain the Borel mass $M_B$ for a fixed $s_0$. The first criterion is to insure the convergence of OPE series. It is done by requiring the $D=10$ terms \big($m_c \langle \bar q q \rangle^3$ and $\langle g_s \bar q \sigma G q \rangle^2$\big) to be less than 10\%, so that the lower limit of $M_B$ can be determined:
%
\begin{equation}
\mbox{Convergence} \equiv \left|\frac{ \Pi^{D=10}(\infty, M_B) }{ \Pi(\infty, M_B) }\right| \leq 10\% \, .
\label{eq:cvg}
\end{equation}
%
We show this function in Fig.~\ref{fig:cvgpole} using the solid curve, and find that the OPE convergence improves with the increase of $M_B$. This criterion leads to $\left(M_B^{min}\right)^2 = 3.27$~GeV$^2$, when setting $s_0 = 24$~GeV$^2$.

The second criterion is to insure the validity of one-pole parametrization. It is done by requiring the pole contribution to be larger than 40\%, so that the upper limit of $M_B$ can be determined:
%
\begin{equation}
\mbox{Pole-Contribution} \equiv \frac{ \Pi(s_0, M_B) }{ \Pi(\infty, M_B) } \geq 40\% \, .
\label{eq:pole}
\end{equation}
%
We show this function in Fig.~\ref{fig:cvgpole} using the dashed curve, and find that it decreases with the increase of $M_B$. This criterion leads to $\left(M_B^{max}\right)^2 = 3.52$~GeV$^2$, when setting $s_0 = 24$~GeV$^2$.

\begin{figure*}[hbt]
\begin{center}
\scalebox{0.626}{\includegraphics{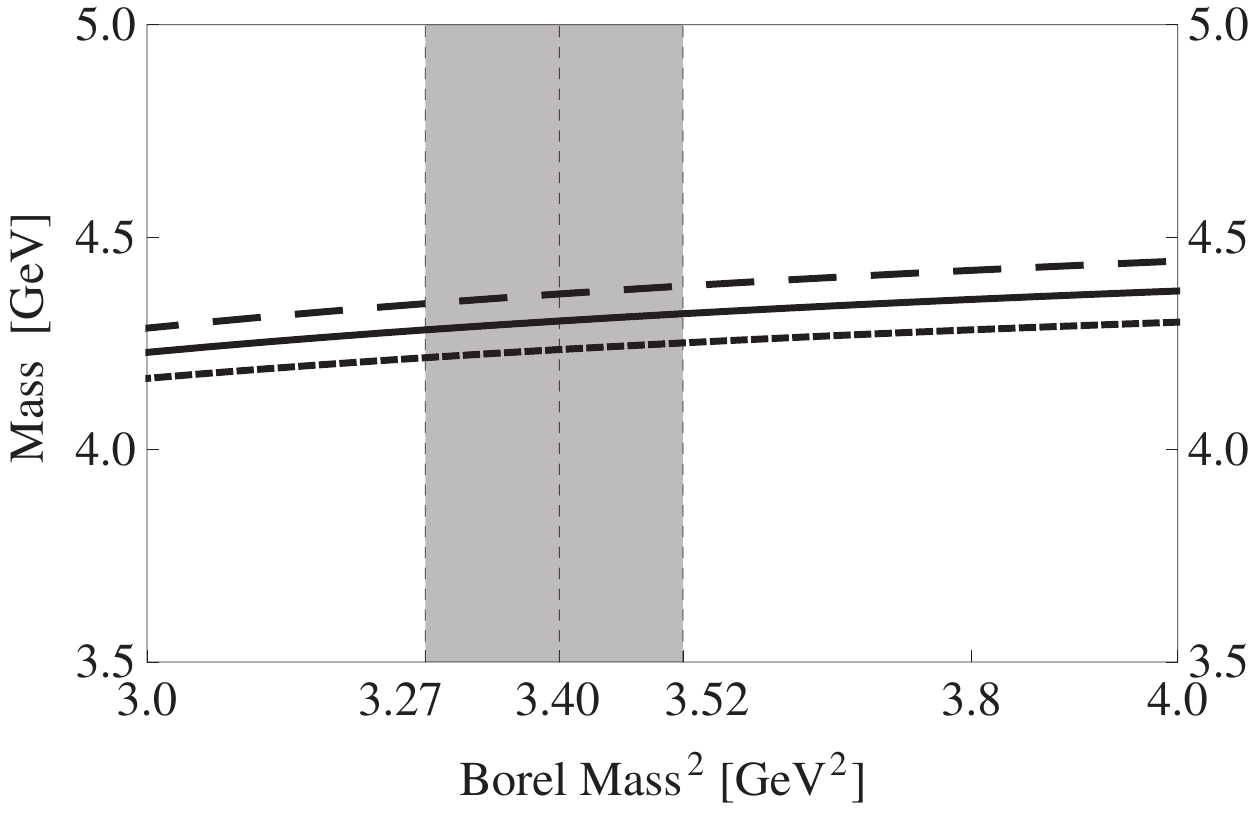}}
~~~
\scalebox{0.6}{\includegraphics{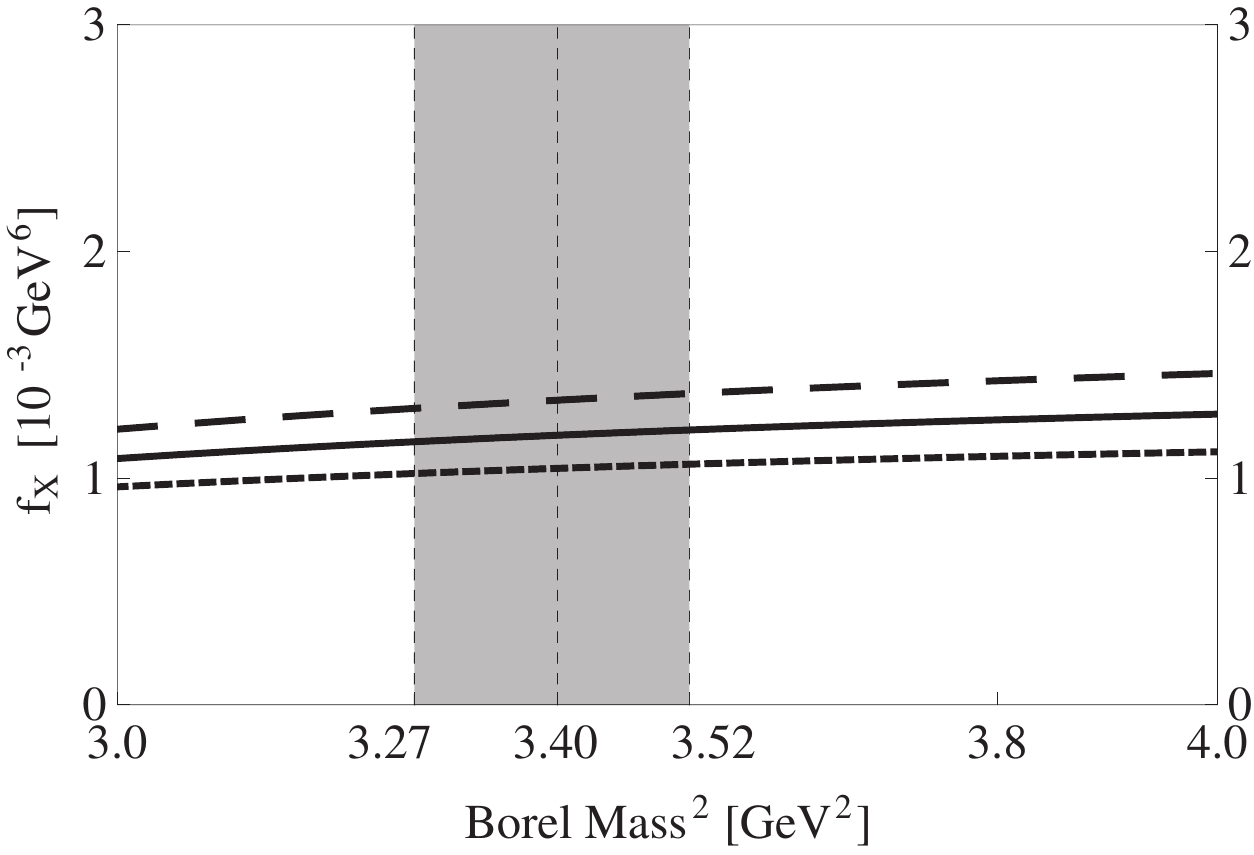}}
\caption{Variations of the mass $M_{X}$ (left) and the decay constant $f_{X}$ (right) with respect to the Borel mass $M_B$, calculated using the current $J_1$.
In both panels, the short-dashed, solid and long-dashed curves are obtained by setting $s_0 = 23$, $24$ and $25$ GeV$^2$, respectively.}
\label{fig:psi9}
\end{center}
\end{figure*}

Altogether we extract the working region of Borel mass to be $3.27$~GeV$^2< M_B^2 < 3.52$~GeV$^2$ for the current $J_1$ with the threshold value $s_0 = 24$~GeV$^2$. We show variations of $M_{X_1}$ and $f_{X_1}$ with respect to the Borel mass $M_B$ in Fig.~\ref{fig:psi9}. They are shown in a broader region $3.0$~GeV$^2\leq M_B^2 \leq 4.0$~GeV$^2$, and they are more stable inside the above Borel window.

Redoing the same procedures by changing $s_0$, we find that there exist non-vanishing Borel windows as long as $s_0 \geq s_0^{min} = 22.4$~GeV$^2$. Accordingly, we choose $s_0$ to be slightly larger with the uncertainty $\pm1.0$~GeV, that is $s_0 = 24.0 \pm 1.0$~GeV$^2$. Altogether our working regions for the current $J_1$ are determined to be $23.0$~GeV$^2\leq s_0\leq 25.0$~GeV$^2$ and $3.27$~GeV$^2\leq M_B^2 \leq 3.52$~GeV$^2$, where we calculate the mass and decay constant of $X_1$ to be:
\begin{eqnarray}
M_{X_1} &=& 4.30^{+0.10}_{-0.10} \mbox{ GeV} \, ,
\\ \nonumber f_{X_1} &=& \left(1.19^{+0.19}_{-0.18}\right) \times 10^{-3}  \mbox{ GeV}^6 \, .
\end{eqnarray}
Here the central values correspond to $M_B^2=3.40$ GeV$^2$ and $s_0 = 24.0$ GeV$^2$. Their uncertainties come from the threshold value $s_0$, the Borel mass $M_B$, the charm quark mass $m_c$, and various QCD sum rule parameters listed in Eqs.~(\ref{paramaters}). This mass value is consistent with the experimental mass of the $P_c(4312)^+$~\cite{Aaij:2019vzc}, supporting it to be the $I = 1/2$ $\bar D \Sigma_c$ molecular state of $J^P=1/2^-$.

\begin{table*}[hpt]
\begin{center}
\renewcommand{\arraystretch}{1.5}
\caption{Masses and decay constants of the $X_{1\cdots7}$, extracted from the currents $J_{1\cdots7}$.}
\begin{tabular}{c | c | c | c | c | c | c | c | c}
\hline\hline
\multirow{2}{*}{Currents} & \multirow{2}{*}{Configuration} & $s_0^{min}$ & \multicolumn{2}{c|}{Working Regions} & \multirow{2}{*}{Pole~[\%]} & \multirow{2}{*}{Mass~[GeV]} & \multirow{2}{*}{~~$f_X$~[GeV$^6$]~~} & \multirow{2}{*}{Candidate}
\\ \cline{4-5} & & [${\rm GeV}^2$] & $s_0~[{\rm GeV}^2]$ & $M_B^2~[{\rm GeV}^2]$ & & &
\\ \hline \hline
$J_1$ & $|\bar D \Sigma_c; 1/2^- \rangle$     & 22.4 & $24.0\pm1.0$ & $3.27$--$3.52$ & $40$--$48$ & $4.30^{+0.10}_{-0.10}$ & $\left(1.19^{+0.19}_{-0.18}\right) \times 10^{-3}$ & $P_c(4312)^+$
\\
$J_2$ & $|\bar D^* \Sigma_c; 1/2^- \rangle$   & 25.5 & $27.0\pm1.0$ & $3.78$--$3.99$ & $40$--$46$ & $4.48^{+0.10}_{-0.10}$ & $\left(2.24^{+0.34}_{-0.30}\right) \times 10^{-3}$ & $P_c(4457)^+$
\\
$J_3$ & $|\bar D^* \Sigma_c; 3/2^- \rangle$   & 24.6 & $26.0\pm1.0$ & $3.51$--$3.72$ & $40$--$46$ & $4.46^{+0.11}_{-0.10}$ & $\left(1.15^{+0.18}_{-0.16}\right) \times 10^{-3}$ & $P_c(4440)^+$
\\
$J_4$ & $|\bar D \Sigma_c^*; 3/2^- \rangle$   & 24.2 & $25.0\pm1.0$ & $3.33$--$3.45$ & $40$--$44$ & $4.43^{+0.10}_{-0.10}$ & $\left(0.65^{+0.11}_{-0.10}\right) \times 10^{-3}$
\\
$J_5$ & $|\bar D^* \Sigma_c^*; 1/2^- \rangle$ & 26.0 & $27.0\pm1.0$ & $3.43$--$3.56$ & $40$--$44$ & $4.51^{+0.10}_{-0.11}$ & $\left(1.12^{+0.19}_{-0.17}\right) \times 10^{-3}$
\\
$J_6$ & $|\bar D^* \Sigma_c^*; 3/2^- \rangle$ & 25.3 & $27.0\pm1.0$ & $3.69$--$3.98$ & $40$--$48$ & $4.52^{+0.11}_{-0.11}$ & $\left(0.85^{+0.14}_{-0.13}\right) \times 10^{-3}$
\\
$J_7$ & $|\bar D^* \Sigma_c^*; 5/2^- \rangle$ & 24.7 & $26.0\pm1.0$ & $3.22$--$3.42$ & $40$--$46$ & $4.55^{+0.15}_{-0.13}$ & $\left(0.65^{+0.11}_{-0.10}\right) \times 10^{-3}$
\\ \hline\hline
\end{tabular}
\label{tab:mass}
\end{center}
\end{table*}

Similarly, we use the spectral densities $\rho_{2\cdots7}(s)$ extracted from the currents $J_{2\cdots7}$ to perform numerical analyses, and calculate masses and decay constants of the $X_{2\cdots7}$. Especially, the sum rule results extracted from the currents $J_6^\alpha$ and $J_7^{\alpha\beta}$ are
\begin{eqnarray}
\nonumber M_{X_6} &=& 4.64^{+0.10}_{-0.10} \mbox{ GeV} \, ,
\\ f_{X_6} &=& \left(1.01^{+0.15}_{-0.14}\right) \times 10^{-3}  \mbox{ GeV}^6 \, ,
\\ \nonumber M_{X_7} &=& 4.64^{+0.14}_{-0.12} \mbox{ GeV} \, ,
\\ \nonumber f_{X_7} &=& \left(0.77^{+0.12}_{-0.11}\right) \times 10^{-3}  \mbox{ GeV}^6 \, .
\end{eqnarray}
These two mass values are both not far from the $\bar D^* \Sigma_c^*$ threshold at $M_{D^*} + M_{\Sigma_c^*} = 4527$~MeV, but a bit larger than that. To get a better description of $\bar D^* \Sigma_c^*$ molecular states that may lie just below the $\bar D^* \Sigma_c^*$ threshold, we slightly release the criterion given in Eq.~(\ref{eq:cvg}) to be
%
\begin{equation}
\mbox{Convergence} \equiv \left|\frac{ \Pi^{D=10}(\infty, M_B) }{ \Pi(\infty, M_B) }\right| \leq 15\% \, .
\label{eq:cvg2}
\end{equation}
%
Now the masses and decay constants extracted from the currents $J_6^\alpha$ and $J_7^{\alpha\beta}$ are modified to be
\begin{eqnarray}
\nonumber M^\prime_{X_6} &=& 4.52^{+0.11}_{-0.11} \mbox{ GeV} \, ,
\\ f^\prime_{X_6} &=& \left(0.85^{+0.14}_{-0.13}\right) \times 10^{-3}  \mbox{ GeV}^6 \, ,
\\ \nonumber M^\prime_{X_7} &=& 4.55^{+0.15}_{-0.13} \mbox{ GeV} \, ,
\\ \nonumber f^\prime_{X_7} &=& \left(0.65^{+0.11}_{-0.10}\right) \times 10^{-3}  \mbox{ GeV}^6 \, .
\end{eqnarray}
Besides, the mass of $|\bar D \Sigma_c^*; 3/2^- \rangle$ is calculated to be $4.43^{+0.10}_{-0.10}$~GeV, which is consistent with, but also a bit larger than, the $\bar D \Sigma_c^*$ threshold at $M_{D} + M_{\Sigma_c^*} = 4385$~MeV. All these divergences indicate that the accuracy of our QCD sum rule results is moderate but not good enough to extract the binding energies of the $\bar D^{(*)} \Sigma_c^{(*)}$ molecular states. Therefore, our results can only suggest but not determine: a) whether these $\bar D^{(*)} \Sigma_c^{(*)}$ molecular states exist or not, and b) whether they are bound states or resonance states. However, in the present study we are more concerned about the ratios, {\it i.e.}, the relative production rates and the relative branching ratios, whose uncertainties can be significantly reduced. Accordingly, the decay constants $f_X$ calculated in this section are input parameters that are more important than the masses $M_X$. Note that the decay constants $f_X$ can also be used within the QCD sum rule method to directly calculate the partial decay widths through the three-point correlation functions, but we shall not do this in the present study.

We summarized all the above sum rule results in Table~\ref{tab:mass}. Our results are consistent with those of Ref.~\cite{Wang:2022ltr}, where the authors applied the same QCD sum rule method to study both the $I=1/2$ $\bar D^{(*)} \Sigma_c^{(*)}$ molecular states as well as the $I=3/2$ ones. Our results support the interpretations of the $P_c(4440)^+$ and $P_c(4457)^+$~\cite{Aaij:2019vzc} as the $I = 1/2$ $\bar D^* \Sigma_c$ molecular states of $J^P=1/2^-$ and $3/2^-$. Again, the accuracy of our sum rule results is not good enough to distinguish/identify them. To better understand them, we move on to study their production and decay properties in the following sections, where we shall find that the $P_c(4440)^+$ and $P_c(4457)^+$ can be better interpreted in our framework as $|\bar D^{*} \Sigma_c; 3/2^- \rangle$ and $|\bar D^{*} \Sigma_c; 1/2^- \rangle$, respectively.

\section{Productions through the current algebra}
\label{sec:production}

In this section we study productions of $\bar D^{(*)} \Sigma_c^{(*)}$ molecular states in $\Lambda_b^0$ decays through the current algebra. We shall calculate their relative production rates, {\it i.e.}, $\mathcal{B}(\Lambda_b^0 \to P_c K^-):\mathcal{B}(\Lambda_b^0 \to P_c^\prime K^-)$ with $P_c$ and $P_c^\prime$ two different states. We refer to Refs.~\cite{Cheng:2015cca,Chen:2015sxa} for more relevant studies.

%
\begin{figure*}[hbt]
\begin{center}
\subfigure[~$\Lambda_b^0 \to D^{(*)-} \Sigma_c^{(*)++} K^-$]{\includegraphics[width=0.45\textwidth]{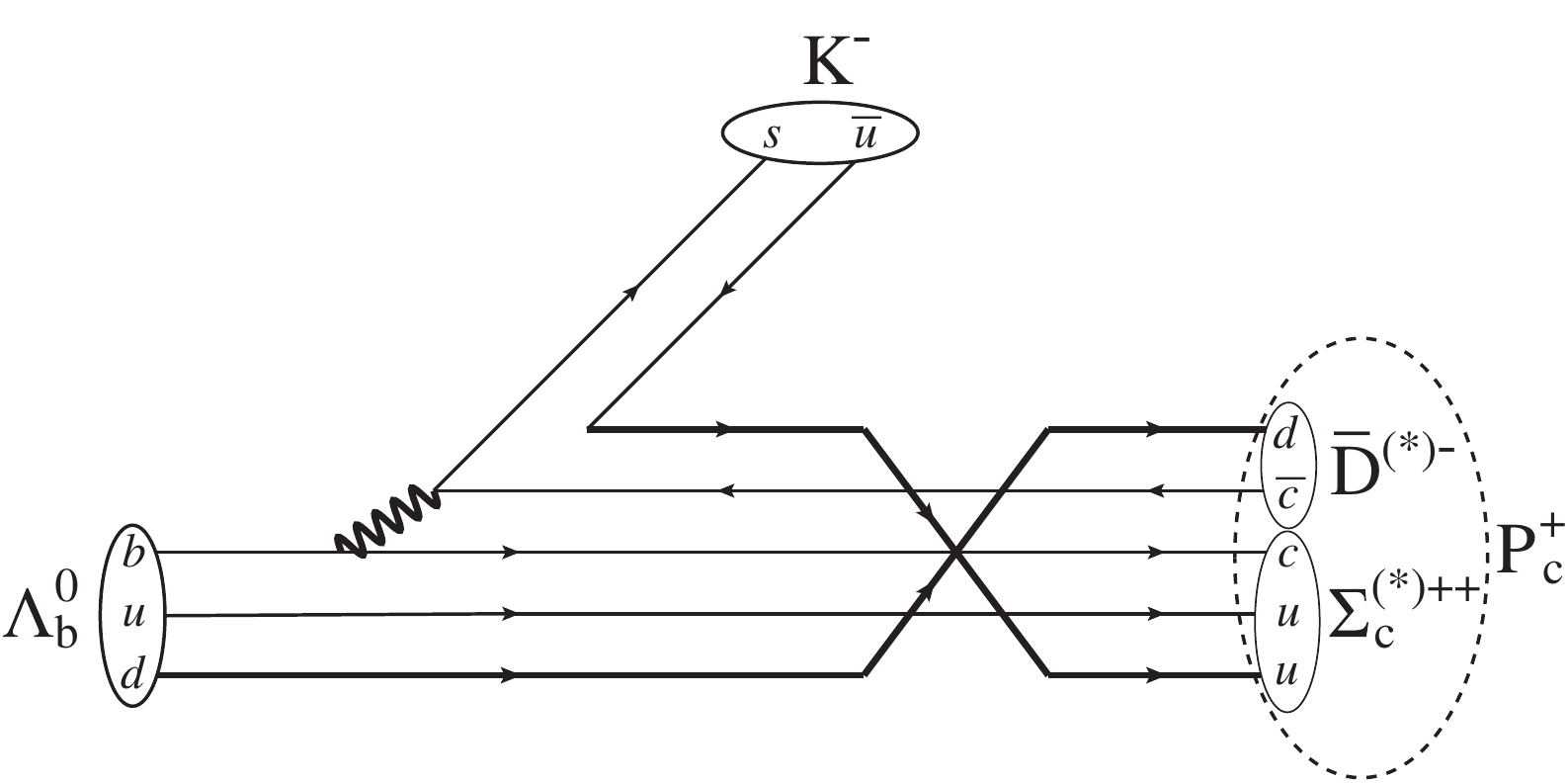}}
~~~~~~~~~
\subfigure[~$\Lambda_b^0 \to \bar D^{(*)0} \Sigma_c^{(*)+} K^-$]{\includegraphics[width=0.45\textwidth]{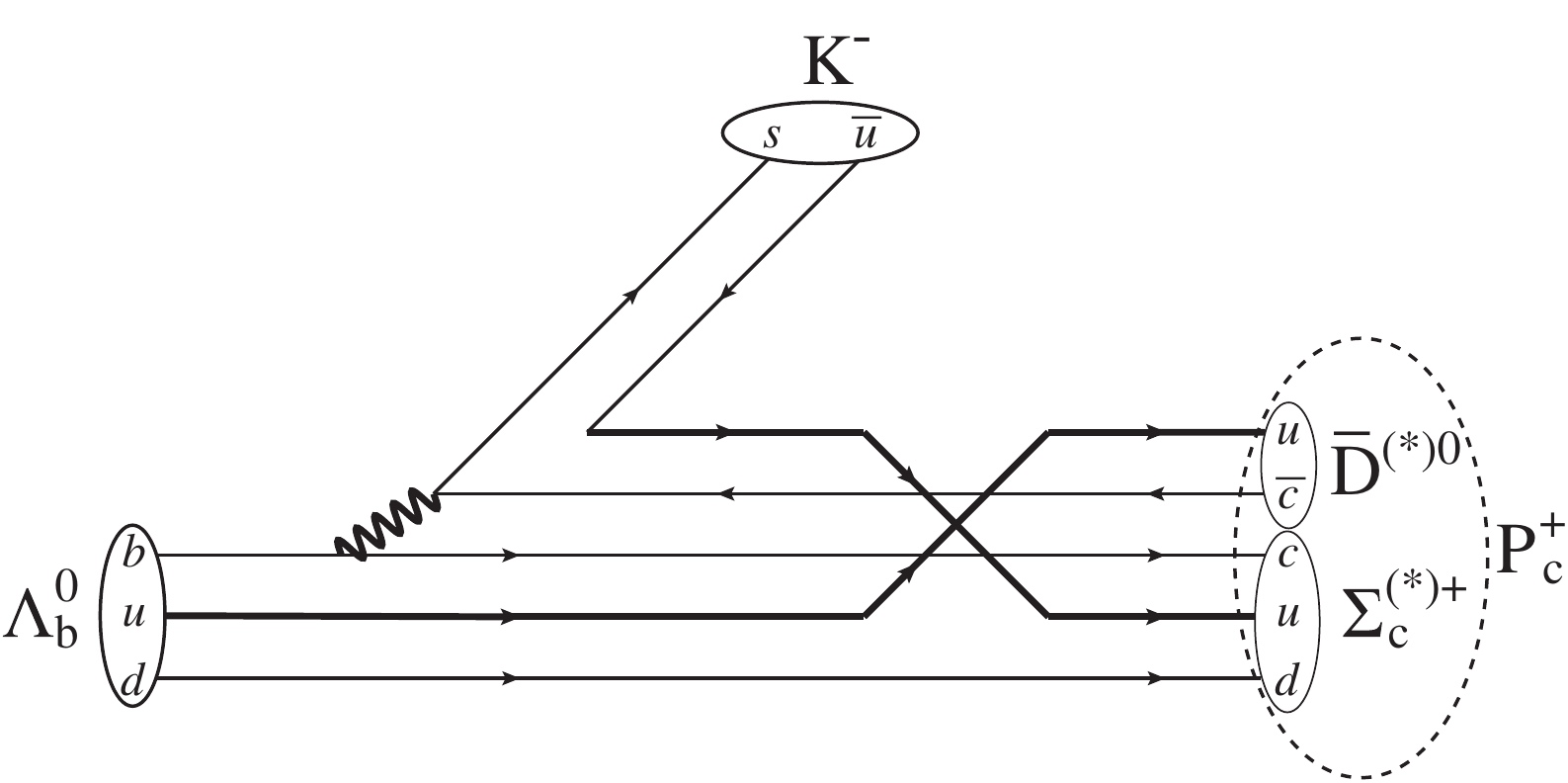}}
\caption{Production mechanisms of $\bar D^{(*)} \Sigma_c^{(*)}$ molecular states in $\Lambda_b^0$ decays.}
\label{fig:production}
\end{center}
\end{figure*}
%

The $P_c(4312)^+$, $P_c(4440)^+$, and $P_c(4457)^+$ were observed by LHCb in the $J/\psi p$ invariant mass spectrum of the $\Lambda_b^0 \to J/\psi p K^-$ decays. The quark content of the initial state $\Lambda_b^0$ is $udb$. In this three-body decay process, first the $b$ quark decays into a $c$ quark by emitting a $W^-$ boson, and the $W^-$ translates into a pair of $\bar c$ and $s$ quarks, both of which are Cabibbo-favored; then they pick up a pair of $\bar u$ and $u$ quarks from the vacuum; finally they hadronize into the three final states $J/\psi p K^-$:
\begin{equation}
\Lambda_b^0 = udb \to ud c~\bar c s \to udc~\bar c s~\bar u u \to J/\psi p K^-  \, .
\end{equation}
Hence, the total quark content of the final states is $udc \bar c s \bar u u$, where the intermediate states $D^{(*)-} \Sigma_c^{(*)++} K^-$ and $\bar D^{(*)0} \Sigma_c^{(*)+} K^-$ can also be produced.

In the present study we study productions of $\bar D^{(*)} \Sigma_c^{(*)}$ molecular states by investigating the mechanisms depicted in Fig.~\ref{fig:production}. Note that the $u$ quark from the vacuum needs to exchange with either the $u$ or $d$ quark of the $\Lambda_b^0$, because the $ud$ pair of the $\Lambda_b^0$ is in a state of $I=0$, while the $\Sigma_c$ and $\Sigma_c^{*}$ both have $I=1$.

As depicted in Fig.~\ref{fig:production}, the weak interaction only involves the initial $b$ quark and the final $c \bar c s$ quarks. Hence, considering the quark pair produced from the vacuum to be $\bar u u + \bar d d$ of $I=0$, the isospin of the whole process is also conserved to be $I=0$:
\begin{eqnarray}
\Lambda_b^0 &\to& ud c~\bar c s~(\bar u u + \bar d d)
\label{eq:isospin}
\\ \nonumber &\to& \sqrt{1\over3} D^{(*)-} \Sigma_c^{(*)++} K^-  + \sqrt{1\over3} \bar D^{(*)0} \Sigma_c^{(*)0} \bar K^0
\\ \nonumber && - \sqrt{1\over6} D^{(*)-} \Sigma_c^{(*)+} \bar K^0 - \sqrt{1\over6} \bar D^{(*)0} \Sigma_c^{(*)+} K^-\, .
\end{eqnarray}
The four fixed isospin factors allow us to consider only the $D^{(*)-} \Sigma_c^{(*)++} K^-$ final state, since the results derived from the $\bar D^{(*)0} \Sigma_c^{(*)+} K^-$ final state are the same. Accordingly, we only need to consider the exchange of the $u$ quark from the vacuum and the $d$ quark from the $\Lambda_b^0$, which is depicted in Fig.~\ref{fig:production}(a).

Summarizing the above discussions, in this section we shall calculate relative production rates of $\bar D^{(*)} \Sigma_c^{(*)}$ molecular states in $\Lambda_b^0$ decays, by investigating the three-body $\Lambda_b^0 \to D^{(*)-} \Sigma_c^{(*)++} K^-$ decays, whose mechanism is depicted in Fig.~\ref{fig:production}(a). We shall develop a Fierz rearrangement to describe this process in Sec.~\ref{sec:producionequation}, and use it to perform numerical analyses in Sec.~\ref{sec:productionnumerical}.

\subsection{Fierz rearrangement}
\label{sec:producionequation}

To describe the production mechanism depicted in Fig.~\ref{fig:production}(a), we use the color rearrangement given in Eq.~(\ref{eq:colorsinglet}) twice to obtain
\begin{eqnarray}
\nonumber \epsilon^{abc} \delta^{de} \delta^{fg} &=& \left( \epsilon^{ebc} \delta^{da} + \epsilon^{aec} \delta^{db} + \epsilon^{abe} \delta^{dc} \right) \times \delta^{fg}
\\ \nonumber &=& \epsilon^{gbc} \delta^{da} \delta^{fe} + \epsilon^{egc} \delta^{da} \delta^{fb} + \epsilon^{ebg} \delta^{da} \delta^{fc}
\\ \nonumber &+& \epsilon^{gec} \delta^{db} \delta^{fa} + \epsilon^{agc} \delta^{db} \delta^{fe} + \epsilon^{aeg} \delta^{db} \delta^{fc}
\\ \nonumber &+& \epsilon^{gbe} \delta^{dc} \delta^{fa} + \epsilon^{age} \delta^{dc} \delta^{fb} + \epsilon^{abg} \delta^{dc} \delta^{fe}  \, .
\\ \label{eq:colortwice}
\end{eqnarray}
Given the initial color structure to be $[\epsilon^{abc} u_a d_b c_c][\delta^{de}\bar c_d s_e][\delta^{fg}\bar u_f u_g]$, we need to fifth one $[\epsilon^{agc}u_a u_g c_c] [\delta^{db} \bar c_d d_b] [\delta^{fe}\bar u_f s_e]$, which corresponds to the $D^{(*)-} \Sigma_c^{(*)++} K^-$ final state.

Besides, we also need to apply the Fierz transformation twice: a) to interchange the $d_b$ and $u_g$ quarks, and b) to interchange the $d_b$ and $s_e$ quarks. Note that the Fierz rearrangement in the Lorentz space is actually a matrix identity. It is valid if each quark field in the initial and final currents is at the same location.

The key formula is as follow:
\begin{widetext}
\begin{eqnarray}
\Lambda_b^0 &\xrightarrow{~~~~~~~~~~~~~}& J_{\Lambda_b^0} = [\epsilon^{abc}u_a^T \mathbb{C} \gamma_5 d_b b_c]
\\ &\xrightarrow{~~~~\rm weak~~~~}&       [\epsilon^{abc}u_a^T \mathbb{C} \gamma_5 d_b \gamma_\rho(1 - \gamma_5) c_c] ~\times~ [\delta^{de}\bar c_d \gamma^\rho(1 - \gamma_5) s_e]
\label{eq:step1}
\\ &\xrightarrow{~~~~\rm QPC~~~~}&        [\epsilon^{abc}u_a^T \mathbb{C} \gamma_5 d_b \gamma_\rho(1 - \gamma_5) c_c] ~\times~ [\delta^{de}\bar c_d \gamma^\rho(1 - \gamma_5) s_e] ~\times~ [\delta^{fg}\bar u_f u_g]
\label{eq:step2}
\\ &\xlongequal{~~~~\rm color~~~~}&       \epsilon^{agc}\delta^{db}\delta^{fe} \times u_a^T \mathbb{C} \gamma_5 d_b \gamma_\rho(1 - \gamma_5) c_c \times \bar c_d \gamma^\rho(1 - \gamma_5) s_e \times \bar u_f u_g + \cdots
\label{eq:step3}
\\ &\xlongequal{{\rm Fierz:}d_b \leftrightarrow u_g}&
-{\delta^{db}\delta^{fe}\over4} \times [\epsilon^{agc}u_a^T \mathbb{C} \gamma_\mu u_g \gamma_\rho(1 - \gamma_5) c_c] \times \bar c_d \gamma^\rho(1 - \gamma_5) s_e \times \bar u_f \gamma^\mu \gamma_5 d_b + \cdots
\label{eq:step5}
\\ &\xlongequal{{\rm Fierz:}d_b \leftrightarrow s_e}&
\label{eq:step6}
+ {1 + \gamma_5 \over16} \times [\epsilon^{agc}u_a^T \mathbb{C} \gamma_\mu u_g \gamma^\mu \gamma_5 c_c] \times [\delta^{db} \bar c_d \gamma_5 d_b] \times [\delta^{fe}\bar u_f \gamma_5 s_e]
\\ \nonumber && +
{(1 + \gamma_5) (g^{\nu\rho} - i \sigma^{\nu\rho}) \over 32} \times [\epsilon^{agc}u_a^T \mathbb{C} \gamma_\mu u_g \gamma^\mu \gamma_5 c_c] \times [\delta^{db} \bar c_d \gamma_\nu d_b] \times [\delta^{fe}\bar u_f \gamma_\rho \gamma_5 s_e]
\\ \nonumber && +
{(1 + \gamma_5) (g^{\alpha\nu} \gamma^\rho + g^{\alpha\rho} \gamma^\nu) \over 16} \times [P_{\alpha \mu}^{3/2}\epsilon^{agc}u_a^T \mathbb{C} \gamma^\mu u_g c_c] \times [\delta^{db} \bar c_d \gamma_\nu d_b] \times [\delta^{fe}\bar u_f \gamma_\rho \gamma_5 s_e] + \cdots
\\ &\xlongequal{~~~~~~~~~~~~~~}&
\label{eq:step7}
+ {1 + \gamma_5 \over8\sqrt2} \times \xi_1 \times [\bar u_a \gamma_5 s_a]
\\ \nonumber && +
{(1 + \gamma_5) (g_{\nu\rho} - i \sigma_{\nu\rho}) \over 16\sqrt2} \left(\xi_3^\nu - {1\over4}\gamma^\nu\gamma_5\xi_2\right) [\bar u_a \gamma^\rho \gamma_5 s_a]
\\ \nonumber && +
{(1 + \gamma_5) (g_{\alpha\nu} \gamma_\rho + g_{\alpha\rho} \gamma_\nu) \over 8\sqrt2} \left(\xi_7^{\alpha\nu} - {1\over9}\gamma^\alpha\gamma_5\xi_6^\nu - {1\over9}\gamma^\nu\gamma_5\xi_6^\alpha + {2\over9} g^{\alpha\nu}\xi_5 \right)
 [\bar u_a \gamma^\rho \gamma_5 s_a]
\\ \nonumber && + \cdots \, .
\end{eqnarray}
\end{widetext}
Its brief explanations are as follows:
\begin{itemize}

\item Eq.~(\ref{eq:step1}) describes the Cabibbo-favored weak decay of $b\to c + \bar{c}s$ via the $V$-$A$ current.

\item Eq.~(\ref{eq:step2}) describes the production of the $\bar u$ and $u$ quark pair from the vacuum via the $^3P_0$ quark pair creation mechanism.

\item In Eq.~(\ref{eq:step3}) we apply the double-color rearrangement given in Eq.~(\ref{eq:colortwice}).

\item In Eq.~(\ref{eq:step5}) we apply the Fierz transformation to interchange the $d_b$ and $u_g$ quarks.

\item In Eq.~(\ref{eq:step6}) we apply the Fierz transformation to interchange the $d_b$ and $s_e$ quarks.

\item In Eq.~(\ref{eq:step7}) we combine the five $u_a u_g c_c \bar c_d d_b$ quarks together so that $D^{(*)-} \Sigma_c^{(*)++}$ molecular states can be produced.

\end{itemize}

In the above expression, we only consider $\xi_{1\cdots7}$ defined in Eqs.~(\ref{def:xi1}-\ref{def:xi7}), which couple to $D^{(*)-} \Sigma_c^{(*)++}$ molecular states through $S$-wave. Actually, there may exist some other currents coupling to these states through $P$-wave, which are not included in the present study, such as
\begin{eqnarray}
\nonumber \xi_6^{\prime \alpha\beta} &=& {1\over\sqrt2}~P^{\alpha\beta,\nu\rho}_{3/2}~[\delta^{ab} \bar c_a \gamma_\nu d_b] ~ P^{3/2}_{\rho\mu} [\epsilon^{cde} u_c^T \mathbb{C} \gamma^\mu u_d c_e]
\\ &=& P^{\alpha\beta,\nu\rho}_{3/2}~ D^{*-}_\nu ~ ~ \Sigma_{c;\rho}^{*++} \, ,
\end{eqnarray}
where $P_{3/2}^{\mu\nu,\rho\sigma}$ is the spin-3/2 projection operator with two antisymmetric Lorentz indices,
\begin{eqnarray}
P_{3/2}^{\mu\nu,\rho\sigma} &=& {1\over2} g^{\mu\rho}g^{\nu\sigma} - {1\over2} g^{\mu\sigma}g^{\nu\rho} + {1\over6} \sigma^{\mu\nu}\sigma^{\rho\sigma}
\\ \nonumber && - {1\over4}g^{\mu\rho}\gamma^\nu\gamma^\sigma + {1\over4}g^{\mu\sigma}\gamma^\nu\gamma^\rho
\\ \nonumber && - {1\over4}g^{\nu\sigma}\gamma^\mu\gamma^\rho + {1\over4}g^{\nu\rho}\gamma^\mu\gamma^\sigma  \, .
\end{eqnarray}
The current $\eta_6^{\prime \alpha\beta}$ couples to $| D^{*-} \Sigma_c^{*++}; 3/2^- \rangle$ through
\begin{equation}
\langle 0| \eta_6^{\prime \alpha\beta} |  D^{*-} \Sigma_c^{*++}; 3/2^- \rangle = i f^T_{6^\prime} (p^\alpha u^\beta - p^\beta u^\alpha) \, ,
\end{equation}
where $u_\alpha$ is the spinor of $| D^{*-} \Sigma_c^{*++}; 3/2^- \rangle$. Besides, it can couple to another state of $J^P = 3/2^+$.

Consequently, $|\bar D \Sigma_c^*; 3/2^- \rangle$ may still be produced in $\Lambda_b^0$ decays, although its directly corresponding current $\xi_4^\alpha$ (and so $J_4^\alpha$) does not appear in Eq.~(\ref{eq:step7}). Besides, omissions of the ``some other possible currents'' produce some theoretical uncertainties.

\subsection{Production analyses}
\label{sec:productionnumerical}

In this subsection we use the Fierz rearrangement given in Eq.~(\ref{eq:step7}) to perform numerical analyses. We shall take into account the isospin factors of Eqs.~(\ref{def:molecule}) and (\ref{eq:isospin}), and directly calculate relative production rates of $I=1/2$ $\bar D^{(*)} \Sigma_c^{(*)}$ molecular states in $\Lambda_b^0$ decays. To do this we need the following couplings to $K^-$:
\begin{eqnarray}
\langle 0 | \bar u_a \gamma_5 s_a | K^-(q) \rangle &=& \lambda_{K} \, ,
\\ \nonumber \langle 0 | \bar u_a \gamma_\mu \gamma_5 s_a | K^-(q) \rangle &=& i q_\mu f_{K} \, ,
\end{eqnarray}
where $f_{K}= 155.6$~MeV~\cite{pdg} and $\lambda_{K} = {f_{K}^2 m_K \over m_u + m_s}$.

We extract from Eq.~(\ref{eq:step7}) the following decay channels:
\begin{enumerate}

\item The decay of $\Lambda_b^0$ into $|\bar D \Sigma_c; 1/2^- \rangle K^-$ is contributed by $\xi_1 \times [\bar u_a \gamma_5 s_a]$:
\begin{eqnarray}
&& \langle \Lambda_b^0(q) ~|~\bar D \Sigma_c; 1/2^-(q_1)~K^-(q_2) \rangle
\\ \nonumber &\approx& -c~ i \lambda_K f_{|\bar D \Sigma_c; 1/2^- \rangle} ~ \bar u_{\Lambda_b^0} \left( {1 + \gamma_5 \over16} \right) u \, ,
\end{eqnarray}
where $u_{\Lambda_b^0}$ and $u$ are spinors of $\Lambda_b^0$ and $|\bar D \Sigma_c; 1/2^- \rangle$, respectively. The decay constant $f_{|\bar D \Sigma_c; 1/2^- \rangle}$ has been calculated in the previous section and given in Table~\ref{tab:mass}. The overall factor $c$ is related to: a) the coupling of $J_{\Lambda_b^0}$ to $\Lambda_b^0$, b) the weak and $^3P_0$ decay processes described by Eqs.~(\ref{eq:step1}) and (\ref{eq:step2}), and c) the isospin factors of Eqs.~(\ref{def:molecule}) and (\ref{eq:isospin}). We shall use the same factor $c$ for all the seven $\bar D^{(*)} \Sigma_c^{(*)}$ molecular states, which can cause a significant theoretical uncertainty not taken into account in the present study.

\item The decay of $\Lambda_b^0$ into $|\bar D^* \Sigma_c; 1/2^- \rangle K^-$ is contributed by $\xi_2 \times [\bar u_a \gamma^\rho \gamma_5 s_a]$:
\begin{eqnarray}
&& \langle \Lambda_b^0(q) ~|~\bar D^* \Sigma_c; 1/2^-(q_1)~K^-(q_2) \rangle
\\ \nonumber &\approx& c~ i f_K f_{|\bar D^* \Sigma_c; 1/2^- \rangle} q_2^\rho \times
\\ \nonumber && \bar u_{\Lambda_b^0} \left( {(1 + \gamma_5) (g_{\nu\rho} - i \sigma_{\nu\rho}) \over 32} \cdot \left(- {1\over4}\gamma^\nu\gamma_5\right) \right) u \, ,
\end{eqnarray}
where $u$ and $f_{|\bar D^* \Sigma_c; 1/2^- \rangle}$ are the spinor and decay constant of $|\bar D^* \Sigma_c; 1/2^- \rangle$, respectively.

\item The decay of $\Lambda_b^0$ into $|\bar D^* \Sigma_c; 3/2^- \rangle K^-$ is contributed by $\xi_3^\nu \times [\bar u_a \gamma^\rho \gamma_5 s_a]$:
\begin{eqnarray}
&& \langle \Lambda_b^0(q) ~|~\bar D^* \Sigma_c; 3/2^-(q_1)~K^-(q_2) \rangle
\\ \nonumber &\approx& c~ i f_K f_{|\bar D^* \Sigma_c; 3/2^- \rangle} q_2^\rho \times
\\ \nonumber && \bar u_{\Lambda_b^0} \left( {(1 + \gamma_5) (g_{\nu\rho} - i \sigma_{\nu\rho}) \over 32}  \right) u^\nu \, ,
\end{eqnarray}
where $u^\nu$ and $f_{|\bar D^* \Sigma_c; 3/2^- \rangle}$ are the spinor and decay constant of $|\bar D^* \Sigma_c; 3/2^- \rangle$, respectively.

\item The decay of $\Lambda_b^0$ into $|\bar D^* \Sigma_c^*; 1/2^- \rangle K^-$ is contributed by $\xi_5 \times [\bar u_a \gamma^\rho \gamma_5 s_a]$:
\begin{eqnarray}
&& \langle \Lambda_b^0(q) ~|~\bar D^* \Sigma_c^*; 1/2^-(q_1)~K^-(q_2) \rangle
\\ \nonumber &\approx& c~ i f_K f_{|\bar D^* \Sigma_c^*; 1/2^- \rangle} q_2^\rho \times
\\ \nonumber && \bar u_{\Lambda_b^0} \left( {(1 + \gamma_5) (g_{\alpha\nu} \gamma_\rho + g_{\alpha\rho} \gamma_\nu) \over 16} \cdot {2\over9} g^{\alpha\nu} \right) u \, ,
\end{eqnarray}
where $u$ and $f_{|\bar D^* \Sigma_c^*; 1/2^- \rangle}$ are the spinor and decay constant of $|\bar D^* \Sigma_c^*; 1/2^- \rangle$, respectively.

\item The decay of $\Lambda_b^0$ into $|\bar D^* \Sigma_c^*; 3/2^- \rangle K^-$ is contributed by $\xi_6^\beta \times [\bar u_a \gamma^\rho \gamma_5 s_a]$:
\begin{eqnarray}
&& \langle \Lambda_b^0(q) ~|~\bar D^* \Sigma_c^*; 3/2^-(q_1)~K^-(q_2) \rangle
\\ \nonumber &\approx& c~ i f_K f_{|\bar D^* \Sigma_c^*; 3/2^- \rangle} q_2^\rho \times
\\ \nonumber && \bar u_{\Lambda_b^0} \Bigg( {(1 + \gamma_5) (g_{\alpha\nu} \gamma_\rho + g_{\alpha\rho} \gamma_\nu) \over 16}
\\ \nonumber && ~~~~~~~~~~~~ \cdot \left( - {1\over9}\gamma^\alpha\gamma_5 g^{\nu\beta} - {1\over9}\gamma^\nu\gamma_5 g^{\alpha\beta} \right) \Bigg) u_\beta \, ,
\end{eqnarray}
where $u_\beta$ and $f_{|\bar D^* \Sigma_c^*; 3/2^- \rangle}$ are the spinor and decay constant of $|\bar D^* \Sigma_c^*; 3/2^- \rangle$, respectively.

\item The decay of $\Lambda_b^0$ into $|\bar D^* \Sigma_c^*; 5/2^- \rangle K^-$ is contributed by $\xi_7^{\alpha\nu} \times [\bar u_a \gamma^\rho \gamma_5 s_a]$:
\begin{eqnarray}
&& \langle \Lambda_b^0(q) ~|~\bar D^* \Sigma_c^*; 5/2^-(q_1)~K^-(q_2) \rangle
\\ \nonumber &\approx& c~ i f_K f_{|\bar D^* \Sigma_c^*; 5/2^- \rangle} q_2^\rho \times
\\ \nonumber && \bar u_{\Lambda_b^0} \left( {(1 + \gamma_5) (g_{\alpha\nu} \gamma_\rho + g_{\alpha\rho} \gamma_\nu) \over 16} \right) u^{\alpha\nu} \, ,
\end{eqnarray}
where $u^{\alpha\nu}$ and $f_{|\bar D^* \Sigma_c^*; 5/2^- \rangle}$ are the spinor and decay constant of $|\bar D^* \Sigma_c^*; 5/2^- \rangle$, respectively.

\end{enumerate}

We shall find that the $P_c(4312)^+$, $P_c(4440)^+$, and $P_c(4457)^+$ can be well interpreted in our framework as $|\bar D \Sigma_c; 1/2^- \rangle$, $|\bar D^{*} \Sigma_c; 3/2^- \rangle$, and $|\bar D^{*} \Sigma_c; 1/2^- \rangle$, respectively. Accordingly, we assume masses of $\bar D^{(*)} \Sigma_c^{(*)}$ molecular states to be:
\begin{eqnarray}
\nonumber M_{|\bar D \Sigma_c; 1/2^- \rangle} &=& M_{P_c(4312)^+} = 4311.9~{\rm MeV} \, ,
\\ \nonumber M_{|\bar D^{*} \Sigma_c; 1/2^- \rangle} &=& M_{P_c(4457)^+} = 4457.3~{\rm MeV} \, ,
\\ \nonumber M_{|\bar D^{*} \Sigma_c; 3/2^- \rangle} &=& M_{P_c(4440)^+} = 4440.3~{\rm MeV} \, ,
\\ M_{|\bar D \Sigma_c^{*}; 3/2^- \rangle} &\approx& M_{D} + M_{\Sigma_c^*} = 4385~{\rm MeV} \, ,
\\ \nonumber M_{|\bar D^{*} \Sigma_c^*; 1/2^- \rangle} &\approx& M_{D^*} + M_{\Sigma_c^*} = 4527~{\rm MeV} \, ,
\\ \nonumber M_{|\bar D^{*} \Sigma_c^*; 3/2^- \rangle} &\approx& M_{D^*} + M_{\Sigma_c^*} = 4527~{\rm MeV} \, ,
\\ \nonumber M_{|\bar D^{*} \Sigma_c^*; 5/2^- \rangle} &\approx& M_{D^*} + M_{\Sigma_c^*} = 4527~{\rm MeV} \, .
\end{eqnarray}
Now we can summarize the above production amplitudes to obtain the following partial decay widths:
\begin{eqnarray}
\nonumber \Gamma(\Lambda_b^0 \to |\bar D \Sigma_c \rangle_{1/2^-} K^-) &=& c^2~6.15 \times 10^{-11}~{\rm GeV}^{17} \, ,
\\[1mm] \nonumber \Gamma(\Lambda_b^0 \to |\bar D^* \Sigma_c \rangle_{1/2^-} K^-) &=& c^2~8.76 \times 10^{-12}~{\rm GeV}^{17} \, ,
\\[1mm] \nonumber \Gamma(\Lambda_b^0 \to |\bar D^* \Sigma_c \rangle_{3/2^-} K^-) &=& c^2~7.52 \times 10^{-12}~{\rm GeV}^{17} \, ,
\\[1mm] \nonumber \Gamma(\Lambda_b^0 \to |\bar D \Sigma_c^* \rangle_{3/2^-} K^-) &=& 0
\\[1mm] \nonumber \Gamma(\Lambda_b^0 \to |\bar D^* \Sigma_c^* \rangle_{1/2^-} K^-) &=& c^2~3.57 \times 10^{-11}~{\rm GeV}^{17} \, ,
\\[1mm] \nonumber \Gamma(\Lambda_b^0 \to |\bar D^* \Sigma_c^* \rangle_{3/2^-} K^-) &=& c^2~1.38 \times 10^{-12}~{\rm GeV}^{17} \, ,
\\[1mm] \Gamma(\Lambda_b^0 \to |\bar D^* \Sigma_c^* \rangle_{5/2^-} K^-) &=& 0 \, .
\label{result:production}
\end{eqnarray}
From these values, we derive the following relative production rates $\mathcal{R}_1(P_c) \equiv {\mathcal{B}\left(\Lambda_b^0 \rightarrow P_c K^- \right) / \mathcal{B}\left(\Lambda_b^0 \rightarrow |\bar D^* \Sigma_c \rangle_{3/2^-} K^- \right)}$, which are summarized in Table~\ref{tab:width} and will be further discussed in Sec.~\ref{sec:summary}:
\begin{widetext}
\begin{eqnarray}
\nonumber && {\mathcal{B}\Bigg(\Lambda_b^0 \rightarrow K^-\Big(
|\bar D \Sigma_c \rangle_{1/2^-}
:|\bar D^* \Sigma_c \rangle_{1/2^-}
:|\bar D^* \Sigma_c \rangle_{3/2^-}
:|\bar D \Sigma_c^* \rangle_{3/2^-}
:|\bar D^* \Sigma_c^* \rangle_{1/2^-}
:|\bar D^* \Sigma_c^* \rangle_{3/2^-}
:|\bar D^* \Sigma_c^* \rangle_{5/2^-}
\Big)\Bigg) \over \mathcal{B}\left(\Lambda_b^0 \rightarrow |\bar D^* \Sigma_c \rangle_{3/2^-} K^- \right)}
\\[2mm] &\approx&
~~~~~~~~~~~~~~~~~~~~~~~~~ 8.2 ~~~~ : ~~~~~\,1.2\,~~~~~ : ~~~~~~~{\bf1}~~~~~~\, : ~~~~~~0~~~~~~ : ~~~~~\,4.8\,~~~~~ : ~~~~~0.18~~~~~ : ~~~~~0~ \, .
\end{eqnarray}
\end{widetext}

\section{Decay properties through the Fierz rearrangement}
\label{sec:decay}

We have applied the Fierz rearrangement~\cite{fierz} of the Dirac and color indices to study decay properties of the $P_c(4312)^+$, $P_c(4440)^+$, and $P_c(4457)^+$ as $\bar D^{(*)} \Sigma_c$ molecular states, based on the currents $J_{1\cdots3}$~\cite{Chen:2020pac}. In this section we follow the same procedures to study decay properties of $\bar D^{(*)} \Sigma_c^*$ molecular states using the currents $J_{4\cdots7}$. We shall study their decays into charmonium mesons and spin-1/2 light baryons as well as charmed mesons and spin-1/2 charmed baryons, such as $J/\psi p$ and $\bar D \Lambda_c$, etc.

We refer to Ref.~\cite{Chen:2020pac} for detailed discussions. This method has been applied to study strong decay properties of the $Z_c(3900)$, $X(3872)$, and $X(6900)$ in Refs.~\cite{Chen:2019wjd,Chen:2019eeq,Chen:2020xwe}, and a similar arrangement of the spin and color indices in the nonrelativistic case has been applied to study decay properties of $XYZ$ and $P_c$ states in Refs.~\cite{Voloshin:2013dpa,Maiani:2017kyi,Voloshin:2018pqn,Voloshin:2019aut,Wang:2019spc,Xiao:2019spy,Cheng:2020nho}.

\subsection{Input parameters}

To study decays of $\bar D^{(*)} \Sigma_c^{*}$ molecular states into charmonium mesons and light baryons, we need to use the $\theta(x)$ currents. We can construct them by combining charmonium operators and light baryon fields, which has been done in Ref.~\cite{Chen:2020pac}. In the present study we need couplings of charmonium operators to charmonium states, which are listed in Table~\ref{tab:coupling}. We also need the Ioffe's light baryon field~\cite{Ioffe:1981kw,Ioffe:1982ce,Espriu:1983hu,Chen:2008qv,Chen:2009sf,Chen:2010ba,Chen:2011rh,Dmitrasinovic:2016hup}:
\begin{eqnarray}
N &=& N_1 - N_2
\\ \nonumber &=& \epsilon^{abc} (u_a^T \mathbb{C} d_b) \gamma_5 u_c - \epsilon^{abc} (u_a^T \mathbb{C} \gamma_5 d_b) u_c \, .
\end{eqnarray}
It couples to the proton through
\begin{equation}
\langle 0 | N | p \rangle = f_p u_p \, ,
\end{equation}
with $u_p$ the Dirac spinor of the proton. The decay constant $f_p$ has been calculated in Ref.~\cite{Chen:2012ex} to be
\begin{equation}
f_p = 0.011  {\rm~GeV}^3 \, .
\label{eq:proton}
\end{equation}

\begin{table*}[hbt]
\begin{center}
\renewcommand{\arraystretch}{1.5}
\caption{Couplings of meson operators to meson states, where color indices are omitted for simplicity. Taken from Ref.~\cite{Chen:2019wjd}.}
\begin{tabular}{ c | c | c | c | c | c}
\hline\hline
~~Operators~~ & ~$I^GJ^{PC}$~ & ~~~~~~Mesons~~~~~~ & ~$I^GJ^{PC}$~ & ~~~Couplings~~~ & ~~~~~~Decay Constants~~~~~~
\\ \hline\hline
$I^{S} = \bar c c$                & $0^+0^{++}$                  & $\chi_{c0}(1P)$      & $0^+0^{++}$ & $\langle 0 | I^S | \chi_{c0} \rangle = m_{\chi_{c0}} f_{\chi_{c0}}$          & $f_{\chi_{c0}} = 343$~MeV~\mbox{\cite{Veliev:2010gb}}
\\ \hline
$I^{P} = \bar c i\gamma_5 c$      & $0^+0^{-+}$                  & $\eta_c$             & $0^+0^{-+}$ & $\langle 0 | I^{P} | \eta_c \rangle = \lambda_{\eta_c}$                      & $\lambda_{\eta_c} = {f_{\eta_c} m_{\eta_c}^2 \over 2 m_c}$
\\ \hline
$I^{V}_\mu = \bar c \gamma_\mu c$ & $0^-1^{--}$                  & $J/\psi$             & $0^-1^{--}$ & $\langle0| I^{V}_\mu | J/\psi \rangle = m_{J/\psi} f_{J/\psi} \epsilon_\mu$  & $f_{J/\psi} = 418$~MeV~\mbox{\cite{Becirevic:2013bsa}}
\\ \hline
\multirow{2}{*}{$I^{A}_\mu = \bar c \gamma_\mu \gamma_5 c$}
                               & \multirow{2}{*}{$0^+1^{++}$}    & $\eta_c$             & $0^+0^{-+}$ & $\langle 0 | I^{A}_\mu | \eta_c \rangle = i p_\mu f_{\eta_c}$                & $f_{\eta_c} = 387$~MeV~\mbox{\cite{Becirevic:2013bsa}}
\\ \cline{3-6}
                               &                              & $\chi_{c1}(1P)$      & $0^+1^{++}$ & $\langle 0 | I^{A}_\mu | \chi_{c1} \rangle = m_{\chi_{c1}} f_{\chi_{c1}} \epsilon_\mu $
                               &  $f_{\chi_{c1}} = 335$~MeV~\mbox{\cite{Novikov:1977dq}}
\\ \hline
\multirow{2}{*}{$I^{T}_{\mu\nu} = \bar c \sigma_{\mu\nu} c$}
                               & \multirow{2}{*}{$0^-1^{\pm-}$}  & $J/\psi$             & $0^-1^{--}$ & $\langle 0 | I^{T}_{\mu\nu} | J/\psi \rangle = i f^T_{J/\psi} (p_\mu\epsilon_\nu - p_\nu\epsilon_\mu) $
                               &  $f_{J/\psi}^T = 410$~MeV~\mbox{\cite{Becirevic:2013bsa}}
\\ \cline{3-6}
                               &                              & $h_c(1P)$            & $0^-1^{+-}$ & $\langle 0 | I^{T}_{\mu\nu} | h_c \rangle = i f^T_{h_c} \epsilon_{\mu\nu\alpha\beta} \epsilon^\alpha p^\beta $
                               &  $f_{h_c}^T = 235$~MeV~\mbox{\cite{Becirevic:2013bsa}}
\\ \hline\hline
$O^{S} = \bar c q$                & $0^{+}$                   & $\bar D_0^{*}$           & $0^{+}$  & $\langle 0 | O^{S} | \bar D_0^{*} \rangle = m_{D_0^{*}} f_{D_0^{*}}$             &  $f_{D_0^{*}} = 410$~MeV~\mbox{\cite{Narison:2015nxh}}
\\ \hline
$O^{P} = \bar c i\gamma_5 q$      & $0^{-}$                   & $\bar D$                & $0^{-}$  & $\langle 0 | O^{P} | \bar D \rangle = \lambda_D$                                &  $\lambda_D = {f_D m_D^2 \over {m_c + m_d}}$
\\ \hline
$O^{V}_\mu = \bar c \gamma_\mu q$ & $1^{-}$                   & $\bar D^{*}$        & $1^{-}$  & $\langle0| O^{V}_\mu | \bar D^{*} \rangle = m_{D^*} f_{D^*} \epsilon_\mu$   &  $f_{D^*} = 253$~MeV~\mbox{\cite{Chang:2018aut}}
\\ \hline
\multirow{2}{*}{$O^{A}_\mu = \bar c \gamma_\mu \gamma_5 q$}
                               & \multirow{2}{*}{$1^{+}$}      & $\bar D$         & $0^{-}$  & $\langle 0 | O^{A}_\mu | \bar D \rangle = i p_\mu f_{D}$                 &  $f_{D} = 211.9$~MeV~\mbox{\cite{pdg}}
\\ \cline{3-6}
                                  &                            & $\bar D_1$                & $1^{+}$  & $\langle 0 | O^{A}_\mu | \bar D_1 \rangle = m_{D_1} f_{D_1} \epsilon_\mu $        &  $f_{D_1} = 356$~MeV~\mbox{\cite{Narison:2015nxh}}
\\ \hline
\multirow{2}{*}{$O^{T}_{\mu\nu} = \bar c \sigma_{\mu\nu} q$}
                               & \multirow{2}{*}{$1^{\pm}$}    & $\bar D^{*}$        & $1^{-}$  &  $\langle 0 | O^{T}_{\mu\nu} | \bar D^{*} \rangle = i f_{D^*}^T (p_\mu\epsilon_\nu - p_\nu\epsilon_\mu) $
                               &  $f_{D^*}^T \approx 220$~MeV
\\ \cline{3-6}
                               &                               &  --                  & $1^{+}$  &  --  &  --
\\ \hline\hline
\end{tabular}
\label{tab:coupling}
\end{center}
\end{table*}

To study decays of $\bar D^{(*)} \Sigma_c^{*}$ molecular states into charmed mesons and charmed baryons, we need to use the $\eta(x)$ and $\xi(x)$ currents. They have been constructed in Sec.~\ref{sec:current} by combining charmed meson operators and charmed baryon fields. In the present study we need couplings of charmed meson operators to charmed meson states, which are also listed in Table~\ref{tab:coupling}. Besides, we need couplings of the charmed baryon fields, $J_{\mathcal{B}}$ defined in Eqs.~(\ref{eq:heavybaryon}), to the ground-state charmed baryons $\mathcal{B} = \Lambda_c/\Sigma_c$:
\begin{equation}
\langle 0 | J_{\mathcal{B}} | \mathcal{B} \rangle = f_{\mathcal{B}} u_{\mathcal{B}} \, .
\end{equation}
Note that we do not investigate decays of $|\bar D^{(*)} \Sigma_c^{*}; J^P\rangle$ into the $\bar D^{(*)} \Sigma_c^*$ final states in the present study, because some of the $J = 3/2$ charmed baryon fields still remain unclear~\cite{Chen:2020pac}. The decay constants $f_{\mathcal{B}}$ have been calculated in Refs.~\cite{Liu:2007fg,Chen:2017sci,Cui:2019dzj} to be
\begin{eqnarray}
f_{\Lambda_c} &=& 0.015 {\rm~GeV}^3 \, ,
\\
\nonumber f_{\Sigma_c} &=& 0.036 {\rm~GeV}^3 \, .
\end{eqnarray}
These values are evaluated using the method of QCD sum rules~\cite{Shifman:1978bx,Reinders:1984sr} within the heavy quark effective theory~\cite{Grinstein:1990mj,Eichten:1989zv,Falk:1990yz}, while the full QCD decay constant $f_p$ for the proton has been given in Eq.~(\ref{eq:proton}). These two different schemes cause some, but not large, theoretical uncertainties.

\subsection{Fierz rearrangement}
\label{sec:fierz}

In this subsection we perform the Fierz rearrangement separately for $\eta_{4\cdots7}$ and $\xi_{4\cdots7}$. The obtained results will be used later to study strong decay properties of $\bar D \Sigma_c^{*}$ and $\bar D^{*} \Sigma_c^{*}$ molecular states.

Before doing this, we note again that the Fierz rearrangement in the Lorentz space is actually a matrix identity. It is valid if each quark field in the initial and final currents is at the same location, {\it e.g.}, we can apply the Fierz rearrangement to transform a non-local current $\eta = [\bar c(x) u(x)] ~ [u(y) d(y) c(y)]$ into the combination of many non-local currents $\theta = [\bar c(x) c(y)] ~ [u(y) d(y) u(x)]$, with all the quark fields remaining at same locations. Keeping this in mind, we shall omit the coordinates in this subsection.

\subsubsection{$\eta \rightarrow \theta$ and $\xi \rightarrow \theta$}

Using the color rearrangement~\cite{Chen:2020pac}
\begin{equation}
\delta^{ab} \epsilon^{cde} = {1\over3}~\delta^{ae} \epsilon^{bcd} - {1\over2}~\lambda^{ae}_n \epsilon^{bcf} \lambda^{fd}_n + {1\over2}~\lambda^{ae}_n \epsilon^{bdf} \lambda^{fc}_n \, ,
\label{eq:color}
\end{equation}
together with the Fierz rearrangement to interchange the $u_b$ and $c_e$ quark fields, we can transform an $\eta$ current into the combination of many $\theta$ currents:
\begin{eqnarray}
\nonumber \eta_4^\alpha &\rightarrow& [\bar c_a \gamma_\mu c_a] \left( - {1\over32} g^{\alpha\mu} - {i\over96} \sigma^{\alpha\mu} \right) N
\\ \nonumber && + ~ [\bar c_a \gamma_\mu \gamma_5 c_a] \left( - {1\over32} g^{\alpha\mu} \gamma_5 - {i\over96} \sigma^{\alpha\mu} \gamma_5 \right) N
\\ \nonumber && + ~ [\bar c_a \sigma_{\mu\nu} c_a] \left( {i\over48} g^{\alpha\mu}\gamma^\nu + {1\over96} \epsilon^{\alpha\mu\nu\rho} \gamma_\rho \gamma_5 \right) N
\\ && + ~ \cdots \, ,
\label{eq:eta4theta}
\\[2.5mm]
\nonumber \eta_5 &\rightarrow& + {1\over8} ~ [\bar c_a c_a] ~ \gamma_5 N + {1\over8} ~ [\bar c_a \gamma_5 c_a] ~ N
\\ \nonumber &&  + {1\over16} ~ [\bar c_a \gamma_\mu c_a] ~ \gamma^\mu \gamma_5 N - {1\over16} ~ [\bar c_a \gamma_\mu \gamma_5 c_a] ~ \gamma^\mu N
\\ &&  + {1\over48} ~ [\bar c_a \sigma_{\mu\nu} c_a] ~ \sigma^{\mu\nu} \gamma_5 N + \cdots \, ,
\label{eq:eta5theta}
\\[2.5mm]
\nonumber \eta_6^\alpha &\rightarrow& [\bar c_a \gamma_\mu c_a] \left( {3\over32} g^{\alpha\mu} + {i\over32} \sigma^{\alpha\mu} \right) N
\\ \nonumber && + ~ [\bar c_a \gamma_\mu \gamma_5 c_a] \left( - {3\over32} g^{\alpha\mu} \gamma_5 - {i\over32} \sigma^{\alpha\mu} \gamma_5 \right) N
\\ && + ~ \cdots \, ,
\label{eq:eta6theta}
\\[2.5mm]
\nonumber \eta_7^{\alpha\beta} &\rightarrow& \Big( {i\over144} \sigma^{\alpha\rho}\epsilon^{\beta\mu\nu\rho} + {1\over72} g^{\alpha\mu}\sigma^{\beta\nu}\gamma_5 - {1\over144} g^{\alpha\beta}\sigma^{\mu\nu}\gamma_5 \Big)
\\ && \times ~ [\bar c_a \sigma_{\mu\nu} c_a] ~  N + \cdots \, .
\label{eq:eta7theta}
\end{eqnarray}
In the above expressions, we have kept all the color-singlet-color-singlet meson-baryon terms depending on the $J=1/2$ light baryon fields, but omitted that: a) the color-octet-color-octet meson-baryon terms, such as $[\lambda^{ae}_n \bar c_a c_e][\epsilon^{bcf}\lambda^{fd}_n u_b u_c d_d]$, and b) terms depending on the $J=3/2$ light baryon fields.

Similarly, we can use Eq.~(\ref{eq:color}) together with the Fierz rearrangement to interchange the $d_b$ and $c_e$ quark fields, and transform a $\xi$ current into the combination of many $\theta$ currents:
\begin{eqnarray}
\nonumber \sqrt2 \xi_4^\alpha &\rightarrow& [\bar c_a \gamma_\mu c_a] \left( {1\over16} g^{\alpha\mu} + {i\over48} \sigma^{\alpha\mu} \right) N
\\ \nonumber && + ~ [\bar c_a \gamma_\mu \gamma_5 c_a] \left( {1\over16} g^{\alpha\mu} \gamma_5 + {i\over48} \sigma^{\alpha\mu} \gamma_5 \right) N
\\ \nonumber && + ~ [\bar c_a \sigma_{\mu\nu} c_a] \left( - {i\over24} g^{\alpha\mu}\gamma^\nu - {1\over48} \epsilon^{\alpha\mu\nu\rho} \gamma_\rho \gamma_5 \right) N
\\ && + ~ \cdots \, ,
\label{eq:xi4theta}
\\[2.5mm]
\nonumber \sqrt2 \xi_5 &\rightarrow& - {1\over4} ~ [\bar c_a c_a] ~ \gamma_5 N - {1\over4} ~ [\bar c_a \gamma_5 c_a] ~ N
\\ \nonumber &&  - {1\over8} ~ [\bar c_a \gamma_\mu c_a] ~ \gamma^\mu \gamma_5 N + {1\over8} ~ [\bar c_a \gamma_\mu \gamma_5 c_a] ~ \gamma^\mu N
\\ &&  - {1\over24} ~ [\bar c_a \sigma_{\mu\nu} c_a] ~ \sigma^{\mu\nu} \gamma_5 N + \cdots \, ,
\label{eq:xi5theta}
\\[2.5mm]
\nonumber \sqrt2 \xi_6^\alpha &\rightarrow& [\bar c_a \gamma_\mu c_a] \left( - {3\over16} g^{\alpha\mu} - {i\over16} \sigma^{\alpha\mu} \right) N
\\ \nonumber && + ~ [\bar c_a \gamma_\mu \gamma_5 c_a] \left( {3\over16} g^{\alpha\mu} \gamma_5 + {i\over16} \sigma^{\alpha\mu} \gamma_5 \right) N
\\ && + ~ \cdots \, ,
\label{eq:xi6theta}
\\[2.5mm]
\nonumber \sqrt2 \xi_7^{\alpha\beta} &\rightarrow& \Big( - {i\over72} \sigma^{\alpha\rho}\epsilon^{\beta\mu\nu\rho} -  {1\over36} g^{\alpha\mu}\sigma^{\beta\nu}\gamma_5
\\ && ~~ + {1\over72} g^{\alpha\beta}\sigma^{\mu\nu}\gamma_5 \Big) ~ [\bar c_a \sigma_{\mu\nu} c_a] ~  N + \cdots \, .
\label{eq:xi7theta}
\end{eqnarray}

\subsubsection{$\eta \rightarrow \eta$ and $\eta \rightarrow \xi$}

Using the color rearrangement
\begin{equation}
\delta^{ab} \epsilon^{cde} = {1\over3}~\delta^{ac} \epsilon^{bde} - {1\over2}~\lambda^{ac}_n \epsilon^{bdf} \lambda^{fe}_n + {1\over2}~\lambda^{ac}_n \epsilon^{bef} \lambda^{fd}_n \, ,
\label{eq:color2}
\end{equation}
together with the Fierz rearrangement to interchange the $u_b$ and $u_c$ quark fields, we can transform an $\eta$ current into the combination of many $\eta$ currents.

Using another color rearrangement:
\begin{equation}
\delta^{ab} \epsilon^{cde} = {1\over3}~\delta^{ad} \epsilon^{cbe} + {1\over2}~\lambda^{ad}_n \epsilon^{bcf} \lambda^{fe}_n - {1\over2}~\lambda^{ad}_n \epsilon^{bef} \lambda^{fc}_n \, ,
\label{eq:color3}
\end{equation}
together with the Fierz rearrangement to interchange the $u_b$ and $d_d$ quark fields, we can transform an $\eta$ current into the combination of many $\xi$ currents.

Altogether, we obtain:
\begin{eqnarray}
\nonumber \eta_4^\alpha &\rightarrow& \left( {1\over16}g^{\alpha\mu} + {i\over48}\sigma^{\alpha\mu} \right) ~ [\bar c_a \gamma_\mu u_a] ~ \Lambda_c^+
\\ \nonumber &+& \left( {i\over384}\sigma^{\alpha\sigma} \epsilon^{\mu\nu\rho\sigma} - {1\over128}\epsilon^{\alpha\mu\nu\rho} \right)  [\bar c_a \sigma_{\mu\nu} u_a] \gamma_\rho \gamma_5 \Sigma_c^{+}
\\ \nonumber &+& \left( {i\sqrt2\over384}\sigma^{\alpha\sigma} \epsilon^{\mu\nu\rho\sigma} - {\sqrt2\over128}\epsilon^{\alpha\mu\nu\rho} \right)
\\ && \times ~ [\bar c_a \sigma_{\mu\nu} d_a] ~ \gamma_\rho \gamma_5 \Sigma_c^{++} + \cdots \, ,
\label{eq:eta4etaxi}
\\[2.5mm]
\nonumber \eta_5 &\rightarrow& - {1\over4} ~ [\bar c_a \gamma_5 u_a] ~ \Lambda_c^+ - {1\over48} ~ [\bar c_a \sigma_{\mu\nu} u_a] ~ \sigma^{\mu\nu}\gamma_5 \Lambda_c^+
\\ \nonumber && - {1\over32} ~ [\bar c_a \gamma_\mu u_a] ~ \gamma^\mu \gamma_5 \Sigma_c^{+} + {1\over32} ~ [\bar c_a \gamma_\mu \gamma_5 u_a] ~ \gamma^\mu \Sigma_c^{+}
\\ \nonumber && - {\sqrt2\over32} [\bar c_a \gamma_\mu d_a] \gamma^\mu \gamma_5 \Sigma_c^{++} + {\sqrt2\over32} [\bar c_a \gamma_\mu \gamma_5 d_a] \gamma^\mu \Sigma_c^{++}
\\ && + ~ \cdots \, ,
\label{eq:eta5etaxi}
\\[2.5mm]
\nonumber \eta_6^\alpha &\rightarrow& \left( {i\over16}g^{\alpha\mu}\gamma^\nu + {1\over32}\epsilon^{\alpha\mu\nu\rho}\gamma_\rho\gamma_5 \right) ~ [\bar c_a \sigma_{\mu\nu} u_a] ~ \Lambda_c^+
\\ \nonumber &+& \left( {1\over96}g^{\alpha\mu}\gamma^\nu\gamma_5 + {1\over96}g^{\alpha\nu}\gamma^\mu\gamma_5 - {1\over192}g^{\mu\nu}\gamma^\alpha\gamma_5 \right)
\\ \nonumber && \times ~[\bar c_a \gamma_\mu u_a] ~  \gamma_\nu \gamma_5 \Sigma_c^{+}
\\ \nonumber &+& \left( {1\over64}g^{\alpha\mu}\gamma^\nu - {1\over64}g^{\alpha\nu}\gamma^\mu - {i\over64}\epsilon^{\alpha \mu \nu \rho}\gamma_\rho\gamma_5 \right)
\\ \nonumber && \times ~[\bar c_a \gamma_\mu \gamma_5 u_a] ~  \gamma_\nu \gamma_5 \Sigma_c^{+}
\\ \nonumber &+& \left( {\sqrt2\over96}g^{\alpha\mu}\gamma^\nu\gamma_5 + {\sqrt2\over96}g^{\alpha\nu}\gamma^\mu\gamma_5 - {\sqrt2\over192}g^{\mu\nu}\gamma^\alpha\gamma_5 \right)
\\ \nonumber && \times ~[\bar c_a \gamma_\mu d_a] ~  \gamma_\nu \gamma_5 \Sigma_c^{++}
\\ \nonumber &+& \left( {\sqrt2\over64}g^{\alpha\mu}\gamma^\nu - {\sqrt2\over64}g^{\alpha\nu}\gamma^\mu - {i\sqrt2\over64}\epsilon^{\alpha \mu \nu \rho}\gamma_\rho\gamma_5 \right)
\\ && \times ~[\bar c_a \gamma_\mu \gamma_5 d_a] ~  \gamma_\nu \gamma_5 \Sigma_c^{++} + \cdots \, ,
\label{eq:eta6etaxi}
\\[2.5mm]
\nonumber \eta_7^{\alpha\beta} &\rightarrow& \Big( {1\over36}g^{\alpha\mu}g^{\beta\nu} - {1\over144}g^{\alpha\beta}g^{\mu\nu} + {i\over144}g^{\alpha\mu}\sigma^{\beta\nu}
\\ \nonumber && + {i\over144}g^{\alpha\nu}\sigma^{\beta\mu} \Big) ~ [\bar c_a \gamma_\mu u_a] ~ \gamma_\nu \gamma_5 \Sigma_c^{+}
\\ \nonumber &+& \Big( {\sqrt2\over36}g^{\alpha\mu}g^{\beta\nu} - {\sqrt2\over144}g^{\alpha\beta}g^{\mu\nu} + {i\sqrt2\over144}g^{\alpha\mu}\sigma^{\beta\nu}
\\ && + {i\sqrt2\over144}g^{\alpha\nu}\sigma^{\beta\mu} \Big) ~ [\bar c_a \gamma_\mu d_a] ~ \gamma_\nu \gamma_5 \Sigma_c^{++} + \cdots \, .
\label{eq:eta7etaxi}
\end{eqnarray}
In the above expressions, we have kept all the color-singlet-color-singlet meson-baryon terms depending on the $J^P=1/2^+$ charmed baryon fields, {\it i.e.}, $J_{\Lambda_c^+}$ and $J_{\Sigma_c^{+/++}}$ defined in Eqs.~(\ref{eq:heavybaryon}). We have omitted that: a) the color-octet-color-octet meson-baryon terms, and b) terms depending on the $J=3/2$ charmed baryon fields.

\subsubsection{$\xi \rightarrow \eta$}

Using Eqs.~(\ref{eq:color2}) and (\ref{eq:color3}) together with the Fierz rearrangement in the Lorentz space, we can transform a $\xi$ current into the combination of many $\eta$ currents (but without $\xi$ currents):
\begin{eqnarray}
\nonumber \sqrt2 \xi_4^\alpha &\rightarrow& \left( - {1\over8}g^{\alpha\mu} - {i\over24}\sigma^{\alpha\mu} \right) ~ [\bar c_a \gamma_\mu u_a] ~ \Lambda_c^+
\\ \nonumber &+& \left( {i\over192}\sigma^{\alpha\sigma} \epsilon^{\mu\nu\rho\sigma} - {1\over64}\epsilon^{\alpha\mu\nu\rho} \right)
\\ && \times ~ [\bar c_a \sigma_{\mu\nu} u_a] ~ \gamma_\rho \gamma_5 \Sigma_c^{+} + \cdots \, ,
\label{eq:xi4eta}
\\[2.5mm]
\nonumber \sqrt2 \xi_5 &\rightarrow& + {1\over2} ~ [\bar c_a \gamma_5 u_a] ~ \Lambda_c^+ + {1\over24} ~ [\bar c_a \sigma_{\mu\nu} u_a] ~ \sigma^{\mu\nu}\gamma_5 \Lambda_c^+
\\ \nonumber && - {1\over16} [\bar c_a \gamma_\mu u_a] \gamma^\mu \gamma_5 \Sigma_c^{+} + {1\over16} [\bar c_a \gamma_\mu \gamma_5 u_a] \gamma^\mu \Sigma_c^{+}
\\ && + ~ \cdots \, ,
\label{eq:xi5eta}
\\[2.5mm]
\nonumber \sqrt2 \xi_6^\alpha &\rightarrow& \left( - {i\over8}g^{\alpha\mu}\gamma^\nu - {1\over16}\epsilon^{\alpha\mu\nu\rho}\gamma_\rho\gamma_5 \right) ~ [\bar c_a \sigma_{\mu\nu} u_a] ~ \Lambda_c^+
\\ \nonumber &+& \left( {1\over48}g^{\alpha\mu}\gamma^\nu\gamma_5 + {1\over48}g^{\alpha\nu}\gamma^\mu\gamma_5 - {1\over96}g^{\mu\nu}\gamma^\alpha\gamma_5 \right)
\\ \nonumber && \times ~[\bar c_a \gamma_\mu u_a] ~  \gamma_\nu \gamma_5 \Sigma_c^{+}
\\ \nonumber &+& \left( {1\over32}g^{\alpha\mu}\gamma^\nu - {1\over32}g^{\alpha\nu}\gamma^\mu - {i\over32}\epsilon^{\alpha \mu \nu \rho}\gamma_\rho\gamma_5 \right)
\\ && \times ~[\bar c_a \gamma_\mu \gamma_5 u_a] ~  \gamma_\nu \gamma_5 \Sigma_c^{+} + \cdots \, ,
\label{eq:xi6eta}
\\[2.5mm]
\nonumber \sqrt2 \xi_7^{\alpha\beta} &\rightarrow& \Big( {1\over18}g^{\alpha\mu}g^{\beta\nu} - {1\over72}g^{\alpha\beta}g^{\mu\nu} + {i\over72}g^{\alpha\mu}\sigma^{\beta\nu}
\\ && + {i\over72}g^{\alpha\nu}\sigma^{\beta\mu} \Big) ~ [\bar c_a \gamma_\mu u_a] ~ \gamma_\nu \gamma_5 \Sigma_c^{+} + \cdots \, .
\label{eq:xi7eta}
\end{eqnarray}

\subsection{Decay analyses}
\label{sec:decayetaxi}

Based on the Fierz rearrangements derived in the previous subsection, we study strong decay properties of $\bar D^{(*)0} \Sigma_c^{*+}$ and $D^{(*)-} \Sigma_c^{*++}$ molecular states in this subsection. As an example, we shall first investigate $|\bar D^0 \Sigma_c^{*+}; 3/2^- \rangle$ through the $\eta_4$ current and the Fierz rearrangements given in Eqs.~(\ref{eq:eta4theta}) and (\ref{eq:eta4etaxi}). Others will be similarly investigated. The obtained results will be combined in Sec.~\ref{sec:isospin} to further study $\bar D^{(*)} \Sigma_c^*$ molecular states of $I=1/2$.

\subsubsection{$\eta_4 \rightarrow \theta / \eta / \xi$}
\label{sec:decayeta4}

%
\begin{figure*}[hbt]
\begin{center}
\subfigure[~\mbox{$\eta \rightarrow \theta$}]{\includegraphics[width=0.32\textwidth]{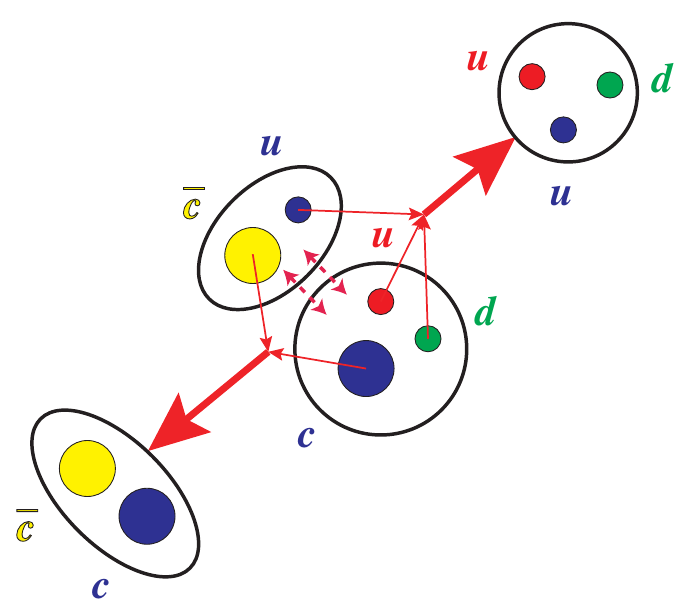}}
\subfigure[~\mbox{$\eta \rightarrow \eta$}]{\includegraphics[width=0.32\textwidth]{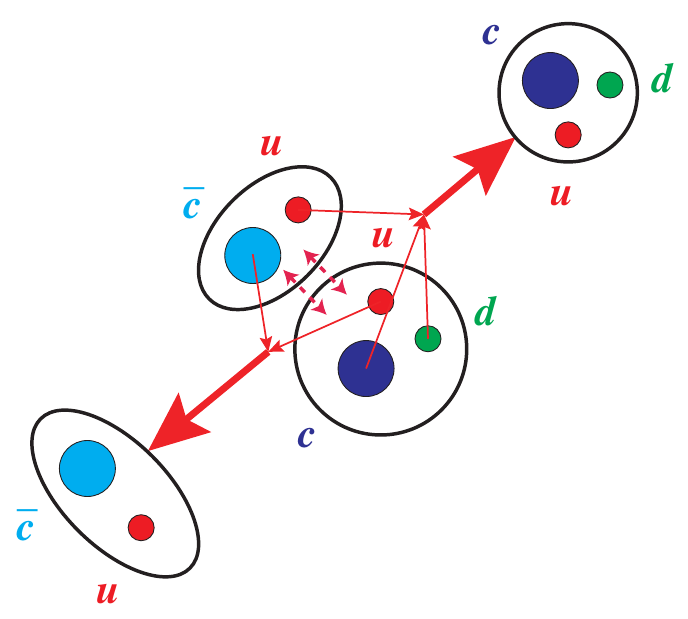}}
\subfigure[~\mbox{$\eta \rightarrow \xi$}]{\includegraphics[width=0.32\textwidth]{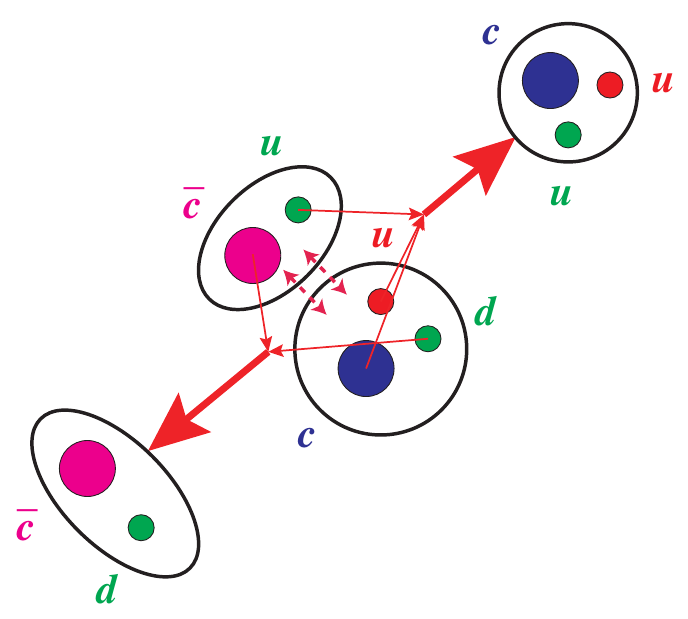}}
\caption{Fall-apart decays of $\bar D^{(*)0} \Sigma_c^{(*)+}$ molecular states, investigated through the $\eta$ currents. There are three possible decay processes: a) $\eta \rightarrow \theta$, b) $\eta \rightarrow \eta$, and c) $\eta \rightarrow \xi$. Their probabilities are the same (33\%), if only considering the color degree of freedom. Taken from Ref.~\cite{Chen:2020pac}.}
\label{fig:decay}
\end{center}
\end{figure*}
%

As an example, we investigate $|\bar D^0 \Sigma_c^{*+}; 3/2^- \rangle$ through the $\eta_4$ current and the Fierz rearrangements given in Eqs.~(\ref{eq:eta4theta}) and (\ref{eq:eta4etaxi}).

First we study Eq.~(\ref{eq:eta4theta}). As depicted in Fig.~\ref{fig:decay}(a), when the $\bar c_a$ and $c_e$ quarks meet each other and the other three quarks meet together at the same time, $|\bar D^0 \Sigma_c^{*+}; 3/2^- \rangle$ can decay into one charmonium meson and one light baryon:
\begin{eqnarray}
\nonumber && \left[\delta^{ab} \bar c_a u_b\right] ~ \left[\epsilon^{cde} u_c d_d c_e\right]
\\ \nonumber &\xlongequal{\rm color}& {1\over3}\delta^{ae} \epsilon^{bcd} ~ \bar c_au_b ~ u_c d_d c_e + \cdots
\\ \nonumber &\xlongequal{\rm Fierz}& {1\over3} ~ \left[\delta^{ae}\bar c_a c_e\right] ~ \left[\epsilon^{bcd} u_c d_d u_b\right] + \cdots .
\\
\label{eq:change}
\end{eqnarray}
Especially, we need to apply the Fierz rearrangement in the first and third steps to interchange both the color and Dirac indices of the $u_b$ and $c_e$ quark fields.

The above decay process can be described by the Fierz rearrangement given in Eq.~(\ref{eq:eta4theta}), from which we extract the following two decay channels that are kinematically allowed:
\begin{enumerate}

\item The decay of $|\bar D^0 \Sigma_c^{*+}; 3/2^- \rangle$ into $\eta_c p$ is contributed by $[\bar c_a \gamma_\mu \gamma_5 c_a]N$:
\begin{eqnarray}
&& \langle \bar D^0 \Sigma_c^{*+}; 3/2^-(q) ~|~\eta_c(q_1)~p(q_2) \rangle
\\ \nonumber &\approx& {a_4}~ i f_{\eta_c} f_p q_1^\mu ~ \bar u^\alpha \left( - {1\over32} g_{\alpha\mu} \gamma_5 - {i\over96} \sigma_{\alpha\mu} \gamma_5 \right) u_p \, ,
\end{eqnarray}
where $u_\alpha$ and $u_p$ are spinors of $|\bar D^0 \Sigma_c^{*+}; 3/2^- \rangle$ and proton, respectively; $a_4$ is an overall factor related to the coupling of $\eta_4$ to $|\bar D^0 \Sigma_c^{*+}; 3/2^- \rangle$ and the dynamical process of Fig.~\ref{fig:decay}(a).

\item The decay of $|\bar D^0 \Sigma_c^{*+}; 3/2^- \rangle$ into $J/\psi p$ is contributed by both $[\bar c_a \gamma_\mu c_a]N$ and $[\bar c_a \sigma_{\mu\nu} c_a]N$:
\begin{eqnarray}
&& \langle \bar D^0 \Sigma_c^{*+}; 3/2^-(q) | J/\psi(q_1,\epsilon_1)~p(q_2) \rangle
\\ \nonumber &\approx& {a_4}~ m_{J/\psi} f_{J/\psi} f_p \epsilon_1^\mu ~ \bar u^\alpha \left( - {1\over32} g_{\alpha\mu} - {i\over96} \sigma_{\alpha\mu} \right) u_p
\\ \nonumber &+& {a_4}~ if^T_{J/\psi} f_p ~ \left(q_1^\mu \epsilon_1^\nu - q_1^\nu \epsilon_1^\mu \right)
\\ \nonumber && ~~~~~~~~~~~ \times \bar u^\alpha \left( {i\over48} g_{\alpha\mu}\gamma_\nu + {1\over96} \epsilon_{\alpha\mu\nu\rho} \gamma^\rho \gamma_5 \right) u_p \, .
\end{eqnarray}

\end{enumerate}

Then we study Eq.~(\ref{eq:eta4etaxi}). As depicted in Fig.~\ref{fig:decay}(b), when the $\bar c_a$ and $u_c$ quarks meet each other and the other three quarks meet together at the same time, $|\bar D^0 \Sigma_c^{*+}; 3/2^- \rangle$ can decay into one charmed meson and one charmed baryon. Similarly, we can study the decay process depicted in Fig.~\ref{fig:decay}(c). These two processes can be described by the Fierz rearrangement given in Eq.~(\ref{eq:eta4etaxi}), from which we extract only one decay channel that is kinematically allowed:
\begin{enumerate}

\item[3.] The decay of $|\bar D^0 \Sigma_c^{*+}; 3/2^- \rangle$ into $\bar D^{*0} \Lambda_c^+$ is contributed by $[\bar c_a \gamma_\mu u_a]\Lambda_c^+$:
\begin{eqnarray}
&& \langle \bar D^0 \Sigma_c^{*+}; 3/2^-(q) ~|~\bar D^{*0}(q_1,\epsilon_1)~\Lambda_c^+(q_2) \rangle
\\ \nonumber &\approx& {b_4}~ m_{D^*} f_{D^*} f_{\Lambda_c} \epsilon_1^\mu ~ \bar u^\alpha \left( {1\over16}g_{\alpha\mu} + {i\over48}\sigma_{\alpha\mu} \right) u_{\Lambda_c} \, ,
\end{eqnarray}
where $u_{\Lambda_c}$ is the Dirac spinor of the $\Lambda_c^+$; $b_4$ is an overall factor related to the coupling of $\eta_4$ to $|\bar D^0 \Sigma_c^{*+}; 3/2^- \rangle$ and the dynamical processes of Fig.~\ref{fig:decay}(b,c).

\end{enumerate}

Simply assuming the mass of $|\bar D^0 \Sigma_c^{*+}; 3/2^- \rangle$ to be about $M_{D} + M_{\Sigma_c^*} \approx 4385$~MeV, we summarize the above decay amplitudes to obtain the following partial decay widths:
\begin{eqnarray}
\nonumber \Gamma(|\bar D^0 \Sigma_c^{*+}; 3/2^- \rangle \to \eta_c p ) &=& a_4^2 ~ 42 ~{\rm GeV}^7 \, ,
\\[1mm] \nonumber \Gamma(|\bar D^0 \Sigma_c^{*+}; 3/2^- \rangle \to J/\psi p ) &=& a_4^2 ~ 60 ~{\rm GeV}^7 \, ,
\\[1mm] \nonumber \Gamma(|\bar D^0 \Sigma_c^{*+}; 3/2^- \rangle \to \bar D^{*0} \Lambda_c^+ ) &=& b_4^2 ~ 1.5 \times 10^{4}~{\rm GeV}^7 \, .
\\ \label{result:eta4}
\end{eqnarray}

There are two different terms, $A \equiv [\bar c_a \gamma_\mu c_a]N$ and $B \equiv [\bar c_a \sigma_{\mu\nu} c_a]N$, both of which can contribute to the decay of $|\bar D^0 \Sigma_c^{*+}; 3/2^- \rangle$ into $J/\psi p$. Let us see their individual contributions:
\begin{eqnarray}
\nonumber \Gamma(|\bar D^0 \Sigma_c^{*+}; 3/2^- \rangle \to J/\psi p )\big|_A &=& a_4^2 ~ 1.0 \times 10^{4}~{\rm GeV}^7 \, ,
\\[1mm]
\nonumber \Gamma(|\bar D^0 \Sigma_c^{*+}; 3/2^- \rangle \to J/\psi p )\big|_B &=& a_4^2 ~ 1.1 \times 10^{4}~{\rm GeV}^7 \, .
\\
\end{eqnarray}
Hence, their contributions are at the same level, but they almost cancel each other out, suggesting that their interference is important. However, the phase angle between them, {\it i.e.}, the phase angle between the two coupling constants $f_{J/\psi}$ and $f_{J/\psi}^T$, can not be well determined in the present study. We shall investigate its relevant (theoretical) uncertainty in Appendix~\ref{app:phase}.

\subsubsection{$\xi_4 \rightarrow \theta / \eta$}
\label{sec:decayxi4}

To study $|D^{-} \Sigma_c^{*++}; 3/2^- \rangle$, we use the $\xi_4$ current and the Fierz rearrangements given in Eqs.~(\ref{eq:xi4theta}) and (\ref{eq:xi4eta}). Assuming its mass to be the same as $|\bar D^{0} \Sigma_c^{*+}; 3/2^- \rangle$, we obtain the following partial decay widths:
\begin{eqnarray}
\nonumber \Gamma(|D^{-} \Sigma_c^{*++}; 3/2^- \rangle \to \eta_c p ) &=& a_4^2 ~ 84 ~{\rm GeV}^7 \, ,
\\[1mm] \nonumber \Gamma(|D^{-} \Sigma_c^{*++}; 3/2^- \rangle \to J/\psi p ) &=& a_4^2 ~ 120 ~{\rm GeV}^7 \, ,
\\[1mm] \nonumber \Gamma(|D^{-} \Sigma_c^{*++}; 3/2^- \rangle \to \bar D^{*0} \Lambda_c^+ ) &=& b_4^2 ~ 3.0 \times 10^{4}~{\rm GeV}^7 \, .
\\ \label{result:xi4}
\end{eqnarray}
Here we have used the same overall factors $a_4$ and $b_4$ as those for the $\eta_4$ current.

\subsubsection{$\eta_5 \rightarrow \theta / \eta / \xi$}
\label{sec:decayeta5}

To study $|\bar D^{*0} \Sigma_c^{*+}; 1/2^- \rangle$, we use the $\eta_5$ current and the Fierz rearrangements given in Eqs.~(\ref{eq:eta5theta}) and (\ref{eq:eta5etaxi}). Assuming its mass to be about $M_{D^*} + M_{\Sigma_c^*} \approx 4527$~MeV, we obtain the following partial decay widths:
\begin{eqnarray}
\nonumber \Gamma(|\bar D^{*0} \Sigma_c^{*+}; 1/2^- \rangle \to \eta_c p ) &=& a_5^2 ~ 3.3 \times 10^{5} ~{\rm GeV}^7 \, ,
\\[1mm] \nonumber \Gamma(|\bar D^{*0} \Sigma_c^{*+}; 1/2^- \rangle \to J/\psi p ) &=& a_5^2 ~ 1.0 \times 10^{4} ~{\rm GeV}^7 \, ,
\\[1mm] \nonumber \Gamma(|\bar D^{*0} \Sigma_c^{*+}; 1/2^- \rangle \to \chi_{c0} p ) &=& a_5^2 ~ 3.2 \times 10^{3} ~{\rm GeV}^7 \, ,
\\[1mm] \nonumber \Gamma(|\bar D^{*0} \Sigma_c^{*+}; 1/2^- \rangle \to \chi_{c1} p ) &=& a_5^2 ~ 1.1 \times 10^{3} ~{\rm GeV}^7 \, ,
\\[1mm] \nonumber \Gamma(|\bar D^{*0} \Sigma_c^{*+}; 1/2^- \rangle \to h_{c} p ) &=& a_5^2 ~ 220 ~{\rm GeV}^7 \, ,
\\[1mm] \nonumber \Gamma(|\bar D^{*0} \Sigma_c^{*+}; 1/2^- \rangle \to \bar D^{0} \Lambda_c^+ ) &=& b_5^2 ~ 3.5 \times 10^{5}~{\rm GeV}^7 \, ,
\\[1mm] \nonumber \Gamma(|\bar D^{*0} \Sigma_c^{*+}; 1/2^- \rangle \to \bar D^{*0} \Lambda_c^+ ) &=& b_5^2 ~ 1.6 \times 10^{4}~{\rm GeV}^7 \, ,
\\[1mm] \nonumber \Gamma(|\bar D^{*0} \Sigma_c^{*+}; 1/2^- \rangle \to \bar D^{0} \Sigma_c^+ ) &=& b_5^2 ~ 1.4 \times 10^{4}~{\rm GeV}^7 \, ,
\\[1mm] \nonumber \Gamma(|\bar D^{*0} \Sigma_c^{*+}; 1/2^- \rangle \to D^{-} \Sigma_c^{++} ) &=& b_5^2 ~ 2.9 \times 10^{4}~{\rm GeV}^7 \, ,
\\[1mm] \nonumber \Gamma(|\bar D^{*0} \Sigma_c^{*+}; 1/2^- \rangle \to \bar D^{*0} \Sigma_c^+ ) &=& b_5^2 ~ 3.3 \times 10^{4}~{\rm GeV}^7 \, ,
\\[1mm] \nonumber \Gamma(|\bar D^{*0} \Sigma_c^{*+}; 1/2^- \rangle \to D^{*-} \Sigma_c^{++} ) &=& b_5^2 ~ 6.6 \times 10^{4}~{\rm GeV}^7 \, ,
\\ \label{result:eta5}
\end{eqnarray}
where $a_5$ and $b_5$ are two overall factors.

\subsubsection{$\xi_5 \rightarrow \theta / \eta$}
\label{sec:decayxi5}

To study $|D^{*-} \Sigma_c^{*++}; 1/2^- \rangle$, we use the $\xi_5$ current and the Fierz rearrangements given in Eqs.~(\ref{eq:xi5theta}) and (\ref{eq:xi5eta}). Assuming its mass to be the same as $|\bar D^{*0} \Sigma_c^{*+}; 1/2^- \rangle$, we obtain the following partial decay widths:
\begin{eqnarray}
\nonumber \Gamma(|D^{*-} \Sigma_c^{*++}; 1/2^- \rangle \to \eta_c p ) &=& a_5^2 ~ 6.5 \times 10^{5} ~{\rm GeV}^7 \, ,
\\[1mm] \nonumber \Gamma(|D^{*-} \Sigma_c^{*++}; 1/2^- \rangle \to J/\psi p ) &=& a_5^2 ~ 2.1 \times 10^{4} ~{\rm GeV}^7 \, ,
\\[1mm] \nonumber \Gamma(|D^{*-} \Sigma_c^{*++}; 1/2^- \rangle \to \chi_{c0} p ) &=& a_5^2 ~ 6.4 \times 10^{3} ~{\rm GeV}^7 \, ,
\\[1mm] \nonumber \Gamma(|D^{*-} \Sigma_c^{*++}; 1/2^- \rangle \to \chi_{c1} p ) &=& a_5^2 ~ 2.1 \times 10^{3} ~{\rm GeV}^7 \, ,
\\[1mm] \nonumber \Gamma(|D^{*-} \Sigma_c^{*++}; 1/2^- \rangle \to h_{c} p ) &=& a_5^2 ~ 450 ~{\rm GeV}^7 \, ,
\\[1mm] \nonumber \Gamma(|D^{*-} \Sigma_c^{*++}; 1/2^- \rangle \to \bar D^{0} \Lambda_c^+ ) &=& b_5^2 ~ 7.0 \times 10^{5}~{\rm GeV}^7 \, ,
\\[1mm] \nonumber \Gamma(|D^{*-} \Sigma_c^{*++}; 1/2^- \rangle \to \bar D^{*0} \Lambda_c^+ ) &=& b_5^2 ~ 3.1 \times 10^{4}~{\rm GeV}^7 \, ,
\\[1mm] \nonumber \Gamma(|D^{*-} \Sigma_c^{*++}; 1/2^- \rangle \to \bar D^{0} \Sigma_c^+ ) &=& b_5^2 ~ 2.9 \times 10^{4}~{\rm GeV}^7 \, ,
\\[1mm] \nonumber \Gamma(|D^{*-} \Sigma_c^{*++}; 1/2^- \rangle \to \bar D^{*0} \Sigma_c^+ ) &=& b_5^2 ~ 6.6 \times 10^{4}~{\rm GeV}^7 \, .
\\ \label{result:xi5}
\end{eqnarray}

\subsubsection{$\eta_6 \rightarrow \theta / \eta / \xi$}
\label{sec:decayeta6}

To study $|\bar D^{*0} \Sigma_c^{*+}; 3/2^- \rangle$, we use the $\eta_6$ current and the Fierz rearrangements given in Eqs.~(\ref{eq:eta6theta}) and (\ref{eq:eta6etaxi}). Assuming its mass to be the same as $|\bar D^{*0} \Sigma_c^{*+}; 1/2^- \rangle$, we obtain the following partial decay widths:
\begin{eqnarray}
\nonumber \Gamma(|\bar D^{*0} \Sigma_c^{*+}; 3/2^- \rangle \to \eta_c p ) &=& a_6^2 ~ 750 ~{\rm GeV}^7 \, ,
\\[1mm] \nonumber \Gamma(|\bar D^{*0} \Sigma_c^{*+}; 3/2^- \rangle \to J/\psi p ) &=& a_6^2 ~ 1.2 \times 10^{5} ~{\rm GeV}^7 \, ,
\\[1mm] \nonumber \Gamma(|\bar D^{*0} \Sigma_c^{*+}; 3/2^- \rangle \to \chi_{c0} p ) &=& a_6^2 ~ 960 ~{\rm GeV}^7 \, ,
\\[1mm] \nonumber \Gamma(|\bar D^{*0} \Sigma_c^{*+}; 3/2^- \rangle \to \bar D^{*0} \Lambda_c^+ ) &=& b_6^2 ~ 4.5 \times 10^{4}~{\rm GeV}^7 \, ,
\\[1mm] \nonumber \Gamma(|\bar D^{*0} \Sigma_c^{*+}; 3/2^- \rangle \to \bar D^{0} \Sigma_c^+ ) &=& b_6^2 ~ 36~{\rm GeV}^7 \, ,
\\[1mm] \nonumber \Gamma(|\bar D^{*0} \Sigma_c^{*+}; 3/2^- \rangle \to D^{-} \Sigma_c^{++} ) &=& b_6^2 ~ 71~{\rm GeV}^7 \, ,
\\[1mm] \nonumber \Gamma(|\bar D^{*0} \Sigma_c^{*+}; 3/2^- \rangle \to \bar D^{*0} \Sigma_c^+ ) &=& b_6^2 ~ 4.3 \times 10^{4}~{\rm GeV}^7 \, ,
\\[1mm] \nonumber \Gamma(|\bar D^{*0} \Sigma_c^{*+}; 3/2^- \rangle \to D^{*-} \Sigma_c^{++} ) &=& b_6^2 ~ 8.7 \times 10^{4}~{\rm GeV}^7 \, ,
\\ \label{result:eta6}
\end{eqnarray}
where $a_6$ and $b_6$ are two overall factors.

\subsubsection{$\xi_6 \rightarrow \theta / \eta$}
\label{sec:decayxi6}

To study $|D^{*-} \Sigma_c^{*++}; 3/2^- \rangle$, we use the $\xi_6$ current and the Fierz rearrangements given in Eqs.~(\ref{eq:xi6theta}) and (\ref{eq:xi6eta}). Assuming its mass to be the same as $|\bar D^{*0} \Sigma_c^{*+}; 1/2^- \rangle$, we obtain the following partial decay widths:
\begin{eqnarray}
\nonumber \Gamma(|D^{*-} \Sigma_c^{*++}; 3/2^- \rangle \to \eta_c p ) &=& a_6^2 ~ 1.5 \times 10^{3} ~{\rm GeV}^7 \, ,
\\[1mm] \nonumber \Gamma(|D^{*-} \Sigma_c^{*++}; 3/2^- \rangle \to J/\psi p ) &=& a_6^2 ~ 2.3 \times 10^{5} ~{\rm GeV}^7 \, ,
\\[1mm] \nonumber \Gamma(|D^{*-} \Sigma_c^{*++}; 3/2^- \rangle \to \chi_{c0} p ) &=& a_6^2 ~ 1.9 \times 10^{3} ~{\rm GeV}^7 \, ,
\\[1mm] \nonumber \Gamma(|D^{*-} \Sigma_c^{*++}; 3/2^- \rangle \to \bar D^{*0} \Lambda_c^+ ) &=& b_6^2 ~ 9.1 \times 10^{4}~{\rm GeV}^7 \, ,
\\[1mm] \nonumber \Gamma(|D^{*-} \Sigma_c^{*++}; 3/2^- \rangle \to \bar D^{0} \Sigma_c^+ ) &=& b_6^2 ~ 71~{\rm GeV}^7 \, ,
\\[1mm] \nonumber \Gamma(|D^{*-} \Sigma_c^{*++}; 3/2^- \rangle \to \bar D^{*0} \Sigma_c^+ ) &=& b_6^2 ~ 8.7 \times 10^{4}~{\rm GeV}^7 \, .
\\ \label{result:xi6}
\end{eqnarray}

\subsubsection{$\eta_7 \rightarrow \theta / \eta / \xi$ and $\xi_7 \rightarrow \theta / \eta$}
\label{sec:decayeta7}

To study $|\bar D^{*0} \Sigma_c^{*+}; 5/2^- \rangle$, we use the $\eta_7$ current and the Fierz rearrangements given in Eqs.~(\ref{eq:eta7theta}) and (\ref{eq:eta7etaxi}), but we do not obtain any non-zero decay channel. This state probably mainly decays into the spin-1 mesons and spin-3/2 baryons, such as $J/\psi N^*$ and $D^* \Sigma_c^*$, etc. However, these final states are not investigated in the present study. The same results are obtained for $|D^{*-} \Sigma_c^{*++}; 5/2^- \rangle$.

\subsection{Isospin analyses}
\label{sec:isospin}

In this subsection we collect the results calculated in the previous subsection to further study decay properties of $\bar D^{(*)} \Sigma_c^*$ molecular states with $I=1/2$.

Combining the results of Sec.~\ref{sec:decayeta4} and Sec.~\ref{sec:decayxi4}, we obtain the following partial decay widths for $|\bar D \Sigma_c^{*}; 3/2^- \rangle$ of $I=1/2$:
\begin{eqnarray}
\nonumber \Gamma(|\bar D \Sigma_c^{*}; 3/2^- \rangle \to \eta_c p ) &=& a_4^2 ~ 130 ~{\rm GeV}^7 \, ,
\\[1mm] \nonumber \Gamma(|\bar D \Sigma_c^{*}; 3/2^- \rangle \to J/\psi p ) &=& a_4^2 ~ 180 ~{\rm GeV}^7 \, ,
\\[1mm] \nonumber \Gamma(|\bar D \Sigma_c^{*}; 3/2^- \rangle \to \bar D^{*0} \Lambda_c^+ ) &=& b_4^2 ~ 4.5 \times 10^{4}~{\rm GeV}^7 \, .
\\
\label{result:etaxi4}
\end{eqnarray}
Combining the results of Sec.~\ref{sec:decayeta5} and Sec.~\ref{sec:decayxi5}, we obtain the following partial decay widths for $|\bar D^* \Sigma_c^{*}; 1/2^- \rangle$ of $I=1/2$:
\begin{eqnarray}
\nonumber \Gamma(|\bar D^{*} \Sigma_c^{*}; 1/2^- \rangle \to \eta_c p ) &=& a_5^2 ~ 9.8 \times 10^{5} ~{\rm GeV}^7 \, ,
\\[1mm] \nonumber \Gamma(|\bar D^{*} \Sigma_c^{*}; 1/2^- \rangle \to J/\psi p ) &=& a_5^2 ~ 3.1 \times 10^{4} ~{\rm GeV}^7 \, ,
\\[1mm] \nonumber \Gamma(|\bar D^{*} \Sigma_c^{*}; 1/2^- \rangle \to \chi_{c0} p ) &=& a_5^2 ~ 9.5 \times 10^{3} ~{\rm GeV}^7 \, ,
\\[1mm] \nonumber \Gamma(|\bar D^{*} \Sigma_c^{*}; 1/2^- \rangle \to \chi_{c1} p ) &=& a_5^2 ~ 3.2 \times 10^{3} ~{\rm GeV}^7 \, ,
\\[1mm] \nonumber \Gamma(|\bar D^{*} \Sigma_c^{*}; 1/2^- \rangle \to h_{c} p ) &=& a_5^2 ~ 670 ~{\rm GeV}^7 \, ,
\\[1mm] \nonumber \Gamma(|\bar D^{*} \Sigma_c^{*}; 1/2^- \rangle \to \bar D^{0} \Lambda_c^+ ) &=& b_5^2 ~ 1.1 \times 10^{6}~{\rm GeV}^7 \, ,
\\[1mm] \nonumber \Gamma(|\bar D^{*} \Sigma_c^{*}; 1/2^- \rangle \to \bar D^{*0} \Lambda_c^+ ) &=& b_5^2 ~ 4.7 \times 10^{4}~{\rm GeV}^7 \, ,
\\[1mm] \nonumber \Gamma(|\bar D^{*} \Sigma_c^{*}; 1/2^- \rangle \to \bar D^{0} \Sigma_c^+ ) &=& b_5^2 ~ 4.8 \times 10^{3}~{\rm GeV}^7 \, ,
\\[1mm] \nonumber \Gamma(|\bar D^{*} \Sigma_c^{*}; 1/2^- \rangle \to D^{-} \Sigma_c^{++} ) &=& b_5^2 ~ 9.6 \times 10^{3}~{\rm GeV}^7 \, ,
\\[1mm] \nonumber \Gamma(|\bar D^{*} \Sigma_c^{*}; 1/2^- \rangle \to \bar D^{*0} \Sigma_c^+ ) &=& b_5^2 ~ 1.1 \times 10^{4}~{\rm GeV}^7 \, ,
\\[1mm] \nonumber \Gamma(|\bar D^{*} \Sigma_c^{*}; 1/2^- \rangle \to D^{*-} \Sigma_c^{++} ) &=& b_5^2 ~ 2.2 \times 10^{4}~{\rm GeV}^7 \, .
\\ \label{result:etaxi5}
\end{eqnarray}
Combining the results of Sec.~\ref{sec:decayeta6} and Sec.~\ref{sec:decayxi6}, we obtain the following partial decay widths for $|\bar D^* \Sigma_c^{*}; 3/2^- \rangle$ of $I=1/2$:
\begin{eqnarray}
\nonumber \Gamma(|\bar D^{*} \Sigma_c^{*}; 3/2^- \rangle \to \eta_c p ) &=& a_6^2 ~ 2.2 \times 10^{3} ~{\rm GeV}^7 \, ,
\\[1mm] \nonumber \Gamma(|\bar D^{*} \Sigma_c^{*}; 3/2^- \rangle \to J/\psi p ) &=& a_6^2 ~ 3.5 \times 10^{5} ~{\rm GeV}^7 \, ,
\\[1mm] \nonumber \Gamma(|\bar D^{*} \Sigma_c^{*}; 3/2^- \rangle \to \chi_{c0} p ) &=& a_6^2 ~ 2.9 \times 10^{3} ~{\rm GeV}^7 \, ,
\\[1mm] \nonumber \Gamma(|\bar D^{*} \Sigma_c^{*}; 3/2^- \rangle \to \bar D^{*0} \Lambda_c^+ ) &=& b_6^2 ~ 1.4 \times 10^{5}~{\rm GeV}^7 \, ,
\\[1mm] \nonumber \Gamma(|\bar D^{*} \Sigma_c^{*}; 3/2^- \rangle \to \bar D^{0} \Sigma_c^+ ) &=& b_6^2 ~ 12~{\rm GeV}^7 \, ,
\\[1mm] \nonumber \Gamma(|\bar D^{*} \Sigma_c^{*}; 3/2^- \rangle \to D^{-} \Sigma_c^{++} ) &=& b_6^2 ~ 24~{\rm GeV}^7 \, ,
\\[1mm] \nonumber \Gamma(|\bar D^{*} \Sigma_c^{*}; 3/2^- \rangle \to \bar D^{*0} \Sigma_c^+ ) &=& b_6^2 ~ 1.4 \times 10^{4}~{\rm GeV}^7 \, ,
\\[1mm] \nonumber \Gamma(|\bar D^{*} \Sigma_c^{*}; 3/2^- \rangle \to D^{*-} \Sigma_c^{++} ) &=& b_6^2 ~ 2.9 \times 10^{4}~{\rm GeV}^7 \, .
\\ \label{result:etaxi6}
\end{eqnarray}
We do not obtain any non-zero decay channel for $|\bar D^* \Sigma_c^{*}; 5/2^- \rangle$ of $I=1/2$. This state probably mainly decays into spin-1 mesons and spin-3/2 baryons, such as $J/\psi N^*$ and $\bar D^* \Sigma_c^*$, etc. However, these final states are not investigated in the present study.

Decay properties of $\bar D^{(*)} \Sigma_c$ molecular states have been investigated in Ref.~\cite{Chen:2020pac}, including $|\bar D \Sigma_c; 1/2^- \rangle$, $|\bar D^{*} \Sigma_c; 1/2^- \rangle$, and $|\bar D^{*} \Sigma_c; 3/2^- \rangle$. There we used them to explain the $P_c(4312)^+$, $P_c(4440)^+$, and $P_c(4457)^+$, respectively. However, we shall find that the $P_c(4440)^+$ and $P_c(4457)^+$ can be better interpreted in our framework as $|\bar D^{*} \Sigma_c; 3/2^- \rangle$ and $|\bar D^{*} \Sigma_c; 1/2^- \rangle$, respectively/inversely.

Accordingly, in this paper we assume masses of $|\bar D \Sigma_c; 1/2^- \rangle$, $|\bar D^{*} \Sigma_c; 1/2^- \rangle$, and $|\bar D^{*} \Sigma_c; 3/2^- \rangle$ to be $M_{P_c(4312)^+} = 4311.9$~MeV, $M_{P_c(4457)^+} = 4457.3$~MeV, and $M_{P_c(4440)^+} = 4440.3$~MeV, respectively. Redoing all the calculations, we summarize the results here, and note that: a) some errors were detected in the results of Ref.~\cite{Chen:2020pac} when calculating $\Gamma(|\bar D^* \Sigma_c; 1/2^- \rangle \to J/\psi p )$, and b) different notations are used here for overall factors.

We extract for $|\bar D \Sigma_c; 1/2^- \rangle$ of $I=1/2$ that:
\begin{eqnarray}
\nonumber \Gamma(|\bar D \Sigma_c; 1/2^- \rangle \to \eta_c p ) &=& a_1^2 ~ 3.2 \times 10^{5}~{\rm GeV}^7 \, ,
\\[1mm] \nonumber \Gamma(|\bar D \Sigma_c; 1/2^- \rangle \to J/\psi p ) &=& a_1^2 ~ 8.5 \times 10^{4}~{\rm GeV}^7 \, ,
\\[1mm] \nonumber \Gamma(|\bar D \Sigma_c; 1/2^- \rangle \to \bar D^{*0} \Lambda_c^+ ) &=& b_1^2 ~ 5.9 \times 10^{4}~{\rm GeV}^7 \, .
\\ \label{result:etaxi1}
\end{eqnarray}
We extract for $|\bar D^{*} \Sigma_c; 1/2^- \rangle$ of $I=1/2$ that:
\begin{eqnarray}
\nonumber \Gamma(|\bar D^{*} \Sigma_c; 1/2^- \rangle \to \eta_c p ) &=& a_2^2 ~ 1.8 \times 10^{5}~{\rm GeV}^7  \, ,
\\[1mm] \nonumber \Gamma(|\bar D^{*} \Sigma_c; 1/2^- \rangle \to J/\psi p ) &=& a_2^2 ~5.1 \times 10^{5}~{\rm GeV}^7  \, ,
\\[1mm] \nonumber \Gamma(|\bar D^{*} \Sigma_c; 1/2^- \rangle \to \chi_{c0} p ) &=& a_2^2 ~ 8.0 \times 10^{3}~{\rm GeV}^7  \, ,
\\[1mm] \nonumber \Gamma(|\bar D^{*} \Sigma_c; 1/2^- \rangle \to \chi_{c1} p ) &=& a_2^2 ~ 200~{\rm GeV}^7  \, ,
\\[1mm] \nonumber \Gamma(|\bar D^{*} \Sigma_c; 1/2^- \rangle \to \bar D^{0} \Lambda_c^+ ) &=& b_2^2 ~ 1.7 \times 10^{6}~{\rm GeV}^7  \, ,
\\[1mm] \nonumber \Gamma(|\bar D^{*} \Sigma_c; 1/2^- \rangle \to \bar D^{*0} \Lambda_c^+ ) &=& b_2^2 ~ 6.0 \times 10^{5}~{\rm GeV}^7  \, ,
\\[1mm] \nonumber \Gamma(|\bar D^{*} \Sigma_c; 1/2^- \rangle \to \bar D^{0} \Sigma_c^+ ) &=& b_2^2 ~ 5.9 \times 10^{4}~{\rm GeV}^7  \, ,
\\[1mm] \nonumber \Gamma(|\bar D^{*} \Sigma_c; 1/2^- \rangle \to D^{-} \Sigma_c^{++} ) &=& b_2^2 ~ 1.2 \times 10^{5}~{\rm GeV}^7 \, .
\\ \label{result:etaxi2}
\end{eqnarray}
We extract for $|\bar D^{*} \Sigma_c; 3/2^- \rangle$ of $I=1/2$ that:
\begin{eqnarray}
\nonumber \Gamma(|\bar D^{*} \Sigma_c; 3/2^- \rangle \to \eta_c p ) &=& a_3^2 ~ 670~{\rm GeV}^7  \, ,
\\[1mm] \nonumber \Gamma(|\bar D^{*} \Sigma_c; 3/2^- \rangle \to J/\psi p ) &=& a_3^2 ~ 1.4 \times 10^{5}~{\rm GeV}^7  \, ,
\\[1mm] \nonumber \Gamma(|\bar D^{*} \Sigma_c; 3/2^- \rangle \to \bar D^{*0} \Lambda_c^+ ) &=& b_3^2 ~ 4.6 \times 10^{4}~{\rm GeV}^7  \, ,
\\[1mm] \nonumber \Gamma(|\bar D^{*} \Sigma_c; 3/2^- \rangle \to \bar D^{0} \Sigma_c^+ ) &=& b_3^2 ~ 1.4 ~{\rm GeV}^7  \, ,
\\[1mm] \nonumber \Gamma(|\bar D^{*} \Sigma_c; 3/2^- \rangle \to D^{-} \Sigma_c^{++} ) &=& b_3^2 ~ 2.7 ~{\rm GeV}^7 \, .
\\ \label{result:etaxi3}
\end{eqnarray}

We use the above partial decay widths to further derive their corresponding relative branching ratios. The obtained results are summarized in Table~\ref{tab:width}, where a new parameter $t \equiv {b_i^2 / a_i^2}$ ($i=1\cdots7$) is introduced to measure which processes happen more easily, the process depicted in Fig.~\ref{fig:decay}(a) or the processes depicted in Fig.~\ref{fig:decay}(b,c). We shall discuss these results in Sec.~\ref{sec:summary}.

\section{Summary and discussions}
\label{sec:summary}

\begin{table*}[hbt]
\begin{center}
\renewcommand{\arraystretch}{1.6}
\caption{Relative branching ratios of $\bar D^{(*)} \Sigma_c^{(*)}$ hadronic molecular states and their relative production rates in $\Lambda_b^0$ decays. In the 2rd-12th columns we show branching ratios relative to the $J/\psi p$ channel, such as ${\mathcal{B}(P_c \to \eta_c p)\over\mathcal{B}(P_c \to J/\psi p)}$ in the 3rd column. The parameter $t \equiv {b_i^2 / a_i^2}$ ($i=1\cdots7$) is introduced to measure which processes happen more easily, the process depicted in Fig.~\ref{fig:decay}(a) or the processes depicted in Fig.~\ref{fig:decay}(b,c). In the 13th column we show the ratio $\mathcal{R}_1(P_c) \equiv {\mathcal{B}\left(\Lambda_b^0 \rightarrow P_c K^- \right) \over \mathcal{B}\left(\Lambda_b^0 \rightarrow |\bar D^* \Sigma_c \rangle_{3/2^-} K^- \right)}$, and in the 14th column we show the ratio $\mathcal{R}_2(P_c) \equiv { \mathcal{B}(\Lambda_b^0 \to P_c K^- \to J/\psi p K^-) \over \mathcal{B}(\Lambda_b^0 \to |\bar D^* \Sigma_c \rangle_{3/2^-} K^- \to J/\psi p K^-) }$. In order to calculate $\mathcal{R}_2$: a) we have simply assumed $t=1$, and b) we have neglected all the spin-3/2 baryons that $P_c$ can decay to, such as the $J/\psi N^*$ and $\bar D \Sigma_c^*$ final states, etc.}
\begin{tabular}{c || c | c | c | c | c || c | c | c | c | c | c || c | c}
\hline\hline
\multirow{2}{*}{Configuration} & \multicolumn{11}{c||}{Decay Channels} & \multicolumn{2}{c}{Productions}
\\ \cline{2-14}
& $J/\psi p$ & $\eta_c p$ & $\chi_{c0} p$ & $\chi_{c1} p$ & $h_c p$
& $\bar D^{0} \Lambda_c^+$ & $\bar D^{*0} \Lambda_c^+$ & $\bar D^{0} \Sigma_c^+$ & $D^{-} \Sigma_c^{++}$ & $\bar D^{*0} \Sigma_c^+$ & $D^{*-} \Sigma_c^{++}$
& \,$\mathcal{R}_1$\, & \,$\mathcal{R}_2$\,
\\ \hline\hline
$|\bar D \Sigma_c; 1/2^- \rangle$         & $1$  & $3.8$ & -- & -- & --             & -- & $0.69t$ & -- & -- & -- & --                                 &  $8.2$   &  $2.0$
\\ \hline
$|\bar D^{*} \Sigma_c; 1/2^- \rangle$     & $1$  & $0.35$ & $0.016$ & $10^{-4}$ &-- & $3.4t$ & $1.2t$ & $0.12t$ & $0.23t$ & -- & --                    &  $1.2$   &  $0.25$
\\ \hline
$|\bar D^{*} \Sigma_c; 3/2^- \rangle$     & $1$  & $0.005$ & -- & -- & --           & -- & $0.34t$ & $10^{-5}t$ & $10^{-5}t$ & -- & --                 &  $\bf1$  &  $\bf1$
\\ \hline
$|\bar D \Sigma_c^*; 3/2^- \rangle$       & $1$  & $0.70$ & -- & -- & --            & -- & $250t$ & -- & -- & -- & --                                  &  --      &  --
\\ \hline
$|\bar D^* \Sigma_c^*; 1/2^- \rangle$     & $1$  & $31$ & $0.30$ & $0.10$ & $0.02$  & $34t$ & $1.5t$ & $0.15t$ & $0.30t$ & $0.35t$ & $0.70t$           &  $4.8$   &  $0.09$
\\ \hline
$|\bar D^* \Sigma_c^*; 3/2^- \rangle$     & $1$  & $0.006$ & -- & $0.008$ & --      & -- & $0.39t$ & $10^{-5}t$ & $10^{-4}t$ & $0.04t$ & $0.08t$       &  $0.18$  &  $0.16$
\\ \hline
$|\bar D^* \Sigma_c^*; 5/2^- \rangle$     &  \multicolumn{5}{c||}{--}               &  \multicolumn{6}{c||}{--}                                        &  --      &  --
\\ \hline\hline
\end{tabular}
\label{tab:width}
\end{center}
\end{table*}

In this paper we systematically investigate the seven possibly existing $\bar D^{(*)} \Sigma_c^{(*)}$ hadronic molecular states of $I=1/2$, including $\bar D \Sigma_c$ of $J^P = {1\over2}^-$, $\bar D^* \Sigma_c$ of $J^P = {1\over2}^-/{3\over2}^-$, $\bar D \Sigma_c^*$ of $J^P = {3\over2}^-$, and $\bar D^* \Sigma_c^*$ of $J^P = {1\over2}^-/{3\over2}^-/{5\over2}^-$.

Firstly, we systematically construct their corresponding interpolating currents, and calculate their masses and decay constants using QCD sum rules. The results are summarized in Table~\ref{tab:mass}, supporting the interpretations of $P_c(4312)^+$, $P_c(4440)^+$, and $P_c(4457)^+$~\cite{Aaij:2019vzc} as the $\bar D \Sigma_c$ and $\bar D^* \Sigma_c$ molecular states. However, the accuracy of our sum rule results is not good enough to distinguish/indentify them. To better understand them, we further study their production and decay properties. The decay constants $f_X$ extracted using QCD sum rules are important input parameters.

Secondly, we use the current algebra to study productions of $\bar D^{(*)} \Sigma_c^{(*)}$ molecular states in $\Lambda_b^0$ decays. We derive the relative production rates
\begin{equation}
\mathcal{R}_1(P_c) \equiv {\mathcal{B}\left(\Lambda_b^0 \rightarrow P_c K^- \right) \over \mathcal{B}\left(\Lambda_b^0 \rightarrow |\bar D^* \Sigma_c \rangle_{3/2^-} K^- \right)} \, ,
\end{equation}
and the obtained results are summarized in Table~\ref{tab:width}.

Thirdly, we use the Fierz rearrangement of the Dirac and color indices to study decay properties of $\bar D^{(*)} \Sigma_c^{*}$ molecular states, including their decays into charmonium mesons and spin-1/2 light baryons as well as charmed mesons and spin-1/2 charmed baryons, such as $J/\psi p$ and $\bar D \Lambda_c$, etc. We calculate their relative branching ratios, and the obtained results are also summarized in Table~\ref{tab:width}. The parameter $t \equiv {b_i^2 / a_i^2}$ ($i=1\cdots7$) is introduced to measure which processes happen more easily, the process depicted in Fig.~\ref{fig:decay}(a) or the processes depicted in Fig.~\ref{fig:decay}(b,c). Generally speaking, the exchange of one light quark with another light quark may be easier than its exchange with another heavy quark~\cite{Landau}, so it can be the case that $t \geq 1$.

In Table~\ref{tab:width} we simply assume $t=1$ to further calculate the ratio $\mathcal{R}_1$ in the $J/\psi p$ mass spectrum, that is
\begin{equation}
\mathcal{R}_2(P_c) \equiv { \mathcal{B}(\Lambda_b^0 \to P_c K^- \to J/\psi p K^-) \over \mathcal{B}(\Lambda_b^0 \to |\bar D^* \Sigma_c \rangle_{3/2^-} K^- \to J/\psi p K^-) } \, .
\end{equation}
In order to calculate this ratio, we have neglected all the spin-3/2 baryons that $P_c$ can decay to, such as the $J/\psi N^*$ and $\bar D \Sigma_c^*$ final states, etc.

Before drawing conclusions, we would like to note:
\begin{itemize}

\item When studying masses and decay constants of $\bar D^{(*)} \Sigma_c^{(*)}$ molecular states through QCD sum rules, we calculate two-point correlation functions at the quark-gluon level as inputs, while masses of charmed mesons and baryons at the hadron level are not used as input parameters. Accordingly, the uncertainty/accuracy is moderate but not enough to extract the binding energy. This means that our sum rule results can only suggest but not determine: a) whether these $\bar D^{(*)} \Sigma_c^{(*)}$ molecular states exist or not, and b) whether they are bound states or resonance states. Instead, we need to assume their existence, then we can use the extracted decay constants to further study their production and decay properties.

\item When studying relative production rates of $\bar D^{(*)} \Sigma_c^{(*)}$ molecular states in $\Lambda_b^0$ decays through the current algebra, we only investigate the hidden-charm pentaquark currents that can couple to these states through $S$-wave, {\it i.e.}, $J_{1\cdots7}$ defined in Eqs.~(\ref{def:current}-\ref{def:xi7}). There may exist some other currents coupling to these states through $P$-wave, which are not taken into account in the present study. Accordingly, $|\bar D \Sigma_c^*; 3/2^- \rangle$ and $|\bar D^* \Sigma_c^*; 5/2^- \rangle$ may still be produced in $\Lambda_b^0$ decays through these ``$P$-wave'' currents. Besides, their omissions produce some theoretical uncertainties.

\item When studying decay properties of $\bar D^{(*)} \Sigma_c^{*}$ molecular states through the Fierz rearrangement, we only consider the leading-order fall-apart decays described by color-singlet-color-singlet meson-baryon currents, but neglect the $\mathcal{O}(\alpha_s)$ corrections described by color-octet-color-octet meson-baryon currents, so there can be other possible decay channels. Besides, we do not consider light/charmed baryon fields of $J=3/2$, so we can not study their decays into the $J/\psi N^*$ and $\bar D \Sigma_c^*$ final states, etc. However, we have kept all the light/charmed baryon fields that couple to ground-state light/charmed baryons of $J^P=1/2^+$, so their decays into these final states are well investigated in this paper.

\end{itemize}

Now we can generally discuss about our uncertainties. The uncertainty of our QCD sum rule results is moderate, while uncertainties of relative branching ratios as well as the two ratios $\mathcal{R}_1$ and $\mathcal{R}_2$ are much larger. In the present study we work under the naive factorization scheme, so our uncertainties are significantly larger than the well-developed QCD factorization scheme~\cite{Beneke:1999br,Beneke:2000ry,Beneke:2001ev}, whose uncertainty is at the 5\% level when investigating conventional (heavy) hadrons~\cite{Li:2020rcg}. On the other hand, in this paper we only calculate the ratios, which significantly reduces our uncertainties. Accordingly, we roughly estimate the uncertainty of relative branching ratios to be at the $X^{+100\%}_{-~50\%}$ level. Due to the omission of the ``$P$-wave'' pentaquark currents, the uncertainty of the ratio $\mathcal{R}_1$ is roughly estimated to be at the $X^{+200\%}_{-~67\%}$ level. We further roughly estimate the uncertainty of the ratio $\mathcal{R}_2$ to be at the $X^{+300\%}_{-~75\%}$ level (or even larger due to the assumption of $t=1$ and the omission of spin-3/2 baryons that $P_c$ can decay to).

Finally, we can draw conclusions using the results summarized in Table~\ref{tab:width}. The LHCb experiment~\cite{Aaij:2019vzc} discovered the $P_c(4312)^+$, $P_c(4440)^+$, and $P_c(4457)^+$, and at the same time measured their relative contributions $\mathcal{R} \equiv \mathcal{B}(\Lambda^0_b \to P_c^+ K^-)\mathcal{B}(P_c^+ \to J/\psi p)/\mathcal{B}(\Lambda^0_b \to J/\psi p K^-)$ to be:
\begin{eqnarray}
\nonumber \mathcal{R}(P_c(4312)^+) &=& 0.30 \pm 0.07 ^{+0.34}_{-0.09} \% \, ,
\\ \mathcal{R}(P_c(4440)^+) &=& 1.11 \pm 0.33 ^{+0.22}_{-0.10} \% \, ,
\label{eq:lhcbratio}
\\ \nonumber \mathcal{R}(P_c(4457)^+) &=& 0.53 \pm 0.16 ^{+0.15}_{-0.13} \% \, ,
\end{eqnarray}
from which we can derive
\begin{eqnarray}
{\mathcal{R}(P_c(4312)^+) \over \mathcal{R}(P_c(4440)^+)} &=& 0.27^{+0.32}_{-0.14} \, ,
\\ \nonumber {\mathcal{R}(P_c(4457)^+) \over \mathcal{R}(P_c(4440)^+)} &=& 0.48^{+0.25}_{-0.25} \, .
\end{eqnarray}
These two values are roughly consistent with our results that
\begin{eqnarray}
\nonumber \mathcal{R}_2(|\bar D \Sigma_c; 1/2^- \rangle) &=& {\mathcal{R}_2(|\bar D \Sigma_c; 1/2^- \rangle) \over \mathcal{R}_2(|\bar D^* \Sigma_c; 3/2^- \rangle)} \approx 2.0 \, ,
\\ \nonumber \mathcal{R}_2(|\bar D^* \Sigma_c; 1/2^- \rangle) &=& {\mathcal{R}_2(|\bar D^* \Sigma_c; 1/2^- \rangle) \over \mathcal{R}_2(|\bar D^* \Sigma_c; 3/2^- \rangle)} \approx 0.25 \, ,
\\
\end{eqnarray}
given their uncertainties to be roughly at the $X^{+300\%}_{-~75\%}$ level.

Therefore, our result supports the interpretations of $P_c(4312)^+$, $P_c(4440)^+$, and $P_c(4457)^+$ as $\bar D \Sigma_c$ of $J^P = {1/2}^-$, $\bar D^* \Sigma_c$ of $J^P = {3/2}^-$, and $\bar D^* \Sigma_c$ of $J^P = {1/2}^-$, respectively. For completeness, we also investigate the interpretations of $P_c(4440)^+$ and $P_c(4457)^+$ as the $\bar D^* \Sigma_c$ molecular states of $J^P = {1/2}^-$ and ${3/2}^-$ respectively, and the results are given in Appendix~\ref{app:inverse}.

Our results suggest that the $\bar D^* \Sigma_c^*$ molecular states of $J^P = 1/2^-$ and $3/2^-$ are also possible to be observed in the $J/\psi p$ invariant mass spectrum of the $\Lambda_b^0 \to J/\psi p K^-$ decays, and their relative contributions are estimated to be
\begin{eqnarray}
\nonumber { \mathcal{B}(\Lambda_b^0 \to  |\bar D^* \Sigma_c^* \rangle_{1/2^-} K^- \to J/\psi p K^-) \over \mathcal{B}(\Lambda^0_b \to J/\psi p K^-) } &\approx&  0.1\% \, ,
\\ \nonumber { \mathcal{B}(\Lambda_b^0 \to  |\bar D^* \Sigma_c^* \rangle_{3/2^-} K^- \to J/\psi p K^-) \over \mathcal{B}(\Lambda^0_b \to J/\psi p K^-) } &\approx&  0.2\% \, .
\\
\end{eqnarray}
Their relative branching ratios to the $\eta_c p$, $\chi_{c0} p$, $\chi_{c1} p$, $h_c p$, $\bar D^{0} \Lambda_c^+$, $\bar D^{*0} \Lambda_c^+$, $\bar D^{0} \Sigma_c^+$, $D^{-} \Sigma_c^{++}$, $\bar D^{*0} \Sigma_c^+$, and $D^{*-} \Sigma_c^{++}$ final states are also given for future experimental searches.

\section*{Acknowledgments}

This project is supported by the National Natural Science Foundation of China under Grants No.~11722540 and No.~12075019,
the Jiangsu Provincial Double-Innovation Program under Grant No.~JSSCRC2021488,
and
the Fundamental Research Funds for the Central Universities.

\appendix

\begin{widetext}
\section{Spectral densities}
\label{app:ope}

In this appendix we list the spectral densities $\rho_{1\cdots7}(s)$ extracted for the currents $J_{1\cdots7}$. In the following expressions, $\mathcal{F}(s) = \FF(s)$, $\mathcal{H}(s) = \HH(s)$, and the integration limits are $\alpha_{min}=\frac{1-\sqrt{1-4m_c^2/s}}{2}$, $\alpha_{max}=\frac{1+\sqrt{1-4m_c^2/s}}{2}$, $\beta_{min}=\frac{\alpha m_c^2}{\alpha s-m_c^2}$, and $\beta_{max}=1-\alpha$.

The spectral density $\rho_{1}(s)$ extracted for the current $J_{1}$ is
\begin{eqnarray}
\nonumber \rho_{1}(s) &=& m_c         \left( \rho^{pert}_{1a}(s) + \rho^{\qq}_{1a}(s) + \rho^{\GGa}_{1a}(s)+ \rho^{\qGqa}_{1a}(s) + \rho^{\qq^2}_{1a}(s)  + \rho^{\qq\qGqa}_{1a}(s)+ \rho^{\qGqa^2}_{1a}(s) + \rho^{\qq^3}_{1a}(s) \right)
\\ \nonumber &+& q\!\!\!\slash ~~ \left( \rho^{pert}_{1b}(s) + \rho^{\qq}_{1b}(s) + \rho^{\GGa}_{1b}(s)+ \rho^{\qGqa}_{1b}(s) + \rho^{\qq^2}_{1b}(s)  + \rho^{\qq\qGqa}_{1b}(s)+ \rho^{\qGqa^2}_{1b}(s) + \rho^{\qq^3}_{1b}(s) \right) \, ,
\\ \label{ope:J1}
\end{eqnarray}
where
\begin{eqnarray}
\nonumber \rho^{pert}_{1a}(s) &=& \dab \Bigg\{ \mathcal{F}(s)^5 \times             \frac{13 (1 - \alpha - \beta)^3}{983040 \pi ^8 \alpha ^5 \beta ^4}                      \Bigg\} \, ,
\non
\rho^{\qq}_{1a}(s) &=& {m_c \qq } \dab \Bigg\{ \mathcal{F}(s)^3 \times       \frac{-(1 - \alpha - \beta)^2}{768 \pi ^6 \alpha ^3 \beta ^3}                   \Bigg\} \, ,
\non
\rho^{\GGa}_{1a}(s) &=& {\GGb } \dab\Bigg\{ m_c^2 \mathcal{F}(s)^2    \times  \frac{13 (1 - \alpha - \beta)^3 \left(\alpha ^3+\beta ^3\right)}{1179648 \pi ^8 \alpha ^5 \beta ^4}
\non && ~~~~~~ + \mathcal{F}(s)^3 \times                   \frac{(\alpha +\beta -1) \left(80 \alpha ^3+\alpha ^2 (206 \beta -79)+\alpha  \left(28 \beta ^2-27 \beta -1\right)-26 (\beta -1)^2 \beta \right)}{2359296 \pi ^8 \alpha ^5 \beta ^3}                 \Bigg\} \, ,
\non
\rho^{\qGqa}_{1a}(s) &=& {m_c\qGqb } \dab \Bigg\{ \mathcal{F}(s)^2 \times   \frac{ (1 - \alpha - \beta) \left(14 \alpha ^2+2 \alpha  (15 \beta -7)+(\beta -1) \beta \right)}{8192 \pi ^6 \alpha ^3 \beta ^3}           \Bigg\} \, ,
\non
\rho^{\qq^2}_{1a}(s)&=& {\qq^2 } \dab \Bigg\{ \mathcal{F}(s)^2 \times     \frac{-29}{1536 \pi ^4 \alpha ^2 \beta }             \Bigg\} \, ,
\non
\rho^{\qq\qGqa}_{1a}(s)&=& {\qq\qGqb } \int^{\alpha_{max}}_{\alpha_{min}}d\alpha  \Bigg\{ \int^{\beta_{max}}_{\beta_{min}}d\beta \Bigg\{ \mathcal{F}(s) \times   \frac{-6 \alpha -29 \beta}{3072 \pi ^4 \alpha ^2 \beta }            \Bigg\}
+  \mathcal{H}(s) \times       \frac{55 }{3072 \pi ^4 \alpha }          \Bigg\} \, ,
\non
\rho^{\qGqa^2}_{1a}(s)&=& {\qGqb^2 } \Bigg\{\int^{\alpha_{max}}_{\alpha_{min}}d\alpha \Bigg\{ \frac{52 \alpha ^2-75 \alpha +29}{12288 \pi ^4 \alpha } \Bigg\}
+ \int^{1}_{0}d\alpha \Bigg\{ m_c^2  \delta\left(s - {m_c^2 \over \alpha(1-\alpha)}\right) \times    \frac{-13}{6144 \pi ^4 \alpha }       \Bigg\}\Bigg\} \, ,
\non
\rho^{\qq^3}_{1a}(s)&=& {m_c\qq^3 } \int^{\alpha_{max}}_{\alpha_{min}}d\alpha \Bigg\{     \frac{13 }{288 \pi ^2}                \Bigg\} \, ,
\non
\rho^{pert}_{1b}(s) &=& \dab \Bigg\{ \mathcal{F}(s)^5 \times                 \frac{13 (1 - \alpha - \beta)^3}{491520 \pi ^8 \alpha ^4 \beta ^4}                 \Bigg\} \, ,
\non
\rho^{\qq}_{1b}(s) &=& {m_c \qq } \dab \Bigg\{ \mathcal{F}(s)^3 \times           \frac{-29 (1 - \alpha - \beta)^2}{12288 \pi ^6 \alpha ^2 \beta ^3}                  \Bigg\} \, ,
\non
\rho^{\GGa}_{1b}(s) &=& {\GGb } \dab\Bigg\{ m_c^2 \mathcal{F}(s)^2    \times     \frac{13 (1 - \alpha - \beta)^3 \left(\alpha ^3+\beta ^3\right)}{589824 \pi ^8 \alpha ^4 \beta ^4}
\non && ~~~~~~ + \mathcal{F}(s)^3 \times                                     \frac{(\alpha +\beta -1) \left(167 \alpha ^2+\alpha  (223 \beta -166)+80 \beta ^2-79 \beta -1\right)}{2359296 \pi ^8 \alpha ^3 \beta ^3}                   \Bigg\} \, ,
\non
\rho^{\qGqa}_{1b}(s) &=& {m_c\qGqb } \dab \Bigg\{ \mathcal{F}(s)^2 \times    \frac{ (1 - \alpha - \beta) \left(110 \alpha ^2+\alpha  (217 \beta -110)+3 (\beta -1) \beta \right)}{32768 \pi ^6 \alpha ^2 \beta ^3}           \Bigg\} \, ,
\non
\rho^{\qq^2}_{1b}(s)&=& {\qq^2 } \dab \Bigg\{ \mathcal{F}(s)^2 \times      \frac{-1}{96 \pi ^4 \alpha  \beta }                \Bigg\} \, ,
\non
\rho^{\qq\qGqa}_{1b}(s)&=& {\qq\qGqb } \int^{\alpha_{max}}_{\alpha_{min}}d\alpha \Bigg\{ \int^{\beta_{max}}_{\beta_{min}}d\beta \Bigg\{ \mathcal{F}(s) \times    \frac{-5 \alpha -15 \beta }{3072 \pi ^4 \alpha  \beta }             \Bigg\}
+  \mathcal{H}(s) \times    \frac{31 }{3072 \pi ^4}        \Bigg\} \, ,
\non
\rho^{\qGqa^2}_{1b}(s)&=& {\qGqb^2 } \Bigg\{ \int^{\alpha_{max}}_{\alpha_{min}}d\alpha \Bigg\{ \frac{30 \alpha ^2-40 \alpha +15}{12288 \pi ^4}     \Bigg\}
+ \int^{1}_{0}d\alpha \Bigg\{  m_c^2 \delta\left(s - {m_c^2 \over \alpha(1-\alpha)}\right) \times    \frac{-5 }{4096 \pi ^4}  \Bigg\} \Bigg\} \, ,
\non
\rho^{\qq^3}_{1b}(s)&=& {m_c\qq^3 } \int^{\alpha_{max}}_{\alpha_{min}}d\alpha \Bigg\{      \frac{13 \alpha }{576 \pi ^2}     \Bigg\} \, .
\end{eqnarray}

The spectral density $\rho_{2}(s)$ extracted for the current $J_{2}$ is
\begin{eqnarray}
\nonumber \rho_{2}(s) &=& m_c         \left( \rho^{pert}_{2a}(s) + \rho^{\qq}_{2a}(s) + \rho^{\GGa}_{2a}(s)+ \rho^{\qGqa}_{2a}(s) + \rho^{\qq^2}_{2a}(s)  + \rho^{\qq\qGqa}_{2a}(s)+ \rho^{\qGqa^2}_{2a}(s) + \rho^{\qq^3}_{2a}(s) \right)
\\ \nonumber &+& q\!\!\!\slash ~~ \left( \rho^{pert}_{2b}(s) + \rho^{\qq}_{2b}(s) + \rho^{\GGa}_{2b}(s)+ \rho^{\qGqa}_{2b}(s) + \rho^{\qq^2}_{2b}(s)  + \rho^{\qq\qGqa}_{2b}(s)+ \rho^{\qGqa^2}_{2b}(s) + \rho^{\qq^3}_{2b}(s) \right) \, ,
\\ \label{ope:J2}
\end{eqnarray}
where
\begin{eqnarray}
\nonumber \rho^{pert}_{2a}(s) &=& \dab \Bigg\{ \mathcal{F}(s)^5 \times             \frac{(1 -\alpha -\beta)^3}{49152 \pi ^8 \alpha ^5 \beta ^4}                      \Bigg\} \, ,
\non
\rho^{\qq}_{2a}(s) &=& {m_c \qq } \dab \Bigg\{ \mathcal{F}(s)^3 \times         \frac{-13 (1 -\alpha -\beta)^2}{3072 \pi ^6 \alpha ^3 \beta ^3}                  \Bigg\} \, ,
\non
\rho^{\GGa}_{2a}(s) &=& {\GGb } \dab\Bigg\{ m_c^2 \mathcal{F}(s)^2    \times     \frac{5 (1 - \alpha - \beta)^3 \left(\alpha ^3+\beta ^3\right)}{294912 \pi ^8 \alpha ^5 \beta ^4}
\non && ~~~~~~ + \mathcal{F}(s)^3 \times                                     \frac{(1 - \alpha - \beta) \left(32 \alpha ^3-\alpha ^2 (16 \beta +31)+\alpha  \left(-14 \beta ^2+15 \beta -1\right)+10 (\beta -1)^2 \beta \right)}{589824 \pi ^8 \alpha ^5 \beta ^3}                 \Bigg\} \, ,
\non
\rho^{\qGqa}_{2a}(s) &=& {m_c\qGqb } \dab \Bigg\{ \mathcal{F}(s)^2 \times     \frac{ (1 - \alpha - \beta) (23 \alpha +2 \beta -2)}{4096 \pi ^6 \alpha ^3 \beta ^2}            \Bigg\} \, ,
\non
\rho^{\qq^2}_{2a}(s)&=& {\qq^2 } \dab \Bigg\{ \mathcal{F}(s)^2 \times            \frac{-5}{192 \pi ^4 \alpha ^2 \beta }                \Bigg\} \, ,
\non
\rho^{\qq\qGqa}_{2a}(s)&=& {\qq\qGqb }  \int^{\alpha_{max}}_{\alpha_{min}}d\alpha \Bigg\{ \int^{\beta_{max}}_{\beta_{min}}d\beta \Bigg\{ \mathcal{F}(s) \times      \frac{7 \alpha -20 \beta}{1536 \pi ^4 \alpha ^2 \beta }                  \Bigg\}
+ \mathcal{H}(s) \times     \frac{11 }{512 \pi ^4 \alpha }             \Bigg\} \, ,
\non
\rho^{\qGqa^2}_{2a}(s)&=& {\qGqb^2 } \Bigg\{ \int^{\alpha_{max}}_{\alpha_{min}}d\alpha \Bigg\{     \frac{26 \alpha ^2-53 \alpha +20}{6144 \pi ^4 \alpha }      \Bigg\}
+ \int^{1}_{0}d\alpha \Bigg\{ m_c^2  \delta\left(s - {m_c^2 \over \alpha(1-\alpha)}\right) \times   \frac{-13}{6144 \pi ^4 \alpha }   \Bigg\} \Bigg\} \, ,
\non
\rho^{\qq^3}_{2a}(s)&=& {m_c\qq^3 } \int^{\alpha_{max}}_{\alpha_{min}}d\alpha \Bigg\{    \frac{23 }{144 \pi ^2}               \Bigg\} \, ,
\non
\rho^{pert}_{2b}(s) &=& \dab \Bigg\{ \mathcal{F}(s)^5 \times             \frac{23 (1 -\alpha -\beta)^3}{245760 \pi ^8 \alpha ^4 \beta ^4}                     \Bigg\} \, ,
\non
\rho^{\qq}_{2b}(s) &=& {m_c \qq } \dab \Bigg\{ \mathcal{F}(s)^3 \times       \frac{-5(1 -\alpha -\beta)^2}{1536 \pi ^6 \alpha ^2 \beta ^3}                \Bigg\} \, ,
\non
\rho^{\GGa}_{2b}(s) &=& {\GGb } \dab\Bigg\{ m_c^2 \mathcal{F}(s)^2    \times    \frac{23(1 - \alpha - \beta)^3 \left(\alpha ^3+\beta ^3\right)}{294912 \pi ^8 \alpha ^4 \beta ^4}
\non && ~~~~~~ + \mathcal{F}(s)^3 \times                                     \frac{(1 - \alpha - \beta) \left(\alpha ^2-\alpha  (11 \beta +1)-24 (\beta -1) \beta \right)}{196608 \pi ^8 \alpha ^3 \beta ^3}     \Bigg\} \, ,
\non
\rho^{\qGqa}_{2b}(s) &=& {m_c\qGqb } \dab \Bigg\{ \mathcal{F}(s)^2 \times    \frac{13 (1 - \alpha - \beta)}{4096 \pi ^6 \alpha  \beta ^2}                  \Bigg\} \, ,
\non
\rho^{\qq^2}_{2b}(s)&=& {\qq^2 } \dab \Bigg\{ \mathcal{F}(s)^2 \times       \frac{-13}{384 \pi ^4 \alpha  \beta }                 \Bigg\} \, ,
\non
\rho^{\qq\qGqa}_{2b}(s)&=& {\qq\qGqb }  \int^{\alpha_{max}}_{\alpha_{min}}d\alpha \Bigg\{ \int^{\beta_{max}}_{\beta_{min}}d\beta \Bigg\{ \mathcal{F}(s) \times     \frac{ - 5 \alpha - 24 \beta}{1536 \pi ^4 \alpha  \beta }              \Bigg\}
+ \mathcal{H}(s) \times    \frac{47 }{1536 \pi ^4}         \Bigg\} \, ,
\non
\rho^{\qGqa^2}_{2b}(s)&=& {\qGqb^2 } \Bigg\{ \int^{\alpha_{max}}_{\alpha_{min}}d\alpha \Bigg\{       \frac{42 \alpha ^2-61 \alpha +24}{6144 \pi ^4}      \Bigg\}
+\int^{1}_{0}d\alpha \Bigg\{  m_c^2 \delta\left(s - {m_c^2 \over \alpha(1-\alpha)}\right) \times  \frac{-7}{2048 \pi ^4}   \Bigg\} \Bigg\} \, ,
\non
\rho^{\qq^3}_{2b}(s)&=& {m_c\qq^3 } \int^{\alpha_{max}}_{\alpha_{min}}d\alpha \Bigg\{    \frac{5 \alpha  }{144 \pi ^2}         \Bigg\} \, .
\end{eqnarray}

The spectral density $\rho_{3}(s)$ extracted for the current $J_{3}$ is
\begin{eqnarray}
\nonumber \rho_{3}(s) &=& m_c         \left( \rho^{pert}_{3a}(s) + \rho^{\qq}_{3a}(s) + \rho^{\GGa}_{3a}(s)+ \rho^{\qGqa}_{3a}(s) + \rho^{\qq^2}_{3a}(s)  + \rho^{\qq\qGqa}_{3a}(s)+ \rho^{\qGqa^2}_{3a}(s) + \rho^{\qq^3}_{3a}(s) \right)
\\ \nonumber &+& q\!\!\!\slash ~~ \left( \rho^{pert}_{3b}(s) + \rho^{\qq}_{3b}(s) + \rho^{\GGa}_{3b}(s)+ \rho^{\qGqa}_{3b}(s) + \rho^{\qq^2}_{3b}(s)  + \rho^{\qq\qGqa}_{3b}(s)+ \rho^{\qGqa^2}_{3b}(s) + \rho^{\qq^3}_{3b}(s) \right) \, ,
\\ \label{ope:J3}
\end{eqnarray}
where
\begin{eqnarray}
\nonumber \rho^{pert}_{3a}(s) &=& \dab \Bigg\{ \mathcal{F}(s)^5 \times       \frac{7 (1 - \alpha - \beta)^3 (\alpha +\beta +4)}{3932160 \pi ^8 \alpha ^5 \beta ^4}                \Bigg\} \, ,
\non
\rho^{\qq}_{3a}(s) &=& {m_c \qq } \dab \Bigg\{ \mathcal{F}(s)^3 \times      \frac{- (1 - \alpha - \beta)^2 (8 \alpha +8 \beta +157)}{147456 \pi ^6 \alpha ^3 \beta ^3}          \Bigg\} \, ,
\non
\rho^{\GGa}_{3a}(s) &=& {\GGb } \dab\Bigg\{ m_c^2 \mathcal{F}(s)^2    \times     \frac{7 (1 - \alpha - \beta)^3 (\alpha +\beta +4) \left(\alpha ^3+\beta ^3\right)}{4718592 \pi ^8 \alpha ^5 \beta ^4}
\non && ~~~~~~ + \mathcal{F}(s)^3 \times                  \Bigg(               \frac{53 \alpha ^5+\alpha ^4 (530 \beta -464)+45 \alpha ^3 \left(22 \beta ^2-16 \beta +17\right)+70 \alpha ^2 \left(8 \beta ^3-3 \beta ^2-5\right)}{28311552 \pi ^8 \alpha ^5 \beta ^3}
\non && ~~~~~~~~~~~~~~~~~~~~~~~~~~~~~~~~~~~~~~~~~~~~~~~~~~~~~~ + \frac{\alpha  (\beta -1)^2 \left(5 \beta ^2+14 \beta -4\right)-42 (\beta -1)^3 \beta  (\beta +4)}{28311552 \pi ^8 \alpha ^5 \beta ^3} \Bigg)    \Bigg\} \, ,
\non
\rho^{\qGqa}_{3a}(s) &=& {m_c\qGqb } \dab \Bigg\{ \mathcal{F}(s)^2 ~    \frac{(1 - \alpha - \beta) \left(42 \alpha ^2+\alpha  (50 \beta +311)+8 \beta ^2+14 \beta -22\right)}{196608 \pi ^6 \alpha ^3 \beta ^2}           \Bigg\} \, ,
\non
\rho^{\qq^2}_{3a}(s)&=& {\qq^2 } \dab \Bigg\{ \mathcal{F}(s)^2 \times      \frac{-4 \alpha -4 \beta -1}{384 \pi ^4 \alpha ^2 \beta }               \Bigg\} \, ,
\non
\rho^{\qq\qGqa}_{3a}(s)&=& {\qq\qGqb } \dab \Bigg\{ \mathcal{F}(s) \times    \frac{4 \alpha ^2+\alpha  (49-1128 \beta )-96 \beta  (4 \beta +1)}{73728 \pi ^4 \alpha ^2 \beta }            \Bigg\}
\non
&+& {\qq\qGqb } \int^{\alpha_{max}}_{\alpha_{min}}d\alpha \Bigg\{ \mathcal{H}(s) \times     \frac{935 }{73728 \pi ^4 \alpha }            \Bigg\} \, ,
\non
\rho^{\qGqa^2}_{3a}(s)&=& {\qGqb^2 }  \int^{\alpha_{max}}_{\alpha_{min}}d\alpha \Bigg\{ \int^{\beta_{max}}_{\beta_{min}}d\beta \Bigg\{ \frac{\alpha -96 \beta}{73728 \pi ^4 \alpha }  \Bigg\}
+    \frac{546 \alpha ^2-1079 \alpha +480}{294912 \pi ^4 \alpha }  \Bigg\}
\non
&+& {\qGqb^2 } \int^{1}_{0}d\alpha \Bigg\{  m_c^2 \delta\left(s - {m_c^2 \over \alpha(1-\alpha)}\right) \times  \frac{-455 }{294912 \pi ^4 \alpha }   \Bigg\} \, ,
\non
\rho^{\qq^3}_{3a}(s)&=& {m_c\qq^3 } \int^{\alpha_{max}}_{\alpha_{min}}d\alpha \Bigg\{   \frac{9}{256 \pi ^2}           \Bigg\} \, ,
\non
\rho^{pert}_{3b}(s) &=& \dab \Bigg\{ \mathcal{F}(s)^5 \times     \frac{9 (1 - \alpha - \beta)^3 (\alpha +\beta +2)}{1310720 \pi ^8 \alpha ^4 \beta ^4}                  \Bigg\} \, ,
\non
\rho^{\qq}_{3b}(s) &=& {m_c \qq } \dab \Bigg\{ \mathcal{F}(s)^3 \times       \frac{-5 (1 - \alpha - \beta)^2}{3072 \pi ^6 \alpha ^2 \beta ^3}             \Bigg\} \, ,
\non
\rho^{\GGa}_{3b}(s) &=& {\GGb } \dab\Bigg\{ m_c^2 \mathcal{F}(s)^2    \times  \frac{3 (1 - \alpha - \beta)^3 (\alpha +\beta +2) \left(\alpha ^3+\beta ^3\right)}{524288 \pi ^8 \alpha ^4 \beta ^4} + \mathcal{F}(s)^3 \times
\non &&                                       \frac{(\alpha +\beta -1) \left(243 \alpha ^3+\alpha ^2 (673 \beta -834)+\alpha  \left(761 \beta ^2-743 \beta +588\right)+331 \beta ^3+103 \beta ^2-437 \beta +3\right)}{28311552 \pi ^8 \alpha ^3 \beta ^3}            \Bigg\} \, ,
\non
\rho^{\qGqa}_{3b}(s) &=& {m_c\qGqb } \dab \Bigg\{ \mathcal{F}(s)^2 \times    \frac{5 (1 - \alpha - \beta) (94 \alpha +3 \beta -3)}{196608 \pi ^6 \alpha ^2 \beta ^2}                      \Bigg\} \, ,
\non
\rho^{\qq^2}_{3b}(s)&=& {\qq^2 } \dab \Bigg\{ \mathcal{F}(s)^2 \times         \frac{-5 (12 \alpha +12 \beta -1)}{6144 \pi ^4 \alpha  \beta }            \Bigg\} \, ,
\non
\rho^{\qq\qGqa}_{3b}(s)&=& {\qq\qGqb } \dab \Bigg\{ \mathcal{F}(s) \times    \frac{- 24 \alpha ^2 - \alpha  (1088 \beta -61) - 4 \beta  (94 \beta -21)}{73728 \pi ^4 \alpha  \beta }        \Bigg\}
\non
&+& {\qq\qGqb } \int^{\alpha_{max}}_{\alpha_{min}}d\alpha \Bigg\{ \mathcal{H}(s) \times    \frac{661 }{73728 \pi ^4}         \Bigg\} \, ,
\non
\rho^{\qGqa^2}_{3b}(s)&=& {\qGqb^2 }  \int^{\alpha_{max}}_{\alpha_{min}}d\alpha \Bigg\{ \int^{\beta_{max}}_{\beta_{min}}d\beta  \Bigg\{ \frac{ - 3 \alpha - 47 \beta }{36864 \pi ^4} \Bigg\}
+    \frac{334 \alpha ^2-663 \alpha +292}{294912 \pi ^4}      \Bigg\}
\non
&+& {\qGqb^2 } \int^{1}_{0}d\alpha \Bigg\{  m_c^2 \delta\left(s - {m_c^2 \over \alpha(1-\alpha)}\right) \times   \frac{-331}{294912 \pi ^4}    \Bigg\} \, ,
\non
\rho^{\qq^3}_{3b}(s)&=& {m_c\qq^3 } \int^{\alpha_{max}}_{\alpha_{min}}d\alpha \Bigg\{      \frac{35 \alpha  }{2304 \pi ^2}    \Bigg\} \, .
\end{eqnarray}

The spectral density $\rho_{4}(s)$ extracted for the current $J_{4}$ is
\begin{eqnarray}
\nonumber \rho_{4}(s) &=& m_c         \left( \rho^{pert}_{4a}(s) + \rho^{\qq}_{4a}(s) + \rho^{\GGa}_{4a}(s)+ \rho^{\qGqa}_{4a}(s) + \rho^{\qq^2}_{4a}(s)  + \rho^{\qq\qGqa}_{4a}(s)+ \rho^{\qGqa^2}_{4a}(s) + \rho^{\qq^3}_{4a}(s) \right)
\\ \nonumber &+& q\!\!\!\slash ~~ \left( \rho^{pert}_{4b}(s) + \rho^{\qq}_{4b}(s) + \rho^{\GGa}_{4b}(s)+ \rho^{\qGqa}_{4b}(s) + \rho^{\qq^2}_{4b}(s)  + \rho^{\qq\qGqa}_{4b}(s)+ \rho^{\qGqa^2}_{4b}(s) + \rho^{\qq^3}_{4b}(s) \right) \, ,
\\ \label{ope:J4}
\end{eqnarray}
where
\begin{eqnarray}
\nonumber \rho^{pert}_{4a}(s) &=& \dab \Bigg\{ \mathcal{F}(s)^5 \times        \frac{13 (1 - \alpha - \beta)^3 (\alpha +\beta +4)}{15728640 \pi ^8 \alpha ^5 \beta ^4}                \Bigg\} \, ,
\non
\rho^{\qq}_{4a}(s) &=& {m_c \qq } \dab \Bigg\{ \mathcal{F}(s)^3 \times             \frac{-(1 - \alpha -\beta)^2 (14 \alpha +14 \beta +43)}{147456 \pi ^6 \alpha ^3 \beta ^3}               \Bigg\} \, ,
\non
\rho^{\GGa}_{4a}(s) &=& {\GGb } \dab\Bigg\{ m_c^2 \mathcal{F}(s)^2    \times     \frac{13 (1 - \alpha - \beta)^3 (\alpha +\beta +4) \left(\alpha ^3+\beta ^3\right)}{18874368 \pi ^8 \alpha ^5 \beta ^4}
\non
&&  + \mathcal{F}(s)^3 \times       \Bigg\{        \frac{341 \alpha ^5+\alpha ^4 (598 \beta +220)-9 \alpha ^3 \left(10 \beta ^2-130 \beta +163\right)}{113246208 \pi ^8 \alpha ^5 \beta ^3}
\non
&&     +    \frac{\alpha ^2 \left(-688 \beta ^3+714 \beta ^2-936 \beta +910\right)-\alpha  (\beta -1)^2 \left(419 \beta ^2+1152 \beta +4\right)-78 (\beta -1)^3 \beta  (\beta +4)}{113246208 \pi ^8 \alpha ^5 \beta ^3}                 \Bigg\} \, ,
\non
\rho^{\qGqa}_{4a}(s) &=& {m_c\qGqb } \dab \Bigg\{
\non
&& \mathcal{F}(s)^2 \times     \frac{(1 - \alpha - \beta) \left(164 \alpha ^3+4 \alpha ^2 (137 \beta +74)+\alpha  \left(382 \beta ^2+693 \beta -460\right)-2 \beta  \left(\beta ^2+4 \beta -5\right)\right)}{1179648 \pi ^6 \alpha ^3 \beta ^3}                  \Bigg\} \, ,
\non
\rho^{\qq^2}_{4a}(s)&=& {\qq^2 } \dab \Bigg\{ \mathcal{F}(s)^2 \times    \frac{- 4 \alpha - 4 \beta -85 }{24576 \pi ^4 \alpha ^2 \beta }             \Bigg\} \, ,
\non
\rho^{\qq\qGqa}_{4a}(s)&=& {\qq\qGqb } \dab \Bigg\{ \mathcal{F}(s) \times    \frac{- 16 \alpha ^2 - \alpha  (40 \beta +46) + \beta  (4 \beta +85)}{147456 \pi ^4 \alpha ^2 \beta }              \Bigg\}
\non
&+& {\qq\qGqb } \int^{\alpha_{max}}_{\alpha_{min}}d\alpha \Bigg\{ \mathcal{H}(s) \times    \frac{509}{147456 \pi ^4 \alpha }              \Bigg\} \, ,
\non
\rho^{\qGqa^2}_{4a}(s)&=& {\qGqb^2 }  \int^{\alpha_{max}}_{\alpha_{min}}d\alpha \Bigg\{  \int^{\beta_{max}}_{\beta_{min}}d\beta   \Bigg\{           \frac{- 4 \alpha + \beta }{147456 \pi ^4 \alpha }                       \Bigg\}
+   \frac{468 \alpha ^2-317 \alpha -89}{589824 \pi ^4 \alpha }          \Bigg\}
\non
&+& {\qGqb^2 } \int^{1}_{0}d\alpha \Bigg\{ m_c^2  \delta\left(s - {m_c^2 \over \alpha(1-\alpha)}\right) \times    \frac{-121 }{294912 \pi ^4 \alpha }      \Bigg\} \, ,
\non
\rho^{\qq^3}_{4a}(s)&=& {m_c\qq^3 } \int^{\alpha_{max}}_{\alpha_{min}}d\alpha \Bigg\{         \frac{13 }{1536 \pi ^2}          \Bigg\} \, ,
\non
\rho^{pert}_{4b}(s) &=& \dab \Bigg\{ \mathcal{F}(s)^5 \times          \frac{13 (1 - \alpha - \beta)^3 (\alpha +\beta +2)}{7864320 \pi ^8 \alpha ^4 \beta ^4}                      \Bigg\} \, ,
\non
\rho^{\qq}_{4b}(s) &=& {m_c \qq } \dab \Bigg\{ \mathcal{F}(s)^3 \times   \frac{- (1 - \alpha - \beta)^2 (112 \alpha +112 \beta +155)}{589824 \pi ^6 \alpha ^2 \beta ^3}                    \Bigg\} \, ,
\non
\rho^{\GGa}_{4b}(s) &=& {\GGb } \dab\Bigg\{ m_c^2 \mathcal{F}(s)^2    \times   \frac{13 (1 - \alpha - \beta)^3 \left(\alpha ^4+\alpha ^3 (\beta +2)+\alpha  \beta ^3+\beta ^3 (\beta +2)\right)}{9437184 \pi ^8 \alpha ^4 \beta ^4}
\non &&  + \mathcal{F}(s)^3 \times                                       \frac{5 (\alpha +\beta -1) \left(136 \alpha ^3+\alpha ^2 (176 \beta +29)+\alpha  \left(40 \beta ^2+69 \beta -166\right)-80 \beta ^2+79 \beta +1\right)}{113246208 \pi ^8 \alpha ^3 \beta ^3}            \Bigg\} \, ,
\non
\rho^{\qGqa}_{4b}(s) &=& {m_c\qGqb } \dab \Bigg\{
\non
&& \mathcal{F}(s)^2 \times   \frac{(1 - \alpha - \beta) \left(1312 \alpha ^3+\alpha ^2 (4400 \beta +358)+\alpha  \left(3088 \beta ^2+1441 \beta -1670\right)-45 (\beta -1) \beta \right)}{4718592 \pi ^6 \alpha ^2 \beta ^3}                    \Bigg\} \, ,
\non
\rho^{\qq^2}_{4b}(s)&=& {\qq^2 } \dab \Bigg\{ \mathcal{F}(s)^2 \times   \frac{- 2 \alpha - 2 \beta - 17}{6144 \pi ^4 \alpha  \beta }                      \Bigg\} \, ,
\non
\rho^{\qq\qGqa}_{4b}(s)&=& {\qq\qGqb } \dab \Bigg\{ \mathcal{F}(s) \times    \frac{- 32 \alpha ^2 - \alpha  (88 \beta +27) + 75 \beta}{147456 \pi ^4 \alpha  \beta }                \Bigg\}
\non
&+& {\qq\qGqb } \int^{\alpha_{max}}_{\alpha_{min}}d\alpha \Bigg\{ \mathcal{H}(s) \times       \frac{437 }{147456 \pi ^4}         \Bigg\} \, ,
\non
\rho^{\qGqa^2}_{4b}(s)&=& {\qGqb^2 }  \int^{\alpha_{max}}_{\alpha_{min}}d\alpha \Bigg\{ \int^{\beta_{max}}_{\beta_{min}}d\beta  \Bigg\{          \frac{-\alpha  }{18432 \pi ^4}                    \Bigg\}
+  \frac{386 \alpha ^2-252 \alpha -75}{589824 \pi ^4}              \Bigg\}
\non
&+& {\qGqb^2 } \int^{1}_{0}d\alpha \Bigg\{ m_c^2  \delta\left(s - {m_c^2 \over \alpha(1-\alpha)}\right) \times    \frac{-209 }{589824 \pi ^4}\Bigg\} \, ,
\non
\rho^{\qq^3}_{4b}(s)&=& {m_c\qq^3 } \int^{\alpha_{max}}_{\alpha_{min}}d\alpha \Bigg\{        \frac{65 \alpha }{9216 \pi ^2}            \Bigg\} \, .
\end{eqnarray}

The spectral density $\rho_{5}(s)$ extracted for the current $J_{5}$ is
\begin{eqnarray}
\nonumber \rho_{5}(s) &=& m_c         \left( \rho^{pert}_{5a}(s) + \rho^{\qq}_{5a}(s) + \rho^{\GGa}_{5a}(s)+ \rho^{\qGqa}_{5a}(s) + \rho^{\qq^2}_{5a}(s)  + \rho^{\qq\qGqa}_{5a}(s)+ \rho^{\qGqa^2}_{5a}(s) + \rho^{\qq^3}_{5a}(s) \right)
\\ \nonumber &+& q\!\!\!\slash ~~ \left( \rho^{pert}_{5b}(s) + \rho^{\qq}_{5b}(s) + \rho^{\GGa}_{5b}(s)+ \rho^{\qGqa}_{5b}(s) + \rho^{\qq^2}_{5b}(s)  + \rho^{\qq\qGqa}_{5b}(s)+ \rho^{\qGqa^2}_{5b}(s) + \rho^{\qq^3}_{5b}(s) \right) \, ,
\\ \label{ope:J5}
\end{eqnarray}
where
\begin{eqnarray}
\nonumber \rho^{pert}_{5a}(s) &=& \dab \Bigg\{ \mathcal{F}(s)^5 \times               \frac{3 (1 - \alpha - \beta)^3}{262144 \pi ^8 \alpha ^5 \beta ^4}               \Bigg\} \, ,
\non
\rho^{\qq}_{5a}(s) &=& {m_c \qq } \dab \Bigg\{ \mathcal{F}(s)^3 \times          \frac{-11(1-\alpha -\beta)^2}{16384 \pi ^6 \alpha ^3 \beta ^3}              \Bigg\} \, ,
\non
\rho^{\GGa}_{5a}(s) &=& {\GGb } \dab\Bigg\{ m_c^2 \mathcal{F}(s)^2    \times    \frac{5 (1 -\alpha -\beta)^3 \left(\alpha ^3+\beta ^3\right)}{524288 \pi ^8 \alpha ^5 \beta ^4}
\non && ~~~~~~ + \mathcal{F}(s)^3 \times                                  \frac{ (1 - \alpha -\beta) \left(4 \alpha ^3+\alpha ^2 (92 \beta -5)+\alpha  \left(94 \beta ^2-95 \beta +1\right)+30 (\beta -1)^2 \beta \right)}{3145728 \pi ^8 \alpha ^5 \beta ^3}                 \Bigg\} \, ,
\non
\rho^{\qGqa}_{5a}(s) &=& {m_c\qGqb } \dab \Bigg\{ \mathcal{F}(s)^2 \times     \frac{45 (1 - \alpha - \beta)}{65536 \pi ^6 \alpha ^2 \beta ^2}                 \Bigg\} \, ,
\non
\rho^{\qq^2}_{5a}(s)&=& {\qq^2 } \dab \Bigg\{ \mathcal{F}(s)^2 \times      \frac{-5}{1024 \pi ^4 \alpha ^2 \beta }                          \Bigg\} \, ,
\non
\rho^{\qq\qGqa}_{5a}(s)&=& {\qq\qGqb }  \int^{\alpha_{max}}_{\alpha_{min}}d\alpha \Bigg\{ \int^{\beta_{max}}_{\beta_{min}}d\beta \Bigg\{ \mathcal{F}(s) \times        \frac{-\alpha +20 \beta}{24576 \pi ^4 \alpha ^2 \beta }                    \Bigg\}
+ \mathcal{H}(s) \times   \frac{97 }{24576 \pi ^4 \alpha }               \Bigg\} \, ,
\non
\rho^{\qGqa^2}_{5a}(s)&=& {\qGqb^2 } \Bigg\{ \int^{\alpha_{max}}_{\alpha_{min}}d\alpha \Bigg\{      \frac{74 \alpha ^2-53 \alpha -20}{98304 \pi ^4 \alpha }          \Bigg\}
+ \int^{1}_{0}d\alpha \Bigg\{  m_c^2 \delta\left(s - {m_c^2 \over \alpha(1-\alpha)}\right) \times   \frac{-37}{98304 \pi ^4 \alpha }     \Bigg\}\Bigg\} \, ,
\non
\rho^{\qq^3}_{5a}(s)&=& {m_c\qq^3 } \int^{\alpha_{max}}_{\alpha_{min}}d\alpha \Bigg\{   \frac{7}{256 \pi ^2}              \Bigg\} \, ,
\non
\rho^{pert}_{5b}(s) &=& \dab \Bigg\{ \mathcal{F}(s)^5 \times              \frac{21 (1-\alpha -\beta)^3}{1310720 \pi ^8 \alpha ^4 \beta ^4}            \Bigg\} \, ,
\non
\rho^{\qq}_{5b}(s) &=& {m_c \qq } \dab \Bigg\{ \mathcal{F}(s)^3 \times   \frac{-5(1-\alpha -\beta)^2}{8192 \pi ^6 \alpha ^2 \beta ^3}               \Bigg\} \, ,
\non
\rho^{\GGa}_{5b}(s) &=& {\GGb } \dab\Bigg\{ m_c^2 \mathcal{F}(s)^2    \times   \frac{7(1 - \alpha -\beta)^3 \left(\alpha ^3+\beta ^3\right)}{524288 \pi ^8 \alpha ^4 \beta ^4}
\non && ~~~~~~ + \mathcal{F}(s)^3 \times                    \frac{(\alpha +\beta -1) \left(17 \alpha ^2-\alpha  (179 \beta +13)-64 \beta ^2+68 \beta -4\right)}{9437184 \pi ^8 \alpha ^3 \beta ^3}           \Bigg\} \, ,
\non
\rho^{\qGqa}_{5b}(s) &=& {m_c\qGqb } \dab \Bigg\{ \mathcal{F}(s)^2 \times       \frac{(35 \alpha -2 \beta +2) (1-\alpha -\beta)}{65536 \pi ^6 \alpha ^2 \beta ^2}                \Bigg\} \, ,
\non
\rho^{\qq^2}_{5b}(s)&=& {\qq^2 } \dab \Bigg\{ \mathcal{F}(s)^2 \times   \frac{-11}{2048 \pi ^4 \alpha  \beta }        \Bigg\} \, ,
\non
\rho^{\qq\qGqa}_{5b}(s)&=& {\qq\qGqb }  \int^{\alpha_{max}}_{\alpha_{min}}d\alpha \Bigg\{ \int^{\beta_{max}}_{\beta_{min}}d\beta \Bigg\{ \mathcal{F}(s) \times    \frac{-13 \alpha +16 \beta }{24576 \pi ^4 \alpha  \beta }      \Bigg\}
+ \mathcal{H}(s) \times    \frac{37 }{8192 \pi ^4}        \Bigg\} \, ,
\non
\rho^{\qGqa^2}_{5b}(s)&=& {\qGqb^2 } \Bigg\{ \int^{\alpha_{max}}_{\alpha_{min}}d\alpha \Bigg\{     \frac{90 \alpha ^2-61 \alpha -16}{98304 \pi ^4}         \Bigg\}
+ \int^{1}_{0}d\alpha \Bigg\{ m_c^2  \delta\left(s - {m_c^2 \over \alpha(1-\alpha)}\right) \times    \frac{-15}{32768 \pi ^4} \Bigg\} \Bigg\} \, ,
\non
\rho^{\qq^3}_{5b}(s)&=& {m_c\qq^3 } \int^{\alpha_{max}}_{\alpha_{min}}d\alpha \Bigg\{     \frac{5 \alpha  }{256 \pi ^2}      \Bigg\} \, .
\end{eqnarray}

The spectral density $\rho_{6}(s)$ extracted for the current $J_{6}$ is
\begin{eqnarray}
\nonumber \rho_{6}(s) &=& m_c         \left( \rho^{pert}_{6a}(s) + \rho^{\qq}_{6a}(s) + \rho^{\GGa}_{6a}(s)+ \rho^{\qGqa}_{6a}(s) + \rho^{\qq^2}_{6a}(s)  + \rho^{\qq\qGqa}_{6a}(s)+ \rho^{\qGqa^2}_{6a}(s) + \rho^{\qq^3}_{6a}(s) \right)
\\ \nonumber &+& q\!\!\!\slash ~~ \left( \rho^{pert}_{6b}(s) + \rho^{\qq}_{6b}(s) + \rho^{\GGa}_{6b}(s)+ \rho^{\qGqa}_{6b}(s) + \rho^{\qq^2}_{6b}(s)  + \rho^{\qq\qGqa}_{6b}(s)+ \rho^{\qGqa^2}_{6b}(s) + \rho^{\qq^3}_{6b}(s) \right) \, ,
\\ \label{ope:J6}
\end{eqnarray}
where
\begin{eqnarray}
\nonumber \rho^{pert}_{6a}(s) &=& \dab \Bigg\{ \mathcal{F}(s)^5 \times      \frac{7 (1 - \alpha  - \beta)^3 (\alpha +\beta +4)}{15728640 \pi ^8 \alpha ^5 \beta ^4}               \Bigg\} \, ,
\non
\rho^{\qq}_{6a}(s) &=& {m_c \qq } \dab \Bigg\{ \mathcal{F}(s)^3 \times       \frac{-5(1 - \alpha - \beta)^2 (8 \alpha +8 \beta +31)}{196608 \pi ^6 \alpha ^3 \beta ^3}           \Bigg\} \, ,
\non
\rho^{\GGa}_{6a}(s) &=& {\GGb } \dab\Bigg\{ m_c^2 \mathcal{F}(s)^2    \times   \frac{7(1 - \alpha - \beta)^3 (\alpha +\beta +4) \left(\alpha ^3+\beta ^3\right)}{18874368 \pi ^8 \alpha ^5 \beta ^4}
\non && + \mathcal{F}(s)^3 \times        \Bigg\{            \frac{-391 \alpha^5 - 2 \alpha ^4 (839 \beta +326) - 3 \alpha ^3 \left(766 \beta ^2-96 \beta -819\right) - 42 (\beta -1)^3 \beta  (\beta +4)}{113246208 \pi ^8 \alpha ^5 \beta ^3}
\non && ~~~~~~~~~~~~~~~~~~~~~ + \frac{-2 \alpha ^2 \left(584 \beta ^3-417 \beta ^2-864 \beta +697\right) - \alpha  (\beta -1)^2 \left(199 \beta ^2+546 \beta +20\right)}{113246208 \pi ^8 \alpha ^5 \beta ^3}      \Bigg\}     \Bigg\} \, ,
\non
\rho^{\qGqa}_{6a}(s) &=& {m_c\qGqb } \dab \Bigg\{ \mathcal{F}(s)^2 \times   \frac{(1 - \alpha - \beta) (110 \alpha +110 \beta +243)}{262144 \pi ^6 \alpha ^2 \beta ^2}         \Bigg\} \, ,
\non
\rho^{\qq^2}_{6a}(s)&=& {\qq^2 } \dab \Bigg\{ \mathcal{F}(s)^2 \times     \frac{4 \alpha +4 \beta -11}{2048 \pi ^4 \alpha ^2 \beta }          \Bigg\} \, ,
\non
\rho^{\qq\qGqa}_{6a}(s)&=& {\qq\qGqb } \dab \Bigg\{ \mathcal{F}(s) \times   \frac{28 \alpha ^2+\alpha  (168 \beta +67)+8 \beta  (11-4 \beta )}{98304 \pi ^4 \alpha ^2 \beta }    \Bigg\}
\non
&+& {\qq\qGqb } \int^{\alpha_{max}}_{\alpha_{min}}d\alpha \Bigg\{ \mathcal{H}(s) \times    \frac{301 }{98304 \pi ^4 \alpha }         \Bigg\} \, ,
\non
\rho^{\qGqa^2}_{6a}(s)&=& {\qGqb^2 } \int^{\alpha_{max}}_{\alpha_{min}}d\alpha \Bigg\{ \int^{\beta_{max}}_{\beta_{min}}d\beta \Bigg\{       \frac{7 \alpha -8 \beta}{98304 \pi ^4 \alpha }      \Bigg\}
+ \frac{342 \alpha ^2-381 \alpha -56}{393216 \pi ^4 \alpha }    \Bigg\}
\non
&+& {\qGqb^2 } \int^{1}_{0}d\alpha \Bigg\{  m_c^2 \delta\left(s - {m_c^2 \over \alpha(1-\alpha)}\right) \times   \frac{-133}{393216 \pi ^4 \alpha }   \Bigg\} \, ,
\non
\rho^{\qq^3}_{6a}(s)&=& {m_c\qq^3 } \int^{\alpha_{max}}_{\alpha_{min}}d\alpha \Bigg\{      \frac{25 }{1024 \pi ^2}       \Bigg\} \, ,
\non
\rho^{pert}_{6b}(s) &=& \dab \Bigg\{ \mathcal{F}(s)^5 \times                \frac{5(1 - \alpha - \beta)^3 (\alpha +\beta +2)}{1048576 \pi ^8 \alpha ^4 \beta ^4}          \Bigg\} \, ,
\non
\rho^{\qq}_{6b}(s) &=& {m_c \qq } \dab \Bigg\{ \mathcal{F}(s)^3 \times     \frac{-(1 - \alpha - \beta)^2 (16 \alpha +16 \beta +5)}{49152 \pi ^6 \alpha ^2 \beta ^3}            \Bigg\} \, ,
\non
\rho^{\GGa}_{6b}(s) &=& {\GGb } \dab\Bigg\{ m_c^2 \mathcal{F}(s)^2    \times  \frac{25 (1 - \alpha - \beta)^3 \left(\alpha ^4+\alpha ^3 (\beta +2)+\alpha  \beta ^3+\beta ^3 (\beta +2)\right)}{6291456 \pi ^8 \alpha ^4 \beta ^4}
\non && + \mathcal{F}(s)^3 \times         \Bigg\{         \frac{(1 - \alpha - \beta)  \left(261 \alpha ^3+\alpha ^2 (1799 \beta -738)+\alpha  \left(1615 \beta ^2-1117 \beta +516\right)\right)}{113246208 \pi ^8 \alpha ^3 \beta ^3}
\non && ~~~~~~~~~~~~~~~~~~~~~~~~~~~~~~~~~~~~~~~~~~~~~~~~~ +     \frac{(1 - \alpha - \beta)  \left(77 \beta ^3+761 \beta ^2-799 \beta -39\right)}{113246208 \pi ^8 \alpha ^3 \beta ^3}       \Bigg\} \Bigg\} \, ,
\non
\rho^{\qGqa}_{6b}(s) &=& {m_c\qGqb } \dab \Bigg\{ \mathcal{F}(s)^2 ~  \frac{ (1 - \alpha - \beta) \left(472 \alpha ^2+\alpha  (488 \beta -68)+16 \beta ^2-11 \beta -5\right)}{786432 \pi ^6 \alpha ^2 \beta ^2}          \Bigg\} ,
\non
\rho^{\qq^2}_{6b}(s)&=& {\qq^2 } \dab \Bigg\{ \mathcal{F}(s)^2 \times     \frac{-20 \alpha -20 \beta -45}{8192 \pi ^4 \alpha  \beta }             \Bigg\} \, ,
\non
\rho^{\qq\qGqa}_{6b}(s)&=& {\qq\qGqb } \dab \Bigg\{ \mathcal{F}(s) \times    \frac{- 24 \alpha ^2 - \alpha  (224 \beta +9) + 12 \beta  (2 \beta +7)}{98304 \pi ^4 \alpha  \beta }           \Bigg\}
\non
&+& {\qq\qGqb } \int^{\alpha_{max}}_{\alpha_{min}}d\alpha \Bigg\{ \mathcal{H}(s) \times \frac{743 }{98304 \pi ^4}     \Bigg\} \, ,
\non
\rho^{\qGqa^2}_{6b}(s)&=& {\qGqb^2 } \int^{\alpha_{max}}_{\alpha_{min}}d\alpha \Bigg\{   \int^{\beta_{max}}_{\beta_{min}}d\beta \Bigg\{     \frac{ (\beta -\alpha )}{16384 \pi ^4}       \Bigg\}
+ \frac{602 \alpha ^2-461 \alpha -108}{393216 \pi ^4}      \Bigg\}
\non
&+& {\qGqb^2 } \int^{1}_{0}d\alpha \Bigg\{ m_c^2  \delta\left(s - {m_c^2 \over \alpha(1-\alpha)}\right) \times  \frac{-353}{393216 \pi ^4} \Bigg\} \, ,
\non
\rho^{\qq^3}_{6b}(s)&=& {m_c\qq^3 } \int^{\alpha_{max}}_{\alpha_{min}}d\alpha \Bigg\{  \frac{35 \alpha  }{9216 \pi ^2}       \Bigg\} \, .
\end{eqnarray}

The spectral density $\rho_{7}(s)$ extracted for the current $J_{7}$ is
\begin{eqnarray}
\nonumber \rho_{7}(s) &=& m_c         \left( \rho^{pert}_{7a}(s) + \rho^{\qq}_{7a}(s) + \rho^{\GGa}_{7a}(s)+ \rho^{\qGqa}_{7a}(s) + \rho^{\qq^2}_{7a}(s)  + \rho^{\qq\qGqa}_{7a}(s)+ \rho^{\qGqa^2}_{7a}(s) + \rho^{\qq^3}_{7a}(s) \right)
\\ \nonumber &+& q\!\!\!\slash ~~ \left( \rho^{pert}_{7b}(s) + \rho^{\qq}_{7b}(s) + \rho^{\GGa}_{7b}(s)+ \rho^{\qGqa}_{7b}(s) + \rho^{\qq^2}_{7b}(s)  + \rho^{\qq\qGqa}_{7b}(s)+ \rho^{\qGqa^2}_{7b}(s) + \rho^{\qq^3}_{7b}(s) \right) \, ,
\\ \label{ope:J7}
\end{eqnarray}
where
\begin{eqnarray}
\nonumber \rho^{pert}_{7a}(s) &=& \dab \Bigg\{ \mathcal{F}(s)^5 \times    \frac{7 (1 - \alpha - \beta)^3 \left(3 \alpha ^2+2 \alpha  (3 \beta +7)+3 \beta ^2+14 \beta +33\right)}{88473600 \pi ^8 \alpha ^5 \beta ^4}               \Bigg\} \, ,
\non
\rho^{\qq}_{7a}(s) &=& {m_c \qq } \dab \Bigg\{ \mathcal{F}(s)^3 \times       \frac{-(1 - \alpha - \beta)^2 (10 \alpha +10 \beta +23)}{73728 \pi ^6 \alpha ^3 \beta ^3}              \Bigg\} \, ,
\non
\rho^{\GGa}_{7a}(s) &=& {\GGb } \dab\Bigg\{
\non && m_c^2 \mathcal{F}(s)^2    \times \Bigg\{ \frac{7(1 - \alpha - \beta)^3 \left(3 \alpha ^5+2 \alpha ^4 (3 \beta +7)+\alpha ^3 \left(3 \beta ^2+14 \beta +33\right)+3 \alpha ^2 \beta ^3\right)}{106168320 \pi ^8 \alpha ^5 \beta ^4}
\non && ~~~~~~~~~~~~~~~~~~~~~~~~~~~~~~~~~~~~~ +  \frac{7(1 - \alpha - \beta)^3 \left(2 \alpha  \beta ^3 (3 \beta +7)+\beta ^3 \left(3 \beta ^2+14 \beta +33\right)\right)}{106168320 \pi ^8 \alpha ^5 \beta ^4} \Bigg\}
\non && + \mathcal{F}(s)^3 \times      \Bigg\{           \frac{(\alpha +\beta -1) \left(252 \alpha ^5-\alpha ^4 (324 \beta +1273)-\alpha ^3 \left(2136 \beta ^2+3771 \beta +3733\right)\right)}{1274019840 \pi ^8 \alpha ^5 \beta ^3}
\non && + \frac{(\alpha +\beta -1) \left(\alpha ^2 \left(2544 \beta ^3+5595 \beta ^2+818 \beta -4817\right)-84 (\beta -1)^2 \beta  \left(3 \beta ^2+14 \beta +33\right)\right)}{1274019840 \pi ^8 \alpha ^5 \beta ^3}
\non && ~~~~~~~~~~~~~~~~~~~~~~~~~ + \frac{(\alpha +\beta -1) \left(-\alpha  \left(1236 \beta ^4+3769 \beta ^3+1717 \beta ^2-6785 \beta +63\right)\right)}{1274019840 \pi ^8 \alpha ^5 \beta ^3}   \Bigg\}   \Bigg\} \, ,
\non
\rho^{\qGqa}_{7a}(s) &=& {m_c\qGqb } \dab \Bigg\{
\non && \mathcal{F}(s)^2 \times      \frac{-530 \alpha ^3 - 105 \alpha ^2 (10 \beta +1) - 102 \alpha  \left(5 \beta ^2+\beta -6\right) + (\beta -1)^2 (10 \beta +23)}{1769472 \pi ^6 \alpha ^3 \beta ^2}                \Bigg\} \, ,
\non
\rho^{\qq^2}_{7a}(s)&=& {\qq^2 } \dab \Bigg\{ \mathcal{F}(s)^2 \times   \frac{- 10 \alpha - 10 \beta - 1}{3072 \pi ^4 \alpha ^2 \beta }        \Bigg\} \, ,
\non
\rho^{\qq\qGqa}_{7a}(s)&=& {\qq\qGqb } \dab \Bigg\{ \mathcal{F}(s) \times   \frac{ 10 \alpha ^2-140 \alpha  \beta +\alpha +3 \beta  (10 \beta +1) }{55296 \pi ^4 \alpha ^2 \beta }               \Bigg\}
\non
&+& {\qq\qGqb } \int^{\alpha_{max}}_{\alpha_{min}}d\alpha \Bigg\{ \mathcal{H}(s) \times   \frac{11 }{3072 \pi ^4 \alpha }      \Bigg\} \, ,
\non
\rho^{\qGqa^2}_{7a}(s)&=& {\qGqb^2 } \int^{\alpha_{max}}_{\alpha_{min}}d\alpha \Bigg\{ \int^{\beta_{max}}_{\beta_{min}}d\beta  \Bigg\{     \frac{5 \alpha + 15 \beta}{110592 \pi ^4 \alpha }            \Bigg\}
+   \frac{108 \alpha ^2-86 \alpha -33}{221184 \pi ^4 \alpha }         \Bigg\}
\non
&+& {\qGqb^2 } \int^{1}_{0}d\alpha \Bigg\{  m_c^2 \delta\left(s - {m_c^2 \over \alpha(1-\alpha)}\right) \times    \frac{-11}{24576 \pi ^4 \alpha }  \Bigg\} \, ,
\non
\rho^{\qq^3}_{7a}(s)&=& {m_c\qq^3 } \int^{\alpha_{max}}_{\alpha_{min}}d\alpha \Bigg\{    \frac{7 }{864 \pi ^2}            \Bigg\} \, ,
\non
\rho^{pert}_{7b}(s) &=& \dab \Bigg\{ \mathcal{F}(s)^5 \times      \frac{7 (1 - \alpha - \beta)^3 \left(6 \alpha ^2+\alpha  (12 \beta +13)+6 \beta ^2+13 \beta +21\right)}{58982400 \pi ^8 \alpha ^4 \beta ^4}       \Bigg\} \, ,
\non
\rho^{\qq}_{7b}(s) &=& {m_c \qq } \dab \Bigg\{ \mathcal{F}(s)^3 \times   \frac{- (1 - \alpha - \beta)^2 (13 \alpha +13 \beta +20)}{73728 \pi ^6 \alpha ^2 \beta ^3}           \Bigg\} \, ,
\non
\rho^{\qq^2}_{7b}(s)&=& {\qq^2 } \dab \Bigg\{ \mathcal{F}(s)^2 \times     \frac{- 13 \alpha - 13 \beta + 2}{3072 \pi ^4 \alpha  \beta }            \Bigg\} \, ,
\non
\rho^{\qq^3}_{7b}(s)&=& {m_c\qq^3 } \int^{\alpha_{max}}_{\alpha_{min}}d\alpha \Bigg\{      \frac{35 \alpha}{5184 \pi ^2}     \Bigg\} \, .
\end{eqnarray}
However, $\rho^{\GGa}_{7b}(s)$, $\rho^{\qGqa}_{7b}(s)$, $\rho^{\qq\qGqa}_{7b}(s)$, and $\rho^{\qGqa^2}_{7b}(s)$ are too complicated to be extracted.
\end{widetext}

\section{Uncertainties due to phase angles}
\label{app:phase}

There are two different terms, $A \equiv [\bar c_a \gamma_\mu c_a]N$ and $B \equiv [\bar c_a \sigma_{\mu\nu} c_a]N$, both of which can contribute to the decay of $|\bar D \Sigma_c^{*}; 3/2^- \rangle$ into $J/\psi p$. Their relevant effective Lagrangians are:
\begin{eqnarray}
\mathcal{L}^A_{\psi p} &=& g_A~\bar P_c^\alpha \left( t_1 g_{\alpha\mu} + t_2 \sigma_{\alpha\mu} \right) N~\psi^\mu \, ,
\\ \nonumber \mathcal{L}^B_{\psi p} &=& g_B~\bar P_c^\alpha \left( t_3 g_{\alpha\mu}\gamma_\nu + t_4 \epsilon_{\alpha\mu\nu\rho} \gamma^\rho \gamma_5 \right) N~\partial^\mu\psi^\nu \, ,
\\
\end{eqnarray}
where $t_i$ are free parameters. These two terms $A$ and $B$ can also contribute to decays of $|\bar D^{*} \Sigma_c; 1/2^- \rangle$ and $|\bar D^{*} \Sigma_c^{*}; 1/2^- \rangle$ into $J/\psi p$. Now the two effective Lagrangians are:
\begin{eqnarray}
\mathcal{L}^C_{\psi p} &=& g_C~\bar P_c \gamma_\mu \gamma_5 N~\psi^\mu \, ,
\\ \mathcal{L}^D_{\psi p} &=& g_D~\bar P_c \sigma_{\mu\nu} \gamma_5 N~\partial^\mu\psi^\nu \, .
\end{eqnarray}
There are two different terms, $C \equiv [\bar c_a \gamma_5 c_a]N$ and $D \equiv [\bar c_a \gamma_\mu \gamma_5 c_a]N$, both of which can contribute to decays of $|\bar D \Sigma_c; 1/2^- \rangle$, $|\bar D^{*} \Sigma_c; 1/2^- \rangle$, and $|\bar D^{*} \Sigma_c^{*}; 1/2^- \rangle$ into $\eta_c p$. Their relevant effective Lagrangians are:
\begin{eqnarray}
\mathcal{L}^E_{\eta_c p} &=& g_E~\bar P_c N~\eta_c \, ,
\\ \mathcal{L}^F_{\eta_c p} &=& g_F~\bar P_c \gamma_\mu N~\partial^\mu\eta_c \, .
\end{eqnarray}

There can be phase angles between $g_A/g_B$, $g_C/g_D$, and $g_E/g_F$, all of which can not be well determined in the present study. In this appendix we rotate these phase angles and redo all the calculations. Their relevant (theoretical) uncertainties are summarized in Table~\ref{tab:uncertainty}.

\begin{table*}[hbt]
\begin{center}
\renewcommand{\arraystretch}{1.6}
\caption{Relative branching ratios of $\bar D^{(*)} \Sigma_c^{(*)}$ hadronic molecular states and their relative production rates in $\Lambda_b^0$ decays. See the caption of Table~\ref{tab:width} for detailed explanations. In this table we take into account the (theoretical) uncertainties due to the phase angles between $g_A/g_B$, $g_C/g_D$, and $g_E/g_F$.}
\begin{tabular}{c || c | c | c | c | c || c | c | c | c | c | c || c | c}
\hline\hline
\multirow{2}{*}{Configuration} & \multicolumn{11}{c||}{Decay Channels} & \multicolumn{2}{c}{Productions}
\\ \cline{2-14}
& $J/\psi p$ & $\eta_c p$ & $\chi_{c0} p$ & $\chi_{c1} p$ & $h_c p$
& $\bar D^{0} \Lambda_c^+$ & $\bar D^{*0} \Lambda_c^+$ & $\bar D^{0} \Sigma_c^+$ & $D^{-} \Sigma_c^{++}$ & $\bar D^{*0} \Sigma_c^+$ & $D^{*-} \Sigma_c^{++}$
& ~$\mathcal{R}_1$~ & ~~$\mathcal{R}_2$~~
\\ \hline\hline
$|\bar D \Sigma_c; 1/2^- \rangle$         & $1$  & $0.5$-$3.8$ & -- & -- & --                     & -- & $0.69t$ & -- & -- & -- & --                                 &  $8.2$   &  $2.0$-$5.0$
\\ \hline
$|\bar D^{*} \Sigma_c; 1/2^- \rangle$     & $0.9$-$1.6$  & $0.3$-$3.1$ & $0.016$ & $10^{-4}$ &-- & $3.4t$ & $1.2t$ & $0.12t$ & $0.23t$ & -- & --                     &  $1.2$   &  $0.2$-$0.4$
\\ \hline
$|\bar D^{*} \Sigma_c; 3/2^- \rangle$     & $1$  & $0.005$ & -- & -- & --           & -- & $0.34t$ & $10^{-5}t$ & $10^{-5}t$ & -- & --                               &  $\bf1$  &  $\bf1$
\\ \hline
$|\bar D \Sigma_c^*; 3/2^- \rangle$       & $1$-$710$  & $0.70$ & -- & -- & --            & -- & $250t$ & -- & -- & -- & --                                          &  --      &  --
\\ \hline
$|\bar D^* \Sigma_c^*; 1/2^- \rangle$     & $1$-$25$  & $3$-$31$ & $0.30$ & $0.10$ & $0.02$  & $34t$ & $1.5t$ & $0.15t$ & $0.30t$ & $0.35t$ & $0.70t$                &  $4.8$   &  $0.1$-$2.4$
\\ \hline
$|\bar D^* \Sigma_c^*; 3/2^- \rangle$     & $1$  & $0.006$ & -- & $0.008$ & --      & -- & $0.39t$ & $10^{-5}t$ & $10^{-4}t$ & $0.04t$ & $0.08t$                     &  $0.18$  &  $0.16$
\\ \hline
$|\bar D^* \Sigma_c^*; 5/2^- \rangle$     &  \multicolumn{5}{c||}{--}               &  \multicolumn{6}{c||}{--}                                                      &  --      &  --
\\ \hline\hline
\end{tabular}
\label{tab:uncertainty}
\end{center}
\end{table*}

\section{Inverse interpretations}
\label{app:inverse}

In this paper we intend to interpret the $P_c(4440)^+$ and $P_c(4457)^+$ as the $\bar D^* \Sigma_c$ molecular states of $J^P = 3/2^-$ and $1/2^-$, respectively. However, they can also be interpreted as the $\bar D^* \Sigma_c$ molecular states of $J^P = 1/2^-$ and $3/2^-$, respectively. Based on the latter interpretations, we assume masses of $\bar D^{(*)} \Sigma_c^{(*)}$ molecular states to be:
\begin{eqnarray}
\nonumber M_{|\bar D \Sigma_c; 1/2^- \rangle} &=& M_{P_c(4312)^+} = 4311.9~{\rm MeV} \, ,
\\ \nonumber M_{|\bar D^{*} \Sigma_c; 1/2^- \rangle} &=& M_{P_c(4440)^+} = 4440.3~{\rm MeV} \, ,
\\ \nonumber M_{|\bar D^{*} \Sigma_c; 3/2^- \rangle} &=& M_{P_c(4457)^+} = 4457.3~{\rm MeV} \, ,
\\ M_{|\bar D \Sigma_c^{*}; 3/2^- \rangle} &\approx& M_{D} + M_{\Sigma_c^*} = 4385~{\rm MeV} \, ,
\\ \nonumber M_{|\bar D^{*} \Sigma_c^*; 1/2^- \rangle} &\approx& M_{D^*} + M_{\Sigma_c^*} = 4527~{\rm MeV} \, ,
\\ \nonumber M_{|\bar D^{*} \Sigma_c^*; 3/2^- \rangle} &\approx& M_{D^*} + M_{\Sigma_c^*} = 4527~{\rm MeV} \, ,
\\ \nonumber M_{|\bar D^{*} \Sigma_c^*; 5/2^- \rangle} &\approx& M_{D^*} + M_{\Sigma_c^*} = 4527~{\rm MeV} \, ,
\end{eqnarray}
and redo all the calculations. We summarize the obtained results in Table~\ref{tab:inverse}. Even considering the uncertainty of $\mathcal{R}_2$ to be at the $X^{+300\%}_{-~75\%}$ level, these results seem not easy to explain the relative contributions $\mathcal{R} \equiv \mathcal{B}(\Lambda^0_b \to P_c^+ K^-)\mathcal{B}(P_c^+ \to J/\psi p)/\mathcal{B}(\Lambda^0_b \to J/\psi p K^-)$ measured by LHCb~\cite{Aaij:2019vzc}, as given in Eqs.~(\ref{eq:lhcbratio}).

\begin{table*}[hbt]
\begin{center}
\renewcommand{\arraystretch}{1.6}
\caption{Relative branching ratios of $\bar D^{(*)} \Sigma_c^{(*)}$ hadronic molecular states and their relative production rates in $\Lambda_b^0$ decays. See the caption of Table~\ref{tab:width} for detailed explanations. In this table we work under then assumption that the $P_c(4440)^+$ and $P_c(4457)^+$ are interpreted as the $\bar D^* \Sigma_c$ molecular states of $J^P = 1/2^-$ and $3/2^-$, respectively.}
\begin{tabular}{c || c | c | c | c | c || c | c | c | c | c | c || c | c}
\hline\hline
\multirow{2}{*}{Configuration} & \multicolumn{11}{c||}{Decay Channels} & \multicolumn{2}{c}{Productions}
\\ \cline{2-14}
& $J/\psi p$ & $\eta_c p$ & $\chi_{c0} p$ & $\chi_{c1} p$ & $h_c p$
& $\bar D^{0} \Lambda_c^+$ & $\bar D^{*0} \Lambda_c^+$ & $\bar D^{0} \Sigma_c^+$ & $D^{-} \Sigma_c^{++}$ & $\bar D^{*0} \Sigma_c^+$ & $D^{*-} \Sigma_c^{++}$
& ~\,$\mathcal{R}_1$\,~ & ~\,$\mathcal{R}_2$\,~
\\ \hline\hline
$|\bar D \Sigma_c; 1/2^- \rangle$         & $1$  & $3.8$ & -- & -- & --             & -- & $0.69t$ & -- & -- & -- & --                                 &  $8.6$   &  $2.1$
\\ \hline
$|\bar D^{*} \Sigma_c; 1/2^- \rangle$     & $1$  & $0.36$ & $0.013$ & -- &--        & $3.4t$ & $1.2t$ & $0.11t$ & $0.22t$ & -- & --                    &  $1.3$   &  $0.28$
\\ \hline
$|\bar D^{*} \Sigma_c; 3/2^- \rangle$     & $1$  & $0.005$ & -- & $10^{-4}$ & --    & -- & $0.35t$ & $10^{-5}t$ & $10^{-5}t$ & -- & --                 &  $\bf1$  &  $\bf1$
\\ \hline
$|\bar D \Sigma_c^*; 3/2^- \rangle$       & $1$  & $0.70$ & -- & -- & --            & -- & $250t$ & -- & -- & -- & --                                  &  --      &  --
\\ \hline
$|\bar D^* \Sigma_c^*; 1/2^- \rangle$     & $1$  & $31$ & $0.30$ & $0.10$ & $0.02$  & $34t$ & $1.5t$ & $0.15t$ & $0.30t$ & $0.35t$ & $0.70t$           &  $5.0$   &  $0.10$
\\ \hline
$|\bar D^* \Sigma_c^*; 3/2^- \rangle$     & $1$  & $0.006$ & -- & $0.008$ & --      & -- & $0.39t$ & $10^{-5}t$ & $10^{-4}t$ & $0.04t$ & $0.08t$       &  $0.19$  &  $0.17$
\\ \hline
$|\bar D^* \Sigma_c^*; 5/2^- \rangle$     &  \multicolumn{5}{c||}{--}               &  \multicolumn{6}{c||}{--}                                        &  --      &  --
\\ \hline\hline
\end{tabular}
\label{tab:inverse}
\end{center}
\end{table*}

\end{document}